\documentclass[sigconf, nonacm, pdfa]{acmart}
\usepackage[a-2b]{pdfx}
\usepackage{multirow}
\usepackage{booktabs}
\usepackage{array}
\usepackage{mathrsfs}
\usepackage{bm}
\usepackage{tabularx}
\usepackage{pifont}
\usepackage{circledsteps}
\usepackage{graphicx,graphics}
\usepackage{subfigure}
\usepackage{enumitem}
\usepackage{caption} 
\usepackage{float} 
\usepackage{bbding}
\usepackage{pifont}
\usepackage{threeparttable}
\usepackage{wasysym}
\usepackage{makecell}
\usepackage{utfsym}
\usepackage{url}
\usepackage{color,xspace}
\usepackage{diagbox}
\usepackage{twemojis}
\usepackage{amsmath, bm}
\usepackage{tablefootnote}
\usepackage{pdfrender}
\usepackage{tikz}
\usepackage{etoolbox}
\newcommand{\circled}[2][]{\tikz[baseline=(char.base)]
    {\node[shape = circle, draw, inner sep = -0.3pt]
    (char) {\phantom{\ifblank{#1}{#2}{#1}}};%
    \node at (char.center) {\makebox[0pt][c]{#2}};}}
\robustify{\circled}

\newcommand*{\thincheckmark}{
  \textpdfrender{
    TextRenderingMode=FillStroke,
    LineWidth=0.5pt, 
  }{\checkmark}%
}

\newtheoremstyle{definition}  % name
{3pt}                         % space above
{3pt}                         % space below
{}                            % body font
{8pt}                            % indent amount
{\itshape}                   % theorem head font
{.}                           % punctuation after theorem head
{.5em}                        % space after theorem head
{\thmname{#1}\thmnumber{ #2}} % theorem head spec
\theoremstyle{definition}
\newtheorem{defn}{Definition}

\newtheoremstyle{example}  % name
{3pt}                         % space above
{3pt}                         % space below
{}                            % body font
{8pt}                            % indent amount
{\itshape}                   % theorem head font
{.}                           % punctuation after theorem head
{.5em}                        % space after theorem head
{\thmname{#1}\thmnumber{ #2}} % theorem head spec
\theoremstyle{example}
\newtheorem{examp}{Example}

\newcolumntype{M}[1]{>{\centering\arraybackslash}m{#1}}
\newcommand\vldbdoi{10.14778/3675034.3675051}
\newcommand\vldbpages{2617 - 2630}
\newcommand\vldbvolume{17}
\newcommand\vldbissue{10}
\newcommand\vldbyear{2024}
\newcommand\vldbauthors{Wei Ni, Xiaoye Miao, Xiangyu Zhao, Yangyang Wu, Shuwei Liang, Jianwei Yin}
\newcommand\vldbtitle{\shorttitle} 
\newcommand\vldbavailabilityurl{https://github.com/WelkinNi/Automatic-Data-Repair}
\newcommand\vldbpagestyle{empty} 

\begin{document}
\title{Automatic Data Repair: Are We Ready to Deploy? }
% [Experiment, Analysis\ \&\ Benchmark]}
% \title{Are Integrity Constraints Important for Data repair?}

%%
%% The "author" command and its associated commands are used to define the authors and their affiliations.

\author{Wei Ni$^{\ast\ddagger1}$, Xiaoye Miao$^{\ast\star2}$, Xiangyu Zhao$^{\ddagger3}$, Yangyang Wu$^{\#4}$, Shuwei Liang$^{\ast\#5}$, Jianwei Yin$^{\ast\#\dagger 6}$}
\affiliation{\institution{$^\ast$Center for Data Science, Zhejiang University, Hangzhou, China}
\institution{$^\star$The State Key Lab of Brain-Machine Intelligence, Zhejiang University, Hangzhou, China}
\institution{$^\ddagger$School of Data Science, City University of Hong Kong, Hong Kong, China}
\institution{$^{\#}$School of Software Technology, Zhejiang University, Ningbo, China}
\institution{$^\dagger$College of Computer Science, Zhejiang University, Hangzhou, China} }
\affiliation{\{$^{1}$wei.ni, $^{2}$miaoxy, $^{4}$zjuwuyy, $^{5}$lsw5221\}@zju.edu.cn \qquad $^{3}$xianzhao@cityu.edu.hk \quad $^{6}$zjuyjw@cs.zju.edu.cn}

% \affiliation{%
%   \country{Iceland}

% \affiliation{%
% \email{vb@rocquencourt.com}


\begin{abstract}
% \zxy{Begin with a strong opening sentence that highlights the significance of the topic. For instance, "The integrity and accuracy of data are paramount in today's data-driven world."}
% \nw{Fixed with this comment}

% \zxy{Clearly state the problem of dirty data and its implications in a concise manner. Mention how it affects data analysis and decision-making, especially in the context of generative models.}
% \nw{Fixed with this comment}
Data quality is paramount in today's data-driven world, especially in the era of generative AI. 
Dirty data with errors and inconsistencies usually leads to flawed insights, unreliable decision-making, and biased or low-quality outputs from generative models.
The study of repairing erroneous data has gained significant importance.
Existing data repair algorithms differ in information utilization, problem settings, and are tested in limited scenarios. 
In this paper, we compare and summarize these algorithms with a driven information-based taxonomy. 
We systematically conduct a comprehensive evaluation of 12 mainstream data repair algorithms on 12 datasets under the settings of various data error rates, error types, and 4 downstream analysis tasks, assessing their error reduction performance with a \emph{novel but practical} metric.  
% \zxy{Emphasize the unique contributions of the paper. For instance, the comprehensive evaluation of 12 mainstream data repair algorithms and the introduction of a new taxonomy.}
% \nw{Fixed with this comment} 
We develop an effective and unified repair optimization strategy that substantially benefits the state of the arts.  
% We extensively inspect the performance of data repair algorithms  
We conclude that, it is always worthy of data repair.
%the \emph{purely clean data} may not necessarily yield the best performance in data analysis tasks. I.  
The clean data does not determine the upper bound of data analysis performance.  
% \zxy{Highlight key findings or results. For example, the observation that pure clean data may not necessarily yield the best performance in data analysis tasks is intriguing and should be emphasized.}
% \nw{Fixed with this comment}
% {\color{red}The data analysis performance is not very sensitive to the error rate of data.}
We provide valuable guidelines, challenges, and promising directions in the data repair domain. %,  for 5 scenarios and 2 main data analysis tasks.
We anticipate this paper enabling researchers and users to well understand and deploy data repair algorithms in practice. 
\end{abstract}
% \vspace{-0.3cm}

\maketitle

\pagestyle{\vldbpagestyle}
\begingroup\small\noindent\raggedright\textbf{PVLDB Reference Format:}\\
\vldbauthors. \vldbtitle. PVLDB, \vldbvolume(\vldbissue): \vldbpages, \vldbyear.
\href{https://doi.org/\vldbdoi}{doi:\vldbdoi}
\endgroup
\begingroup
\renewcommand\thefootnote{}\footnote{\noindent
This work is licensed under the Creative Commons BY-NC-ND 4.0 International License. Visit \url{https://creativecommons.org/licenses/by-nc-nd/4.0/} to view a copy of this license. For any use beyond those covered by this license, obtain permission by emailing \href{mailto:info@vldb.org}{info@vldb.org}. Copyright is held by the owner/author(s). Publication rights licensed to the VLDB Endowment. \\
\raggedright Proceedings of the VLDB Endowment, Vol. \vldbvolume, No. \vldbissue\ %
ISSN 2150-8097. \\
\href{https://doi.org/\vldbdoi}{doi:\vldbdoi} \\
}\addtocounter{footnote}{-1}\endgroup

\ifdefempty{\vldbavailabilityurl}{}{
\vspace{.3cm}
\begingroup\small\noindent\raggedright\textbf{PVLDB Artifact Availability:}\\
The source code, data, and/or other artifacts have been made available at \url{\vldbavailabilityurl}.
\endgroup
}

\section{Introduction}

% High-quality data serves as the cornerstone for effective analysis and decision-making. 
% Regrettably, the pervasive presence of erroneous information within datasets renders them incomplete, inconsistent, and inaccurate, ultimately compromising the accuracy of data analysis, leading to huge economic costs~\cite{Krishnan16activeclean, Ilyas15trendsdc}.

% Automatic data cleaning plays an important role in preventing compromises to accuracy and mitigating economic costs arising from the ubiquitous erroneous information within the data

% \zxy{Ensure that the introduction flows logically. Start with the problem, its implications, the current solutions, their limitations, and then introduce what this paper brings to the table.}
% \nw{Checked with this comment}

% \zxy{Clearly define the problem of dirty data and its implications. Mention how it can lead to fraud, inaccurate decision-making, and legal issues. This will set the stage for the importance of automatic data cleaning.}
% \nw{Fixed with this comment}
The global application landscape relies heavily on data, while ubiquitously erroneous information involved in data often compromises the data analysis performance, leading to a huge economic cost~\cite{Krishnan16activeclean, Ilyas15trendsdc, Chu16datacleaning, wu23imputationtoolbox}. 
Dirty data may trigger the spread of fraud information, inaccurate decision-making, and even legal repercussions~\cite{Kadir20dataprivacy, Zhang2020ManuallyDE, miao23imputationsurvey, miao18incompletesurvey, tang2023verifai}.
Among the strategies for efficiently mitigating erroneous information and ensuring data quality, \emph{automatic data cleaning}~\cite{hu2019automatic, Zhang2020ManuallyDE}  has become a \emph{pivotal} solution, especially in the age of generative artificial intelligence (GAI), where the large volume of high-quality training data for powerful GAI models like ChatGPT~\cite{ouyang2022trainingllm} and Midjourney~\cite{borji2023midjourney} currently often resort to \emph{resource-intensive manual} data cleaning~\cite{zhao2023survey}.

The promising automatic data cleaning task is generally conducted in two consecutive stages: \emph{error detection} and \emph{error repair}.
The error detection stage is to identify all wrong data values or rule violation sets, while the error repair stage is to correct these wrong values into latent right ones~\cite{Ziawasch16detecting, Mahdavi20baran, raha19mahdavi}.
Existing error detection studies~\cite{Pham21spade, raha19mahdavi, Heidari19holodetect, Neutatz19ed2} have achieved obvious advancements, with an average F1 score exceeding 0.85 on real datasets~\cite{Pham21spade, raha19mahdavi}. 
In contrast, the error repair is more complex and challenging. 
Various techniques for error repair have been employed, such as machine learning~\cite{Yakout13scared, Rekatsinas17holoclean, Mahdavi20baran}, transfer learning~\cite{Mahdavi20baran, Hasan21wikierrorcorrection}, boosting algorithm~\cite{Krishnan17boostclean} and few-shot learning~\cite{Mahdavi20baran}. 
However, within the advanced detection results, the \emph{final} data repair performance of current methods still falls short of the expected ideal, with an average F1 score below 0.7 in real scenarios~\cite{Rekatsinas17holoclean, Mahdavi20baran, Giannakopoulou20relaxation}.
Thus, it is crucial and urgent to analyze and explore how to deploy \emph{end-to-end} data cleaning approaches with effective \emph{data repair} algorithms in real-life scenarios.
% \zxy{Consider adding a diagram or flowchart that summarizes the main challenges in automatic data repair and how this paper addresses them.}
% \nw{Checked with this comment, current limitations are stated as follows, and adding a diagram may not significantly enhance the understanding, while further taking up the limited space.}

Table~\ref{tab:comparison of methods} compares and summarizes existing surveys of end-to-end data cleaning study.
They either analyze the data repair process without experiments~\cite{Chu16datacleaning, Ilyas15trendsdc, Ilyas19datacleaning, Krishnan16survey},
or verify the impact of data repair on downstream tasks while overlooking the inclusion of certain important algorithms~\cite{Li21cleanml, Abdelaal23rein}.
%However, they are confined to a subset of experiment evaluation, error elimination analysis, downstream model influence, data and rule types.
The limitations and analysis perspectives come from the following \emph{three} aspects. 
% ~\cite{Abdelaal23rein} present a unified experiments of current methods both on the cleaning performance and downstream impact analysis.
% However, they are confined to a subset of algorithms, downstream model influence, optimizations, various rule types, as well as the actual impact on the original data after repair process. 

% Furthermore, while some comparative experiments have been conducted, they come with certain limitations.

%\noindent 
\textbf{L1: Metric shortfalls in evaluation.}
Fewer data errors typically lead to higher data quality and reduced biases in subsequent analyses, which constitute the \emph{main goals} of the data cleaning task.
Prior data cleaning surveys~\cite{Ilyas15trendsdc, Ilyas19datacleaning, Krishnan16survey, Li21cleanml, Abdelaal23rein} lack an evaluation of \emph{error reduction} in final data repair results, as indicated in Table~\ref{tab:comparison of methods}. 
They often evaluate the data repair results with metrics such as \emph{precision} and \emph{recall}, which are based on the proportion of correctly repaired cells, disregarding the absolute quantity changes.
Thus, such metrics result in a \emph{biased} evaluation.  
% Two issues arise in the calculation of these metrics.
%Since the calculation of precision/recall is .
For example, suppose there is a dataset that initially encompasses 10 erroneous cells.
One data cleaning approach repairs 100 cells, of which 80 are correct (including the initial 10 error cells). 
Despite 70 of 80 values being identical to their originals, they are still flagged as erroneous values, and replaced by a value within calculation, thereby should be considered as a repaired cell. 
It means that, the dataset currently contains 20 erroneous cells, indicating a \emph{doubles} amount of erroneous cells after the repair process.
While the precision and recall of this repair both reach as high as 0.8 and 1.0, respectively. 
The correct repair of more initially right data increases the precision, according to the mediant inequality, but it cannot benefit the data quality.
As a result, a higher precision/recall value does not represent a larger degree of error reduction. 
Furthermore, for distance-based metrics like Jaccard distance or mean squared error (MSE), they only measure the distance between the repaired and clean data, while \emph{ignoring} the distance between the original dirty data and clean data.
In a situation where the initial distance between dirty and clean data is 0.1, which then increases to 0.15 after repair.
Simply stating the distance as 0.15 fails to directly indicate an increase in data errors.
It cannot convey the actual \emph{improvement degree} in data quality, either.
Incorporating error reduction degree into evaluations will align repair algorithms more closely with improving data quality, and offer a direct perspective on data quality changes.
% The error reduction performance is one of the most important factors when choosing the proper data repair algorithms in real-world scenarios.
% Prior experimental comparisons of different algorithms mainly focus on evaluating the repair performance of the algorithms~\cite{Chu13holistic, Ebaid13nadeef, Rekatsinas17holoclean, Mahdavi20baran, Abdelaal23rein, Chu13holistic, Khayyat15bigdansing}. 
%This inadequacy arises from two factors. 
% {\color{purple}The higher precision/recall/F1 score means that, finding more errors or xxx, while it does not indicate the fraction of erroneous information in the whole data is decreasing after data repair.}}
%First, metrics like \emph{F1 score} inherently fail to present the error elimination information directly.
%Second, the calculations of these metrics may potentially include correctly repaired cells from the \emph{initially correct} data, resulting in higher performance on these metrics. However, these repaired cells have no actual impact on improving the quality of the data.

%\noindent 
% \textbf{L2: Limited complex scenarios exploration.}
\textbf{L2: Insufficient scenarios exploration in evaluation.}
The effectiveness of data repair algorithms differs in different scenarios. %depends on the data characteristics like error rate and type.
As depicted in Table~\ref{tab:comparison of methods}, existing surveys lack the study of data repair algorithms' capacity under various \emph{error types}, in either the data repair performance evaluation or the assessments of its impact on downstream tasks~\cite{Chu16datacleaning, Abdelaal23rein, Li21cleanml, Ilyas19datacleaning, Krishnan16survey}. 
Meanwhile, the \emph{error rate} analysis is absent in downstream impact evaluation, even though it is included in repair performance assessments.
It results in an inadequate understanding of the proper deployment specific to distinct error types and error rates. 
It is vital and practical to comprehensively evaluate data repair algorithms with a diverse range of error types and rates in both repair and downstream tasks. 
% by only examining random string variations or original errors in the datasets~\cite{Li21cleanml, Abdelaal23rein}, which also manifests across research studies. 
% Moreover, numerous data repair algorithms are based on \emph{distinct} types of rules like functional dependency and denial constraint~\cite{Chu13holistic, Khayyat15bigdansing, Fan09RMRules, Hao17novelcost, Rezig21horizon, Beskales13relative}, underscoring their importance.
% While experiments in current surveys only provide functional dependency information and may ignore the broader context of the data~\cite{Abdelaal23rein, }.
% , which solely provide dependency relation information between attributes 

\begin{table}[tp]
% \vspace*{-0.12in}
\centering
\footnotesize
\caption{Comparison of previous surveys and ours.}
\begin{threeparttable}
% \tablefootnote{DRE and DME refer to data repair evaluation and downstream model evaluation.}
\vspace*{-0.15in}
% \begin{tabular}{|c|c|c|c|c|c|c|c|c|}
\setlength{\tabcolsep}{1.1mm}
{\begin{tabular}{|c|c|c|c|c|c|c|}
\hline
  \multirow{2}{*}{\textbf{Survey study}} & \multirow{2}{*}{\textbf{\makecell[c]{Exp. \\ evaluation}}} & \multirow{2}{*}{\textbf{\makecell[c]{Error reduc. \\ evaluation}}} & \multicolumn{2}{c|}{\textbf{Error rate}} & \multicolumn{2}{c|}{\textbf{Error type}}   \\ \cline{4-7}
     &  &  & DRE* & DME* & DRE* & DME* \\ \hline
 Ilyas et al.~\cite{Ilyas15trendsdc, Ilyas19datacleaning}   & \twemoji{multiply}  & \twemoji{multiply}    & \twemoji{multiply} & \twemoji{multiply} & \twemoji{multiply} & \twemoji{multiply} \\ \hline
 Krishnan et al.~\cite{Krishnan16survey}                    & \twemoji{multiply}    & \twemoji{multiply}    & \twemoji{multiply} & \twemoji{multiply} & \twemoji{multiply} & \twemoji{multiply} \\ \hline
  Chu et al.~\cite{Chu16datacleaning}                       & \twemoji{multiply}    & \twemoji{multiply}    & \twemoji{multiply} & \twemoji{multiply} & \twemoji{multiply} & \twemoji{multiply} \\ \hline
 Li et al.~\cite{Li21cleanml}                               & \thincheckmark    & \twemoji{multiply}      & \twemoji{multiply} & \twemoji{multiply} & \twemoji{multiply} & \thincheckmark \\ \hline
 Abdelaal et al.~\cite{Abdelaal23rein}                      & \thincheckmark    & \twemoji{multiply}      & \thincheckmark & \twemoji{multiply} & \twemoji{multiply} & \twemoji{multiply} \\ \hline
 Ours                                                       & \thincheckmark    & \thincheckmark      & \thincheckmark   &  \thincheckmark  & \thincheckmark   & \thincheckmark   \\ 
 \hline
\end{tabular}}
\begin{tablenotes}
    \footnotesize
    \item[*]  DRE and DME refer to data repair and downstream model evaluation. 
\end{tablenotes}
\end{threeparttable}
\label{tab:comparison of methods}
\vspace*{-0.15in}
\end{table}

% Ziawasch et al.~\cite{Ziawasch16detecting}  & \twemoji{check mark}  & \twemoji{multiply}  & \twemoji{multiply} & \twemoji{check mark} & \twemoji{check mark} \\ \hline

%\noindent 
% \textbf{L3: Absence analysis of data repair influence on downstream model    complex data scenarios and strategy and guidelines.}
\textbf{L3: Lacking algorithm deployment guidelines.}
Blindly deploying data repair algorithms may potentially hurt data quality and task performance.
Although existing surveys discuss data repair algorithm performance~\cite{Chu16datacleaning, Ilyas15trendsdc, Ilyas19datacleaning, Krishnan16survey}, \emph{three} critical questions regarding algorithm deployment remain unanswered: (i) Is data repair consistently beneficial irrespective of the error rate? (ii) Based on the observation that, the suboptimal performance of data repair primarily results from incorrect repairs, how to develop a practical strategy to mitigate incorrect repairs based on error reduction evaluation?  (iii)  How to select the proper repair method in different scenarios based on task objectives? 
The data repair process plays a crucial role in the cleaning task, and addressing these issues will provide practical guidance to deploy data cleaning algorithms.
Therefore, in this survey, we conduct an exhaustive comparative analysis and experimental evaluation of 12 state-of-the-art data repair algorithms.
%This evaluation is conducted based on a novel taxonomy and across various complex data scenarios.
Our main contributions are described below.
\begin{itemize}[leftmargin=*,itemsep=2pt,topsep=2pt]
\setlength{\itemsep}{0pt}
\setlength{\parsep}{0pt}
\setlength{\parskip}{0pt}

\item We systematically evaluate \emph{twelve} mainstream and state-of-the-art data repair algorithms based on a driven information-based taxonomy of data repair algorithms, including \emph{cstr-driven} data repair algorithms, \emph{data-driven} ones, and \emph{hybrid-driven} ones.  
We also state the properly applied scenarios and graphically represent the workflow within each category.

\item We define a novel and effective \emph{metric} named \emph{Error Drop Rate} to evaluate the error reduction condition of final data repair results. 
Surprisingly, in the majority of cases, most data repair algorithms tend to introduce more errors rather than eliminate erroneous information. We propose a \emph{universal} optimization strategy that leverages error detection techniques to prevent the alteration of original right values into wrong ones. 

\item Extensive experiments on \emph{twelve} real datasets under various error rates/types and data scales demonstrate both the performance of \emph{twelve} data repair algorithms and the impact on \textit{four} common downstream tasks. Particularly, except for five algorithms with publicly available codes, we implement \emph{seven} additional data repair algorithms for comprehensive comparisons. 
We investigate the effect of \emph{semantic} and \emph{syntactic} data errors (that simulate complex practical scenarios) on the algorithm performance.
%\emph{Inner} errors arise from {\color{blue} incorrect value allocation and \emph{outer} errors are out-of-domain values.}}

% add findings related to the L3 and other limitations 
% using clean data is not the best for downstream tasks.Strategically repairing data may obtain better performance than using the completely cleaning data.It signifies that, integrating data cleaning and downstream decision making is more promising than the typical two- step framework of data cleaning and decision making.
\item We conclude several interesting findings: 
\emph{i)} Repair algorithms often \emph{increase} errors. Contrary to expectations, most existing data repair algorithms tend to raise the error rate in the data rather than eliminate it;
\emph{ii)} Using the clean data is \emph{not} always the best. Downstream analysis models trained on purely clean data may perform worse than models trained using dirty data.
%with additional errors from certain repair algorithms can outperform models using clean data. 
% {\color{blue}It may signify that  some ``suitable errors'' could regularize model training and improve generalization;}
\emph{iii)} Data repair could provide universal benefits for downstream tasks. With proper algorithm selection, data repair boosts downstream tasks in the vast majority of cases;
\emph{iv)} \emph{Semantic} errors pose a more stubborn challenge. 
Semantic errors within the data are harder to eliminate and have a more negative impact on downstream data analysis than \emph{syntactic} ones,
highlighting the need for greater consideration in real-world applications. 
% even lead to the introduction of additional errors by existing repair algorithms
%We conclude interesting findingsdations valuable research guidelines and  and    considering  \emph{five} scenarios factored by error reduction, downstream task performance, and time limit.} 
 
%We discuss research challenges in data repair and outline promising future directions, especially in light of the advancements in large language models (LLMs).  
\end{itemize}
% to achieve results outperforming \emph{state-of-the-art} algorithms.

% \zxy{Conclude the introduction with a statement that encourages the reader to delve deeper into the paper. Clearly outline what the paper will cover. }
% \nw{Fixed with this comment}

In the rest of this paper, Section~\ref{sec:preliminaries} presents the 
preliminaries, while the new taxonomy and algorithm analysis are covered in Section~\ref{sec:methods}. Comprehensive evaluations are offered in Section~\ref{sec:experiment}. Section~\ref{sec:discussion} discusses current challenges and future directions, followed by related work in Section~\ref{sec:relatedwork}. Finally, Section~\ref{sec:conclusions} concludes the study.
% We introduce the preliminaries in Section ~\ref{sec:preliminaries}. 
% Section~\ref{sec:methods} describes data repair algorithms with a new taxonomy.
% Section~\ref{sec:experiment} provides a comprehensive experimental evaluation of data repair methods under both real and simulated complex scenarios, as well as valuable guidelines. 
% We offer current challenges, and future directions in Section ~\ref{sec:discussion}, related work in Section~\ref{sec:relatedwork}, and conclusions in Section~\ref{sec:conclusions}, respectively. 

% However, as the deployment of automatic data repair systems becomes more prevalent, it brings forth a series of crucial questions and challenges. Are these systems truly ready to be integrated into critical decision-making processes? Can they consistently and reliably repair a wide range of errors across various domains? What are the potential risks and limitations associated with their deployment, and how can these be mitigated? 
% This paper delves into these essential inquiries, aiming to assess the readiness of automatic data repair methods for real-world deployment.

% \input{6.Related_Work}
% \vspace*{0.05in}
\section{Preliminaries} \label{sec:preliminaries}

In this section, we first introduce the problem definition.  %of data repair.
We then describe the constraints widely used in automatic repair algorithms. 

% \subsection{Rules for Data Repair}
Data cleaning mainly serves two primary goals: eliminating errors and ensuring data consistency, with the latter being a sub-goal of the former.
In this survey, we aim to evaluate \emph{error reduction} performance over the final data repair results, as stated below.

% Considering that error elimination aligns more closely with attaining high-quality data, 

\begin{defn}
\label{defn: problem-definition}
  %\textbf{Holistic Repair.} 
  \textbf{Problem Statement.}
  Given an instance $I$ of relation $R$ characterized by a set of attributes $Attrs=\{A_1, A_2, \cdots, A_n\}$. Each $t \in I$ comprises a set of cells represented by $Cells[t] = {A_i[t]}$ corresponding to a distinct attribute in $Attrs$.
  For each cell $c$, its unknown true value is denoted by $v_{c}^{*}$, its initial observed value is $v_{c}$, and its estimated true value is $\hat{v}_c$.
  A data repair algorithm aims to make $\hat{v}_c$ equal to $v_{c}^{*}$ for all cells $c$.
  % , where $v_c \neq v_{c}^{*}$.
\end{defn}

This error reduction evaluation defined in \emph{Definition} \ref{defn: problem-definition} encompasses correcting all erroneous cells within a dataset. It shares the same definition of \emph{holistic repair}, as studied in most recently proposed data repair algorithms~\cite{Rekatsinas17holoclean, Mahdavi20baran, raha19mahdavi, Pham21spade}.

%Based on the instance $I$ and relation $R$ defined previously, the problem definition is outlined as follows.

% The rule serves as a critical guideline in the data repair process. It usually encodes expert experience and field knowledge of the data. 
Expert experience and field knowledge of the data serve as a critical guideline in the data repair process.
These guidelines manifest as rules and constraints (cstrs).
Rules provide explicit instructions for modifying erroneous values, making data repair a deterministic process~\cite{Fan12master, Hao17novelcost}.
However, real-world scenarios often lack external sources of authoritative information, such as master data or expert insights.
This absence poses a significant challenge in employing rules in the data repair process~\cite{Hao17novelcost}. 
In contrast, constraints are designed to capture the relationships between specific attributes or values.
Constraints have a \emph{wider} range of applications, since professional external information may not always be accessible.

Numerous automatic data repair algorithms are built based on constraints like \emph{functional dependency} (FD) \cite{Papenbrock15FdEvaluation} and \emph{denial constraint} (DC) \cite{dcfinder19Pena}, as stated below. 
%{\color{blue} Based on the definition of instance $I$, relation $R$ and the attribute set $Attrs$ in Definition~\ref{defn: problem-definition}.}
% Given an instance $I$ of a relation $R$, which is characterized by a set of attributes $Attrs=\{A_1, A_2, \ldots, A_n\}$ that describe $I$ as a collection of tuples.
%FD and DC could be stated as follows.

\begin{defn}
\textbf{Functional Dependency.} 
A functional dependency is a statement of the form $X \rightarrow A$ where $X \subseteq Attrs$ and $A \in Attrs$, denoting that the values of attribute set $X$ uniquely determine the values of attribute $A$ across all tuples in $I$.   $X$ and $A$ are the left-hand and right-hand attributes, respectively. 
\end{defn}
% And if there is more than one attribute on the right side, then they can be split into multiple FDs, according to each attribute on the right side.

\begin{defn}
\textbf{Denial Constraint.}
% Denial constraint (DC) defines a set of predicates for which its predicates cannot hold true simultaneously. 
A denial constraint $\varphi$ is a statement of the form $ \forall t_\alpha, t_\beta \in I: \neg\left(p_1 \wedge \cdots \wedge p_m\right)$ where $p_i$ is in the following form: $t_\alpha.A_x \ \phi \ t_\beta.A_y \text{ or } t_\alpha.A_x \  \phi \ c,$
$A_x,A_y \in Attrs, \ c$ is a constant, and $\phi$ is a built-in operator, i.e., $\neq,\ =,\ \leq, \ <, \ \geq, \ >$.
\end{defn}

\begin{examp}  
  A sample instance from real-world tax data ~\cite{Arocena15bart} is presented in Table~\ref{example1}, along with two real FDs and two DCs: 
  
  % $F_1: FirstName \rightarrow Gender$, 

  $F_1: City \rightarrow State$,
  
  $DC_1: \forall t_1, t_2 \in I:$
  $ \neg(t_1.City=t_2.City  \wedge t_1.State \neq t_2.State)$, 
  
  $DC_2: \forall t_1, t_2 \in I:$
$ \neg(t_1.City=t_2.City \wedge t_1.Salary \geq t_2.Salary \wedge t_1.Rate \leq t_2.Rate)$.
  
% The first FD $F_1$ means if the values of $FirstName$ are the same, then the values in $Gender$ should also be the same. 
The FD $F_1$ means if the values of $City$ are the same, then the values of $State$ should also be the same.
% The meaning of $DC_1$ is for all tuple pairs in $R$, the condition that the values of $City$ are the same while the values in $State$ are different should not exist.
Meanwhile, $DC_1$ asserts that no two tuples in $R$ can have identical $City$ values paired with differing $State$ values.
$DC_2$ indicates that within a given $City$, a higher $Salary$ should not correspond to a lower $Rate$ of tax.
In this case, $F_1$ and $DC_1$ are equivalent, expressing the same meaning.
The erroneous cells violating these constraints are colored in {\color{gray}gray}.
\end{examp}

An FD in the above form $X \rightarrow A$ can be equally expressed in the following DC format:
$\forall t_\alpha, t_\beta \in I,$ $\neg(t_\alpha.X_0=t_\beta.X_0 \wedge \cdots \wedge t_\alpha.X_k =t_\beta.X_k \wedge t_\alpha.A \neq t_\beta.A)$, 
$X_0, \cdots, X_k \in X.$ 
While DC has greater expressive ability than FD.
For instance, the monotonic relationship between values in $DC_2$ cannot be captured by FD, underscoring the expressivity gap between them. 
Considering the expressive ability and broader application spectrum, we mainly focus on DC-based data repair algorithms when evaluating the repair solutions adopting constraints.

\vspace*{0.05in}
Based on constraints, to achieve data consistency, previous work~\cite{Beskales13relative, Chu13holistic} also offers other repair definitions, which are significant in algorithm understanding and deployment process:
\begin{defn}
  \textbf{Tolerant Repair.}
  Given a set of constraints $\Sigma$ over a relation schema $R$, along with a set of modified constraints $\Sigma'$, and two instances $I$ and $I'$ of $R$. $(\Sigma', I')$ is a tolerant repair of $I$ if $I'$ adheres to the modified constraint set $\Sigma'$. %, denoted as $I' \vDash \Sigma'$.
\end{defn}

\begin{defn}
  \textbf{Consistency Repair.}
  Given a set of constraints $\Sigma$ over a relation $R$, and two instances $I$ and $I'$ of $R$.
  $I'$ is a consistency repair of $I$ if $I'$ adheres to the constraint set $\Sigma$. %, denoted as $I \vDash \Sigma$.
\end{defn}

\begin{table}[tbp]
%\vspace*{-0.12in}
% \small
\setlength{\tabcolsep}{0.7mm}
  \caption{Relation of tax.}
  \label{example1}
  \vspace*{-0.15in}
 \resizebox{\columnwidth}{!}{
  \begin{tabular}{|c|c|c|c|c|c|c|c|}
    \hline
        \textbf{Tuple} & \textbf{FirstName} & \textbf{LastName} & \textbf{Gender} & \textbf{City} & \textbf{State} &\textbf{Salary} & \textbf{Rate} \\ 
    \hline
        $t_1$ &Weiming  & Posthoff & {\color{gray} Female}    & Vera    & Okla. & 45,000 & 6.25\\ \hline
        $t_2$ &Weiming  & Zongtian & {\color{gray} Male}    & Okemos     & Mich. &10,000 & 3.9\\ \hline
        $t_3$ &Shivraj  & Alpin  & Female  & Eastham & {\color{gray} Mass.} & 90,000 & {\color{gray}5.3}\\ \hline
        $t_4$ &Yurdaer  & Thackray   & Female    & Eastham     & {\color{gray} LA.} & 30,000 & {\color{gray}5.2}\\
    \hline
  \end{tabular}
  }
  \vspace*{-0.15in}
\end{table}

\section{Data Repair Algorithms}  \label{sec:methods}

In this section, we elaborate 12 mainstream data repair algorithms based on a new \emph{information-driven} taxonomy.
We present a general and effective strategy to optimize these data repair algorithms. 
%Then we introduce each of them in detail. 

\subsection{Algorithm Taxonomy}
% \zxy{Clearly define the criteria for the taxonomy of data repair algorithms. For instance, when classifying algorithms as rule-driven, data-driven, data\&rule-driven, and model-driven, provide concise definitions for each category.}
%\nw{Fixed with this comment}
% BRIEF INTRODUCTION OF TAXONOMY
% rule-driven: all rely on info provided by rules
% data-driven: rely on data info, no rule provided
% rule and data-driven: using both rule info and data info not mentioned in the rules or repair rules based on data info.
Automatic data repair, which typically relies on data and/or constraint information, aims to enhance data quality or improve the performance of downstream tasks.

To offer insights into how they utilize constraints and data
according to the driven information, we categorize existing data repair algorithms into three groups,  i.e., \emph{constraint(cstr)-driven}, \emph{data-driven}, and \emph{hybrid-driven} methods. 
The \emph{cstr-driven} methods repair data based on constraints, typically employing \emph{equivalent classes} (which are sets of cells that should satisfy the same constraint conditions).
The \emph{data-driven} algorithms only utilize data distribution information (without constraints) to repair data.
For \emph{hybrid-driven} algorithms, except for constraints, they also leverage data information not covered in constraints to facilitate data repair. 

This taxonomy is useful for providing clear guidelines on using constraints and data, and offering targeted methods for different error types (cstr-driven for semantic, data-driven for syntactic, and hybrid for general errors).
% and catering to the varying needs in deployment, from strictly regulated ones like healthcare and finance to more flexible fields such as machine learning.
It also aligns with varying needs in deployment. Cstr-driven methods for regulated scenarios like healthcare and finance, and adaptable data-driven or hybrid methods for fields like machine learning tasks.

Moreover, there are \emph{three} primary repair objectives within these algorithms, namely \emph{data consistency}, \emph{holistic repair}, and \emph{model performance boost}.
Data consistency encompasses \emph{consistency repair}, which is applied to data with strict constraints like, geographic data where city names rely on states, and \emph{tolerant repair}, used when constraints are either outdated or incorrect. \emph{Holistic repair} focuses on maximizing error removal, crucial in fields like cybersecurity or healthcare. Lastly, the aim of \emph{model performance boost} is to refine training data to enhance outcomes of data-driven tasks, such as image classification.
Regarding these objectives, cstr-driven algorithms universally aim to achieve consistency repair. Data-driven approaches are for holistic repair and enhancing model performance. While the hybrid-driven methods are oriented toward tolerant repair or holistic repair.

% Specifically, data consistency contains two types: \emph{consistency repair} and \emph{tolerant repair}.
% In practice, consistency repair is mostly pursued in contexts with well-defined constraints such as geographic data where city values determine the state values. Tolerant repair, on the other hand, is more appealing when the rules or constraints are outdated or have errors.
% Second, holistic repair aims to eliminate errors to the greatest extent possible. 
% It is suitable for error-sensitive scenarios, like cybersecurity or medical diagnoses. 
% The third objective is to seek \emph{model performance boost}.
% Instead of focusing on the data itself, this objective involves repairing training data to improve data analytical tasks such as image classification.

% It is most relevant in specific data-driven analytical tasks like image classification. 
% of downstream analysis models. 
% , both of which ensure the data strictly adheres to given rules on the trustworthy assumption.
% The consistency repair assumes \emph{full} trust in the rules, enabling data modification to conform to given rules.  
% While the tolerant repair assumes a \emph{relative} trust between rules and data, necessitating the repair of both data and rules to attain consistency. 

Guided by the taxonomy, we select methods based on the utilization of constraints and data. 
In \emph{cstr-driven} methods, we examine widely cited Holistic~\cite{Chu13holistic} and Nadeef~\cite{Ebaid13nadeef} for mainstream graph-based and statistic-based data modeling with DCs. 
For \emph{data-driven} ones, methods predominantly encompass statistical boosting and machine learning (ML)-based techniques, with BoostClean~\cite{Krishnan17boostclean} as the only example for the former, Baran~\cite{Mahdavi20baran} and Scare~\cite{Yakout13scared} pioneered for ML applications in data repair. \emph{Hybrid-driven} tools like Relative~\cite{Beskales13relative} and Unified~\cite{Chiang11unified} are representative in heuristic-based methods, while HoloClean~\cite{Rekatsinas17holoclean} merges constraints with data using ML models. 
Then from a scalability viewpoint, \emph{cstr-driven} techniques face time cost challenges; BigDaning~\cite{Khayyat15bigdansing} alleviate this with methods to reduce redundancy and group data. Horizon~\cite{Rezig21horizon} proposes a fresh strategy to reduce time costs using FDs; meanwhile, Daisy~\cite{Giannakopoulou20relaxation} showcases data repair's practicality for query tasks. Lastly, MLNClean~\cite{Ge22mlnclean} introduces a novel weighted constraint application using Markov logic networks, showcasing enhanced performance and novel constraint application methods.
 
% When the rules are insufficient or outdated, the hybrid method is more attractive.
% analyze the three groups, e.g., the relationship among them, the respective application scenarios, etc.

\begin{figure}[t]
\vspace*{-0.16in}
\center
  \includegraphics[width=0.83\linewidth]{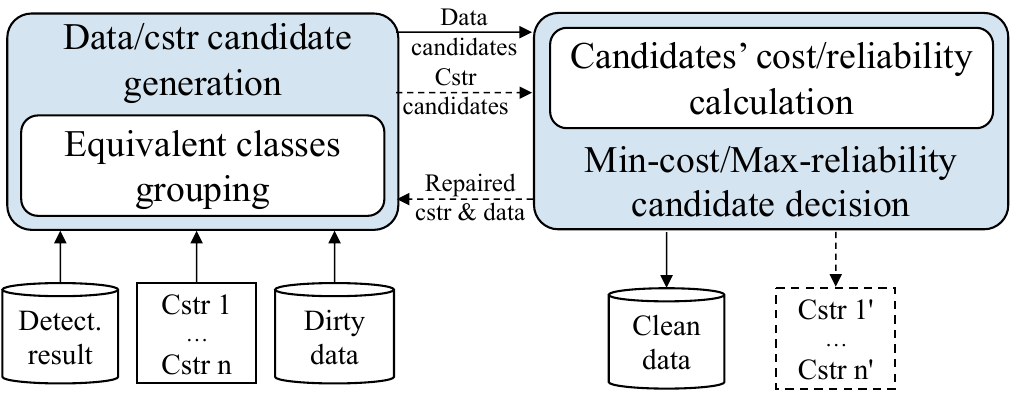}
  \vspace*{-0.1in}
  \caption{Workflow of \emph{cstr/hybrid-driven} repair algorithms. %The black line depicts the data repair process of \emph{rule-driven} algorithms; green and blue lines represent \emph{extra} steps employed by certain \emph{rule\&data-driven} methods to repair rules and data respectively; ED refers to error detection.
  }
 \label{fig:DPW-1}
\vspace*{-0.2in} 
\end{figure}

\subsection{Algorithm Description}

Each data repair algorithm falls into one of the three categories.  
We summarize the workflow of each group of data repair algorithms, and describe the representative algorithms belonging to the group.

\subsubsection{Cstr-driven Data Repair Algorithms}
%\textbf{Workflow.} 

Figure~\ref{fig:DPW-1} depicts the workflow of \emph{cstr-driven} algorithms  (not including dashed arrows and modified constraints in the figure).  It takes a dirty dataset with constraints and the \emph{constraints violation} detection results as inputs. It generally includes two phases to get the clean data: getting the possible correct candidates for \emph{constraint violation} cells and repairing the data based on cost minimization or reliability maximization. 

Specifically, data candidate generation of constraint violation cells is based on \emph{equivalent classes},  
% the dataset is partitioned into equivalent classes based on rules, 
which shows that a set of attribute values should satisfy the same condition. 
For example, in Table~\ref{example1}, the $State$ value of $t_3$ and $t_4$ should be identical due to the same value of $City$ based on $F_2$.
Thus, $(t_3.City, t_3.State)$ and $(t_4.City, t_4.State)$ form an equivalent class, indicating that their values \emph{should} be identical.
All values in the equivalent classes could be the data repair candidates. 

In addition, the choice of candidates for repair is determined by candidates within minimal calculated costs like edit distance~\cite{Hao17novelcost} and cardinality~\cite{Chu13holistic,Ebaid13nadeef}, or maximal reliability metrics such as frequency~\cite{Giannakopoulou20relaxation} and overall data support degree~\cite{Rezig21horizon}.  
% {\color{red}the detection of cells that violate these rules} 

\textbf{Holistic}~\cite{Chu13holistic}.
\textsf{Holistic} is a classical repair algorithm addressing DC violations.
When generating data candidates, it encodes all equivalent classes 
% derived from different DCs 
in a \emph{conflict hypergraph}, where nodes and hyperedges represent violation cells and associated constraints, respectively. 
By analyzing nodes' interactions, \textsf{Holistic} generates a variety of repair candidates. 
To determine candidates, it first calculates the applied costs of all candidates, and then heuristics are employed to achieve cost minimum and data consistency. 
Due to the necessity of comparing tuple pairs to find all equivalent classes,  the time complexity of \textsf{Holistic} escalates to $O(|I|^2)$.  
This poses a significant challenge when dealing with large datasets. 
% Additionally, to repair all violations regardless of the actual cost model, \textsf{Holistic} introduces a novel algorithm along with several heuristics to achieve both minimal cardinality and minimal distance repair.

\textbf{BigDansing}~\cite{Khayyat15bigdansing}. 
To alleviate the high time cost of repair algorithms, \textsf{BigDansing} proposes two key acceleration strategies.
The first one is removing data irrelevant to the given constraints from the repair process.
% , thus avoiding unnecessary computation costs. 
The second one is grouping equivalent classes with the \emph{same keys} to narrow down the candidates' number.  
Thus, they accelerate the tuple comparison process and reduce the actual time cost of data repair.
% , though the time complexity remains at the same level.
% is a scalable and user-friendly framework for data repair in large-scale data environments utilizing DCs. 
% To enhance efficiency, 
% Furthermore, \textsf{BigDansing} offers methods to implement existing repair algorithms in distributed settings, including the \textsf{Holistic} algorithm mentioned above. 

\textbf{Horizon}~\cite{Rezig21horizon}.
% From another perspective, \textsf{Horizon} leverages {\color{red}the direction feature} of attributes in FDs to avoid exhaustively tuples comparison in grouping equivalent classes.
To avoid exhaustively tuple comparisons in grouping equivalent classes, \textsf{Horizon} constructs a directed \emph{FD pattern graph} with multiple hierarchies, each corresponding to the attributes in FDs.
Directed edges connect nodes across hierarchies, linking values from the left-hand attributes to right-hand ones indicated by FDs.
To group equivalent classes and generate data candidates, the graph is traversed  with a time complexity \emph{linear} to the number of edges. 
\textsf{Horizon} finally determines candidates for constraint violation cells with the highest support score.  
% {\color{red} Notably, although the traversal complexity is \emph{linear} to the number of edges, which is quadratic with the number of distinct values.
% While the number of distinct values can equal the size of instance $I$ in the worst case, in practice, it is typically much smaller, resulting in significant time savings.}
% }
% as the values of left-hand attributes uniquely determine the right-hand attributes, 
%, as shown in Table~\ref{tab:Summary}.
% Traditional data repair algorithms have a time complexity of $O(I^2)$,  resulting in excessive time when dealing with millions of records.
% \textsf{Horizon} is a linear-time efficient FD repair algorithm, outperforming the time complexity of previous algorithms, which typically exhibit a complexity of $O(I^2)$.
% After receiving the FDs and dirty data, \textsf{Horizon} initially constructs a cost model to retain frequent patterns and generates an \emph{FD pattern graph} to analyze rule quality and interactions.
 
\textbf{Nadeef}~\cite{Ebaid13nadeef}.
Unlike prior methods leveraging a certain constraint type, \textsf{Nadeef} focuses on utilizing various constraints by compiling and managing them in a unified format. It enables users to specify not only constraints like FDs and DCs but also other constraints. Based on the candidates generated with equivalent classes, \textsf{Nadeef} minimizes the cardinality cost with two data repair algorithms tailored to emphasize efficiency or effectiveness.
% Based on various predefined rules, \textsf{Nadeef} offers two data repair algorithms tailored to different scenarios.
% dynamic semantics and its capability to address quality issues specific to each application.
% Additionally, \textsf{NADEEF} can employ customized repair algorithms by the users to accommodate real-world scenarios.
% The first one prioritizes the accuracy.
% It converts all violations and potential data changes to a variable-weighted conjunctive normal form (CNF).
% This CNF set is then input to a weighted MAX-SAT solver that computes repairs while minimizing their overall cost.
% The second one is designed to be more efficient.
% It leverages the concept of equivalence classes.
% It merges data that violates constraints into multiple equivalence classes and assigns a distinct value to each equivalence class

\textbf{MLNClean}~\cite{Ge22mlnclean}.
Previous algorithms often instantiate the given constraints by corresponding values to repair data, which may not be trustworthy, thus lacking robustness. 
To mitigate this issue, \textsf{MLNClean} leverages Markov logic networks to learn the trustworthy degree of each \emph{instantiated constraint}, thereby enhancing the robustness.
Based on the learned instantiated constraints, \textsf{MLNClean} generates multiple data candidates.
Subsequently, a \emph{reliability} score of each data candidate is calculated with the trustworthy degree.
To decide the proper candidates, it first selects candidates with maximal reliability scores, thus generating multiple repaired data versions.  
Then, \textsf{MLNClean} designs a \emph{fusion} score to resolve conflicts across different data versions and determine the final repairs.
 
% Within each data version, errors are initially addressed using a reliability score, followed by the elimination of conflicting values across different data versions using a fusion score.
% contributes to addressing the limitations of generalized rules by leveraging Markov Logic Networks (MLNs) to infer \emph{instantiated rules} and generate multiple data versions for cleaning.
% a hybrid data cleaning framework that integrates Markov Logic Networks (MLNs).
% The primary contribution of \textsf{MLNClean} is 
% The framework comprises two stages: pre-processing and two-stage data cleaning.
% In the pre-processing stage, possible \emph{instantiated rules} are inferred according to MLNs, and a two-layer index structure is constructed to generate multiple data versions for subsequent cleaning based on equivalence classes.
% In the second stage, errors within each data version are initially cleaned using a reliability score, and then conflicting values among different data versions are eliminated using a fusion score.

\textbf{Daisy}~\cite{Giannakopoulou20relaxation}.
Unlike previous methods concerning offline data repair, \textsf{Daisy} aims to repair results from online queries. It first generates candidates from the query data based on equivalent classes indicated by the given DCs. To evaluate candidates’ reliability, \textsf{Daisy} computes their conditional probability using the frequency appearing with other attribute values. Finally, \textsf{Daisy} merges distinct conditional probabilities to derive the repair candidates. 
% performs on-demand probabilistic repair for DC violations. It strategically places {\color{red}the cleaning step in the logical plan} based on data and query conditions. \textsf{Daisy} then converts query results into probabilistic outcomes by replacing errors with candidate fixes and their associated probabilities.
% Then \textsf{Daisy} cleans the result of each query operator affected by the rules in the logical plan.
% works in two steps: a logical step and an execution step. 
% In the logical step, \textsf{Daisy} 
% and incorporates a cost model to optimize the placement of cleaning operators by identifying the data subset.
% is an approach which integrates with exploratory Select-Project-Join (SPJ) and aggregate queries for data cleaning purposes.
% By identifying the data subset that impacts result cleanliness, \textsf{Daisy} 

\begin{figure}[tbp]
\vspace*{-0.17in}
\center
  \includegraphics[width=0.84\linewidth]{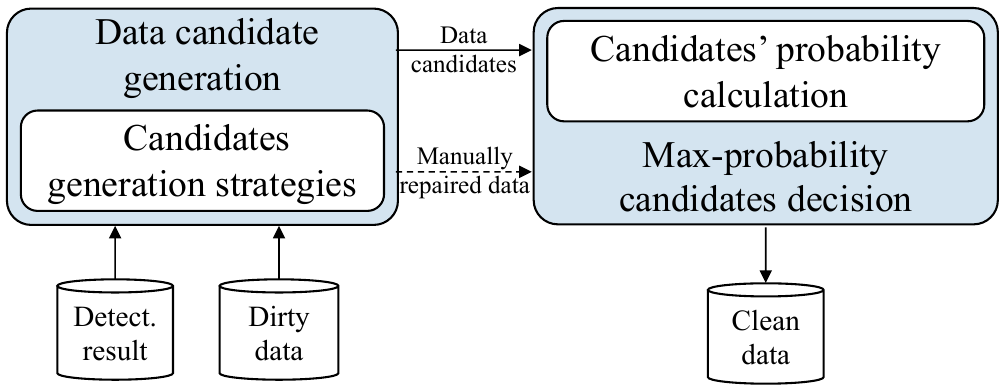}
  \vspace*{-0.11in}
  \caption{Workflow of \emph{data-driven} repair algorithms.} %Solid lines and dotted lines represent different data repair workflows.
 \label{fig:DPW-2}
\vspace*{-0.2in}
\end{figure}

\begin{table*}[t]
\vspace*{-0.06in}
  \caption{Summary and analysis of existing data repair algorithms.}
  \label{tab:Summary}
  \vspace*{-0.13in}
  \fontsize{7pt}{0.85\baselineskip}\selectfont
  \setlength{\tabcolsep}{0.9mm}
  \centering \footnotesize
  %\resizebox{\linewidth}{!}{
  \begin{tabular}{|c|c|c|c|c|c|c|c|c|}
  \hline
    \textbf{Cats.} &\textbf{Algorithms} & \textbf{Base model} & \textbf{Human config.} & \textbf{Repair goal} & \textbf{Candidate source} & \textbf{Candidate evaluation} & \textbf{Time complexity} & \textbf{Scala. Strategy}	\\
    \hline
    \multirow{6}{*}{\rotatebox{90}{\makecell[c]{Cstr-driven}}} 
    & Holistic~\cite{Chu13holistic}    & Graph-based               & C+Cost       & Consistency repair   & Equivalent class                 & Cardinality/Distance cost            & $O(|\Sigma|\cdot |I|^2 \cdot |A|)$ & None \\ \cline{2-9}
    & BigDansing~\cite{Khayyat15bigdansing}  & Graph-based               & C+Cost       & Consistency repair   & Equivalent class                 & Cardinality/Distance cost                                & $O(|\Sigma|\cdot |I|^2 \cdot |A|)$ & RR+Group. \\ \cline{2-9}
    & Horizon~\cite{Rezig21horizon}     & Graph-based               & C       & Consistency repair   & Equivalent class                 & Support score                                 & $O(|I|^2 \cdot |A|^2)$ & VA \\ \cline{2-9}
    & Nadeef~\cite{Ebaid13nadeef}      & Stats-based               & C+Cost       & Consistency repair   & Equivalent class                 & Cardinality cost                                & $O(|\Sigma|\cdot |I|^2 \cdot |A|)$ & None \\ \cline{2-9}
    & MLNClean~\cite{Ge22mlnclean}    & Stats-based               & C+LD       & Consistency repair   & Equivalent class                 & Reliability score                                 & $O(|\Sigma|\cdot |I|^2 \cdot |A|)$ & VA \\ \cline{2-9}
    & Daisy~\cite{Giannakopoulou20relaxation}       & Stats-based              & C       & Consistency repair   & Equivalent class                 & Probability                                 & $O(|\Sigma|\cdot |I|^2\cdot|A|)$ & None \\ \hline
    \multirow{3}{*}{\rotatebox{90}{\makecell[c]{Data-\\driven}}} 
    & Scare~\cite{Yakout13scared}       & ML-based               & LD+ML model     & Holistic repair      & Domain                & Probability                                & $O(|I|\cdot log|I|\cdot|A|^2))$ & None \\ \cline{2-9}
    & Baran~\cite{Mahdavi20baran}       & ML-based               & LD+ML model  & Holistic repair      & Domain+Str variation  & Probability                                 & $O(|I|^2\cdot|A|^2))$ & PS \\ \cline{2-9}
    & BoostClean~\cite{Krishnan17boostclean}  & Stats. boosting & Rep libs.      & Model perf. boost    & Mean+Mode+Median          & Model performance                                 & $O(|I|\cdot|A|)$ & Rep libs.\\
    \hline
    \multirow{3}{*}{\rotatebox{90}{\makecell[c]{Hybrid-\\driven}}} 
    & Unified~\cite{Chiang11unified}     & Heuristic-based         & C+4 Paras       & Tolerant repair      & Equivalent class                 & Description length cost                                & $O(|\Sigma|\cdot|I|^2\cdot|A|)$ & None \\ \cline{2-9}
    & Relative~\cite{Beskales13relative}    & Heuristic-based         & C+1 Para       & Tolerant repair      & Equivalent class                 & Cardinality cost                                & $O(|I|^2\cdot|A|^{|\Sigma||I|})$ & None \\ \cline{2-9}
    & HoloClean~\cite{Rekatsinas17holoclean}   & ML-based         & C+13 Paras+ML       & Holistic repair      & Domain                & Probability                                 & $O(|\Sigma|\cdot|I|^2\cdot|A|)$ & DP+TP+PS \\ \hline
  \end{tabular}
  %}
  \vspace*{-0.1in}
\end{table*}

\vspace*{-0.05in}
\subsubsection{Data-driven Data Repair Algorithms}
% \textbf{Workflow.}
The workflow of \emph{data-driven} data repair algorithms is plotted in Figure~\ref{fig:DPW-2}.
It takes dirty data and the \emph{error} detection results as inputs, and extracts data distribution information.
Data candidates are generated for error cells based on configured strategies. 
Then, the probability of a candidate  is calculated using ML models~\cite{Mahdavi20baran} or the likelihood benefit formulas~\cite{Yakout13scared}.
Note that, when computing the probability of each candidate, some algorithms may involve few manually repaired data, as indicated by the dashed line.
Finally, the candidates with the highest probabilities are decided as the repair decisions. 

\textbf{Scare}~\cite{Yakout13scared}.
To achieve holistic repair, \textsf{Scare} incorporates ML models into the data repair process.
To generate data candidates, a set of \emph{classifiers} is obtained by learning from attribute values detected as clean. 
These classifiers are utilized to predict potential candidates for values of dirty attributes.
Then, the joint probability of candidates is computed based on their co-occurring frequency.
Finally, \textsf{Scare} determines the final tuple repair by selecting the candidate values of dirty attributes with the highest joint probability. 
% \textsf{Scare} involves two main phases: the updates generation phase and the tuple repair selection phase. 
% These predicted values are considered candidate tuple repairs and stored in a temporary repair storage. 
% In the updates generation phase, 
% The primary goal of \textsf{Scare} is to maximize the benefits of replacement data.
% while also considering the dependability of the acquired classifiers.

\textbf{Baran}~\cite{Mahdavi20baran}.
 Same to \textsf{Scare}, \textsf{Baran} also pursues holistic repair.
When generating data candidates, instead of solely relying on dirty data information, \textsf{Baran} provides \emph{more comprehensive} candidate generation strategies based on various contexts like vicinity, domain, and string variation.
It thus significantly enhances the likelihood of including the correct values.
Within manually cleaned data and generated candidates of few wrong data, \textsf{Baran} trains a \emph{classifier} to determine the repair candidates for other error cells with the highest probabilities.  
% Specifically, it pretrains the models using Wikipedia revisions to enhance the candidate generation performance.
% Furthermore, \textsf{Baran} integrates transfer learning techniques to further enhance its performance. 
% is an innovative error correction system that considers a wide range of errors to conduct effective repair. 
% Unlike traditional error correction solutions that rely on hand-crafted rules or master data with plenty of configurations, \textsf{Baran} aims to derive corrections from a limited number of example repairs, achieved by capturing the full error-to-right pattern for each wrong value.
% The core idea of \textsf{Baran} is to generate a comprehensive array of candidates, thereby enhancing the likelihood of including the correct value.
% To accomplish this, \textsf{Baran} employs multiple error correction models grounded in various contexts, including factors like value, vicinity, and domain. 
% These models collectively generate multiple correction candidates for each data error.
% Then, with few labeled data and the generated candidates, \textsf{Baran} trains a classifier to determine the ultimate repair for other error cells.
% Furthermore, \textsf{Baran} integrates transfer learning techniques to further enhance its performance. Specifically, it pretrains the models using Wikipedia revisions to enhance their proficiency.

\textbf{BoostClean}~\cite{Krishnan17boostclean}.
\textsf{BoostClean} aims to boost the performance of \emph{downstream analysis} models.
Unlike the previous two methods, \textsf{BoostClean} follows an \emph{iterative} process.
It first generates data candidates using strategies like mean, mode, and median. 
Repaired data stemming from these candidates then trains the downstream analysis model. \textsf{BoostClean} computes 
The candidates' probability is computed by assessing the trained model's performance on validation data. This iterative process continues until maximum probability is reached, which translates into optimized model performance and enhanced quality of data repairs.

% The downstream analysis model is trained using the repaired data generated by these candidates. 
% \textsf{BoostClean} calculates the data candidates' probability by evaluating the performance of the trained model on validation data. 
% This iterative process continues until the highest probability is achieved, leading to enhanced downstream model performance and improved data repairs.

\vspace*{-0.09in}
\subsubsection{Hybrid-driven Data Repair Algorithms}
 % There exist two primary workflows for \emph{rule\&data-driven} methods.
Figure~\ref{fig:DPW-1} also depicts the workflow of \emph{hybrid-driven} methods (including the part with dashed arrows and modified constraints).  %They can repair both data and rules.
After the equivalent classes grouping process, \emph{hybrid-driven} algorithms generate \emph{both} constraint \emph{and} data candidates.
The repair algorithm computes the costs of data and \emph{constraint} candidates to determine the minimum-cost candidates.
These rectified constraints and data are then applied iteratively to further enhance repair, until the optimal state or cost minimum is reached, as indicated by the dashed arrows.
% The second workflow is similar to the workflow of \emph{rule-driven} methods, aiming to reach \emph{holistic repair}.
% The distinctions emerge in two aspects.
% Firstly, prior to cost calculation for candidates, the algorithm identifies likely correct data for conducting statistical learning.
% Secondly, in this workflow, all values within the attribute domain are considered as candidates, eliminating the need for an equivalent classes grouping process.

\textbf{Unified}~\cite{Chiang11unified}. \textsf{Unified} aims to achieve \emph{tolerant repair} with minimal description length (DL)~\cite{rissanen1978modeling} for the entire data with FDs.
It first calculates the overall DL of each FD and its equivalent classes.
The constraints are then processed one by one in the decreasing order of DL.
For each FD, \textsf{Unified} generates both constraint and data candidates based on the equivalent classes.
To determine the final repair, the candidates are evaluated by DL-based cost.
Candidates with minimal cost are decided for the repair.
The repaired constraints and data are then taken into consideration when processing the next constraint, continuing until reaching the overall minimal DL.

\textbf{Relative}~\cite{Beskales13relative}.
Similar to \textsf{Unified}, \textsf{Relative} aims to achieve \emph{tolerant repair} with FDs. 
% For each candidate, repairs are made aiming to meet a specified threshold. 
To generate both constraint and data candidates,  \textsf{Relative} first explores FD modification space to generate constraint candidates, following grouping equivalent classes.
For each constraint candidate, \textsf{Relative} endeavors to repair data with the specified threshold. % $\tau$.  
Finally, from the candidates that meet threshold conditions, the ones with the least cardinality costs are determined as the final repairs.  
% , meanwhile ensuring the existence of a data repair satisfying the modified rule set.
% when FDs may not be entirely correct due to data updates or other occasions. 
% operates within the second workflow mentioned above
% It offers another solution for data cleaning when FDs may not be entirely correct due to data updates or other occasions.
% \textsf{Relative} argues that considering the notion of relative trust is crucial, as users need to decide to what extent they should rely on the data and FDs to achieve consistency.

\textbf{HoloClean}~\cite{Rekatsinas17holoclean}.
Though utilizing both constraint and data information, \textsf{HoloClean} focuses on \emph{holistic repair}, thus lacking the steps with dashed arrows and modified constraints in Figure~\ref{fig:DPW-1}. 
After the detection process, \textsf{HoloClean} skips the equivalent classes grouping, treating all values in the dirty data as data candidates.
For candidate determination, \textsf{HoloClean} extracts quantitative statistics of constraint and identifies probable correct values with \emph{Naive Bayesian} model.
Utilizing these statistics and data, \textsf{HoloClean} learns a statistical model and calculates candidate probabilities via \emph{DeepDive} framework~\cite{Shin15deepdive}.
Ultimately, data candidates with highest probability are selected as final repairs.

\vspace{-0.04in}
\subsubsection{Discussion} 
Table~\ref{tab:Summary} summarizes these data repair algorithms from several aspects. The scalable strategy involves reducing redundancy (RR), parallel strategy (PS), domain pruning (DP), tuple partition (TP), and value-based approach (VA). The human configuration includes cost function (Cost), matching learning (ML) model, constraints (C), repair libraries (Rep libs.), parameter (Para), and labeled data (LD)
The time complexity is estimated based on the constraint set size $|\Sigma|$, instance size $|I|$, and attribute set size $|A|$.

It is shown in Table~\ref{tab:Summary} that \emph{cstr-driven} and some \emph{hybrid-driven} algorithms mainly use equivalent classes as candidate sources, minimizing differences between candidates and original dirty data. However, these methods mainly consider constraint equivalence, potentially overlooking other logical or domain-specific relationships. This leads to possible omission of latent correct values and inadequate data error elimination.
In contrast, unconstrained methods like \textsf{HoloClean} and \emph{data-driven} algorithms make more liberal use of information, including domain and string variations. This increases the likelihood of including latent clean data in the candidate pool. However, \textsf{BoostClean}'s reliance on mean, mode, and median values falls short in identifying latent clean data, resulting in inferior performance in experiments.
% {\color{red}Better to connect with the experimental results.}

The evaluation of data repair candidates hinges on the \emph{minimization} of costs and \emph{maximization} of certain scores. 
Among them, probability is heavily influenced by the applied models, forming the basis of data-driven and some hybrid-driven methods. 
Other evaluation metrics depend on heuristic approaches and are constrained by available information within equivalent classes, as employed by \emph{cstr-driven} and some \emph{hybrid-driven} methods.
These metrics can be customized to prioritize specific elements, making the derived repair decisions easy to explain. 
However, these evaluates presume the majority of the data is accurate, posing challenges when dealing with significantly erroneous datasets.

% The evaluation of data repair candidates mainly involves the \emph{minimization} of cardinality or distance costs, and  \emph{maximization} of support score, reliability score, and probability.
% Among them, probability is heavily influenced by the applied models, which determine the capacity to learn meaningful patterns from the data based on various assumptions.
% \emph{Data-driven} and some \emph{hybrid-driven} methods mainly evaluate candidates based on probability metric.
% The other evaluation metrics depend on heuristic approaches and are usually constrained by the information available within equivalent classes, as employed by \emph{cstr-driven} and some \emph{hybrid-driven} methods.
% From another perspective, They can be tailored to prioritize specific variables such as cardinality or distance, depending on the requirements of the task.
% Hence, the derived repair decisions are straightforward to explain, due to the clear definition of costs and scores.
% It is crucial to note that, all of these evaluation metrics operate under the assumption that the majority of the data is accurate. 
% In other words, current data repair methods may face challenges when confronted with datasets containing a substantial number of errors. 
% {\color{red}Better to connect with the experimental results.}

The time complexity analysis is primarily based on the instance size $|I|$, the constraint set size $|\Sigma|$, and the number of attributes $|A|$. 
Most data repair methods maintain an $O(|I|^2)$  complexity due to necessary tuple comparisons.
The real execution time is subject to various factors such as error and data distribution.
% , and the content being compared significantly influences the time cost.
% Since the comparison of distinct values rather than tuples involves a much lower operation cost, leading to significant time savings.
% Time costs can be mitigated by comparing distinct values rather than tuples, resulting in measurable time savings.
Besides, methods like \textsf{MLNClean} and \textsf{Horizon} focus on operating on the values, rather than the tuples, often exhibiting shorter running time in the experiments.
For \textsf{Scare}, its time complexity is at the $O(|I|\cdot \log|I|)$ level, but the classifier training process involved in candidate generation can be time-consuming.
\textsf{BoostClean} operates with a time complexity at the $O(|I|)$ level, owing to the limited candidate sources. 
\textsf{Relative} stands out as an exception. Although it exhibits a time complexity at $O(|I|^2)$ level, its exploration of the FD modification space results in exponential time complexity with $O(|A|)$, which is also confirmed in our experimental study. 
% {\color{red}Better to connect with the experimental results.}

For human configurations, 
they can be classified into two categories of human cost: for elements such as constraints, Rep libs, and labeled data, in-depth knowledge is imperative; for ML models, parameters, and cost functions, the emphasis is on practical deployment experience.
Algorithms such as \textsf{Horizon}, \textsf{BoostClean}, and \textsf{Daisy} necessitate minimal knowledge and deployment expenses. While \textsf{HoloClean} demands extensive knowledge and thorough deployment analysis, suggesting a higher cost. 
Moreover, \textsf{BigDansing}, \textsf{Holistic}, \textsf{Nadeef}, \textsf{Relative}, and \textsf{Unified} require additional deployment experience beyond the standard constraints. 
\textsf{MLNClean} also calls for additional specialized knowledge. 
\textsf{Scare} and \textsf{Baran} require labeled data and their implementation of ML models may incur additional training time costs.

\begin{table}[t]
\vspace*{-0.08in}
\fontsize{8pt}{0.9\baselineskip}\selectfont
\centering
\vspace*{-0.1in}
\caption{Case study of EDR.}
\vspace*{-0.15in}
\setlength{\tabcolsep}{1.4mm}
\begin{tabular}{|c|c|c|c|c|c|c|c|}
\hline
\textbf{Case} & \bm{$\#d_{w}$} & \bm{$\#d_{w2r}$} & \bm{$\#d_{w2w}$} & \bm{$\#d_{r2w}$} & \textbf{EDR} & \textbf{Precision} & \textbf{Recall} \\ \hline
case 1 &100 & 10 &  0 & 10 & 0    & 0.50 & 0.10    \\ \hline
case 2 & 100 & 10 & 90 & 0  & 0.10 & 0.10 & 0.10   \\ \hline
case 3 &100 & 10 & 10 & 10 & 0    & 0.33 & 0.10    \\ \hline
\end{tabular}
\label{tab:analysis_pre}
\vspace*{-0.18in}
\end{table}

\vspace{-0.04in}
\subsection{Optimization Strategy} 
In this subsection, we propose an effective strategy to optimize the data repair algorithms. 
We also present a new metric to fairly evaluate the final data repair results. 

The primary cause of suboptimal in existing repair algorithms is \emph{incorrect repair} on \emph{initial right} data. 
For methods aiming at consistency repair and tolerant repair, this stems mainly from two factors. 
First, during repair candidate selection, \emph{faulty} information is used as constraint violation cells are not classified as incorrect or correct. 
Second, the selection process may result in \emph{local-optimal} repairs due to the sole use of equivalent classes. For algorithms targeting \emph{holistic repair}, despite considering a more comprehensive data range, the error detection performance may fall short compared to a dedicated error detection model.

% A major cause of suboptimal performance in existing data repair algorithms is \emph{incorrect repair}. 
% Obviously, incorrectly repairing erroneous data does not increase the error rate, while incorrectly repairing correct data \emph{does}. For repair methods aiming at consistency repair and tolerant repair, incorrectly repairing correct data is mainly attributed to two factors.
% First, \emph{wrong} information is utilized in the repair candidate selection process, since constraint violation cells are not explicitly specified as erroneous or accurate. 
% Second, the candidate selection may yield \emph{local-optimal} rather than global-optimal repairs, due to the sole consideration of equivalent class information. 
% For data repair methods aiming at \emph{holistic repair} with error detection modules, despite considering a broader range of information in the detection process, the performance may not be as ideal as that of the specifically designed error detection model. 
As a result,  we propose to adopt \textsf{Raha}~\cite{raha19mahdavi} as an \emph{optimization strategy} to further improve existing data repair algorithms, for its state-of-the-art (sota) error detection performance~\cite{raha19mahdavi,Pham21spade,Abdelaal23rein}.  
Specifically, the error detection results derived by \textsf{Raha} can be employed to ensure that,  the identified correct data remains unchanged during the data repair process, thereby effectively mitigating the risk of erroneously changing \emph{right} data.
Although inconsistencies may arise, our experiments validate its efficiency.
% It is inspired by the fact that 
% contemporary state-of-the-art (SOTA) error detection methods excel at accurately identifying correct data.
% Considering the potential harm caused by incorrectly repairing accurate data, 

Moreover, we attempt to more accurately measure the ultimate data repair performance. 
As analyzed earlier, metrics like \emph{precision} and \emph{recall} are \emph{biased}, and distance-based metrics fail to indicate the \emph{improvement degree} or conditions of data error increasing.
We denote the initial right data repaired into right and wrong ones by $\#d_{r2r}$ and $\#d_{r2w}$, respectively. 
The number of initial wrong data repaired into right and wrong ones are denoted by $\#d_{w2r}$ and $\#d_{w2w}$, respectively.
Hence, precision is calculated as:  $pre=\frac{\#d_{r2r}+\#d_{w2r}}{\#d_{r2r}+\#d_{w2r}+\#d_{r2w}+\#d_{w2w}}$. 
Obviously, $\#d_{r2r}$ increases precision according to the mediant inequality, and $\#d_{w2w}$ distorts the precision calculation, making a high precision not guarantee error reduction, and vice versa,
as cases shown in Table~\ref{tab:analysis_pre}.
Further, the \emph{F1} score, as the harmonic mean of precision and recall, cannot explicitly reflect the error reduction degree of data.
% among them, $\#d_{r2r}$ and $\#d_{w2w}$ have no impact on the data error rate, while $\#d_{w2w}$ decreases the precision and 
% Recall only reflects the effectiveness of eliminating original errors, disregarding extra introduced errors by repair algorithms. 
% The precision calculation is based on proportions and includes cells unrelated to data quality.

To this end, we introduce a new definition termed \emph{Error Drop Rate (EDR)} to measure the error reduction performance of data repair algorithms.
It disregards irrelevant data quality factors, and directly evaluates the error reduction degree.
For generality, we defined it as follows.
\begin{equation}
\begin{aligned}
EDR=\frac{dis^{d2c}-dis^{r2c}}{dis^{d2c}}.
\end{aligned}
\label{eq:edr}
\vspace*{-0.05in}
\end{equation}
Here, $dis^{d2c}$ indicates the distance between dirty and clean data, whereas $dis^{r2c}$ is the distance from repaired data to clean data.
Aligning with the problem statements, in the experiments, in the experiments, $dis^{d2c}$ equals to $d_w$, and $dis^{r2c}$ equals to $d_w-(d_{w2r}-d_{r2w})$.
% While overlooking $d_{r2r}$ might underestimate the algorithms' repair capability, the advantages of the new EDR metrics are numerous and valuable.
Though ignoring $d_{r2r}$ and $d_{w2w}$ somehow neglects the algorithms' repair ability, the benefits of EDR are multiple. 
Firstly, EDR allows effective evaluation of data quality enhancements by contrasting pre and post repaired data distance. Secondly, the normalization process enables equal comparison on repair effectiveness irrespective of initial data range and quality. Lastly, EDR provides a straightforward depiction of error reduction/raising (with positive and negative values), and its monotonic feature signifies that increased values align with larger data quality improvements.
% it employs $\#d_{w2r}$, $\#d_{r2w}$, and the initial wrong data count denoted by $\#d_{w}$, to calculate \emph{EDR}. 
% The exclusion of $\#d_{w2w}$ and $\#d_{r2r}$ from the calculation allows \textit{EDR} to more accurately depict the error reduction performance, since they do not impact the data error rate.

% Moreover, \textit{EDR} offers a more straightforward interpretation. Negative and positive \textit{EDR} values signify a reduction or increase in the data error rate, with higher \textit{EDR} values indicating a more substantial reduction in errors.
 
% Further explain why EDR works and is meaningful, and highlight its advantages. 
% The new metric  \emph{EDR} provides more straightforward insight into error changes and better reflects error reduction performance.

% {\color{red}This approach parallels established precision, recall, and F1 score calculations,} with \emph{DEC} corresponding to \emph{false positives} and \emph{IEC} representing \emph{true negatives}.

\section{Experimental Evaluation} \label{sec:experiment}
% \subsection{Method Robustness Evaluation}
% \paragraph{P1:} 

\begin{table}[tbp]
\vspace*{-0.08in}
\fontsize{7.5pt}{0.85\baselineskip}\selectfont
\vspace*{-0.1in}
  \caption{The used datasets in repair evaluation. }
  \label{tab:Dataset}
  \vspace{-0.15in}
  \centering
    % \setlength{\tabcolsep}{1.4mm}
  %\resizebox{\columnwidth}{!}{
  \begin{tabular}{|c|r|r|r|r|r|}
  \hline
        \textbf{Name}      & \makecell[c]{\textbf{\#Tuples}}  & \makecell[c]{\textbf{\#Attrs}} & \makecell[c]{\textbf{Error rate}} & \makecell[c]{\textbf{Error types}} & \makecell[c]{\textbf{\#Cstrs}} \\
    \hline
        Hospital  & 1,000  &  20   & 3\%        & T, VAD & 15 \\ \hline
        Flights   & 2,376  &   7   & 30\%       & MV, FI, VAD & 6 \\ \hline
        Beers     & 2,410  &  11   & 16\%       & MV, FI, VAD & 5 \\ \hline
        Rayyan    & 1,000  &  11   & 9\%        & MV, T, FI, VAD & 10 \\ \hline
        Tax       & 200,000  &  15   & 4\%        & T, FI, VAD & 8 \\ 
    \hline
  \end{tabular}
 % }
  \vspace{-0.13in}
\end{table}

\begin{table*}[ht]
% \vspace*{-0.03in}
  \small
  \centering
  \caption{Data repair performance comparison on real-world datasets.}
  \vspace*{-0.15in}
  \label{tab:repair_real_data}
  \setlength{\tabcolsep}{0.3mm}
  \resizebox{\linewidth}{!}{
  \begin{tabular}{|c|c|rrr|rrr|rrr|rrr|rrr|}
  \hline
    \multirow{2}{*}{\textbf{Cat.}} 
    & \multirow{2}{*}{\textbf{Algos.}} 
    & \multicolumn{3}{c|}{\textbf{\makecell[c]{Hospital}}} & \multicolumn{3}{c|}{\textbf{\makecell[c]{Flights}}} & \multicolumn{3}{c|}{\textbf{\makecell[c]{Beers}}} & \multicolumn{3}{c|}{\textbf{\makecell[c]{Rayyan}}} & \multicolumn{3}{c|}{\textbf{\makecell[c]{Tax-10k}}}  \\ \cline{3-17}
    & & \makecell[c]{EDR} & \makecell[c]{F1} & \makecell[c]{Hybrid} & \makecell[c]{EDR} & \makecell[c]{F1} & \makecell[c]{Hybrid} & \makecell[c]{EDR} & \makecell[c]{F1} & \makecell[c]{Hybrid} & \makecell[c]{EDR} & \makecell[c]{F1} & \makecell[c]{Hybrid} & \makecell[c]{EDR} & \makecell[c]{F1} & \makecell[c]{Hybrid}   \\ 
    \hline 
    \multirow{6}{*}{\rotatebox{90}{\makecell[c]{Rule-driven}}} 
    & BigDansing & $-$0.0897\ding{199} & 0.6239\ding{197} & 0.0298\ding{199} & $-$0.1665\circled{\scriptsize{11}} & 0.3772\ding{197}   &  0.4193\ding{201} & $-$0.0111\ding{197} & 0.0802\ding{194} & 0.0185\ding{195} & $-$0.4540\ding{196}  & 0.0259\ding{194}  &  0.4010\ding{197} & $-$1.1938\ding{196} &  0.1463\ding{193} &  0.3058\ding{199}     \\ \cline{2-17}
    & Holistic  & $-$0.0039\ding{198} & 0.6403\ding{195} &  0.0282\ding{197} & $-$0.1335\ding{199} & 0.3912\ding{195}  & 0.4014\ding{198}  &  $-$0.0110\ding{195} & 0.0688\ding{195}  & 0.0186\ding{196}  & $-$0.9614\ding{199}  & 0.0047\ding{195} &  0.4616\ding{199}  & $-$1.1938\ding{196} &  0.1463\ding{193} & 0.0099\ding{195}      \\ \cline{2-17}
    & Horizon   & 0.0530\ding{196} & 0.5661\ding{198} & 0.0264\ding{195}  & 0.1148\ding{193} & 0.3869\ding{196}  & 0.3290\ding{193}  &  $-$0.0110\ding{195} & 0.0688\ding{195}  & 0.0219\ding{197}  & $-$0.9614\ding{199}  & 0.0047\ding{195} & 0.3528\ding{196}   & $-$50.957\ding{199} &  0.1134\ding{195} & 0.0162\ding{196}    \\ \cline{2-17}
    & Nadeef    & $-$1.7996\ding{201} & 0.0713\ding{201} & 0.1074\ding{201}   & $-$0.0528\ding{198} & 0\ding{200}  & 0.3828\ding{197}  &  $-$0.4783\ding{199} & 0.0094\ding{199}  & 0.1101\ding{200}  & $-$2.5367\circled{\scriptsize{11}}  & 0\ding{197} &  0.9503\circled{\scriptsize{11}} & $-$55.387\ding{200} &  0.0009\ding{199} & 0.2074\ding{198}   \\ \cline{2-17}
    & MLNClean    & 0.4322\ding{195} & 0.7240\ding{193} & 0.0152\ding{193}  & $-$0.0126\ding{197} & 0.0051\ding{198} &  0.3576\ding{196} &  0.0482\ding{193} & 0.1191\ding{193}  &   \textbf{0.0057}\ding{192} & $-$0.6042\ding{197}  & 0\ding{197} & 0.4128\ding{198}   & $-$0.1147\ding{195} &  \textbf{0.1927}\ding{192} & 0.0053\ding{194}      \\ \cline{2-17}
    & Daisy    & 0\ding{197} & 0\circled{\scriptsize{11}} &  0.0284\ding{198} & 0\ding{195} & 0\ding{200} & 0.3519\ding{195} & 0\ding{194} & 0\ding{200}  & 0.0152\ding{194} & 0\ding{193}  & 0\ding{197} & 0.2652\ding{194}  & n/a &  n/a & n/a   \\
    \hline
    \multirow{3}{*}{\rotatebox{90}{\makecell[c]{Data-\\driven}}} 
    & Scare  & $-$0.5350\ding{200} & 0.1511\ding{200} &  0.0566\ding{200}  & $-$0.1364\ding{200} & 0.0002\ding{199}  & 0.4017\ding{199}  &  $-$0.5238\ding{200}  & 0.0015\ding{200}  & 0.1036\ding{199}  & $-$0.0886\ding{194}  & 0\ding{197} &  0.3000\ding{195}  & $-$7.4933\ding{198} &  0.0013\ding{198} &  0.0337\ding{197}     \\ \cline{2-17}
    & Baran & 0.4872\ding{194} & 0.6299\ding{196} &  0.0156\ding{194}  & \textbf{0.4910}\ding{192} & \textbf{0.6369}\ding{192} & \textbf{0.1894}\ding{192}  & \textbf{0.7245}\ding{192} & \textbf{0.7514}\ding{192}  & 0.0073\ding{193}  & \textbf{0.7403}\ding{192}  & \textbf{0.8415}\ding{192} &  \textbf{0.1005}\ding{192}  & \textbf{0.0160}\ding{192} &  0.0634\ding{196} & 0.0050\ding{193}      \\ \cline{2-17}
    & BoostClean   & $-$5.7132\circled{\scriptsize{11}} & 0.3310\ding{199} &  0.3152\circled{\scriptsize{11}}  & $-$0.0028\ding{196} & 0\ding{200} & 0.5088\circled{\scriptsize{11}}  &  $-$0.7174\ding{201} & 0\ding{200}  & 0.1581\ding{201}  & $-$0.6220\ding{198}  & 0\ding{197} &   0.4869\ding{200}  & 0\ding{193} &  0\ding{200} &  \textbf{0.0049}\ding{192}   \\ 
    \hline
    \multirow{3}{*}{\rotatebox{90}{\makecell[c]{Hybrid- \\ driven}}} 
    & Unified  & \textbf{0.6012}\ding{192} & \textbf{0.7826}\ding{192} &  0.0268\ding{196}  & 0.0415\ding{194} & 0.5579\ding{193} & 0.3414\ding{194}  & $-$0.1221\ding{198} & 0.0106\ding{198}  & 0.0538\ding{198}  & $-$0.1862\ding{195}  & 0\ding{197} &  0.2612\ding{193}  & n/a &  n/a & n/a      \\ \cline{2-17}
    & Relative   & n/a	 & n/a & n/a & n/a & n/a & n/a & n/a & n/a & n/a & n/a & n/a & n/a & n/a & n/a & n/a  \\ \cline{2-17}
    & HoloClean & 0.4872\ding{193} & 0.6515\ding{194} & \textbf{0.0137}\ding{192}   & $-$0.1390\ding{201} & 0.4711\ding{194} & 0.4019\ding{200}  &  $-$3.7310\circled{\scriptsize{11}} & 0.0652\ding{197}  & 0.7141\circled{\scriptsize{11}}  & $-$1.8897\ding{201} & 0.6467\ding{193} &  0.7674\ding{201} & $-$0.0213\ding{194}  & 0.0417\ding{197} & 0.7020\ding{200}      \\ 
    \hline
  \end{tabular}
  }
 \vspace*{-0.08in}
\end{table*}

\begin{table*}[ht]
  \small
  \centering
  \caption{The case study for in-depth metric analysis.}
  \vspace*{-0.15in}
  \label{tab:case_study}
  \setlength{\tabcolsep}{0.4mm}
  \resizebox{\linewidth}{!}{
  \begin{tabular}{|c|r|r|r|r|r|r|r|r|r|r|r|r|r|r|r|r|r|r|r|r|r|r|r|r|r|}
  \hline
    \multirow{2}{*}{\textbf{Algos.}} 
    & \multicolumn{5}{c|}{\textbf{\makecell[c]{Hospital}}} & \multicolumn{5}{c|}{\textbf{\makecell[c]{Flights}}} & \multicolumn{5}{c|}{\textbf{\makecell[c]{Beers}}} & \multicolumn{5}{c|}{\textbf{\makecell[c]{Rayyan}}} & \multicolumn{5}{c|}{\textbf{\makecell[c]{Tax-10k}}}  \\ \cline{2-26}
    & \makecell[c]{$\#d_{w}$} & \makecell[c]{$\#d_{w2r}$} & \makecell[c]{$\#d_{w2w}$} & \makecell[c]{$\#d_{r2r}$} & \makecell[c]{$\#d_{r2w}$} & \makecell[c]{$\#d_{w}$} & \makecell[c]{$\#d_{w2r}$} & \makecell[c]{$\#d_{w2w}$} & \makecell[c]{$\#d_{r2r}$} & \makecell[c]{$\#d_{r2w}$} & \makecell[c]{$\#d_{w}$} & \makecell[c]{$\#d_{w2r}$} & \makecell[c]{$\#d_{w2w}$} & \makecell[c]{$\#d_{r2r}$} & \makecell[c]{$\#d_{r2w}$} & \makecell[c]{$\#d_{w}$} & \makecell[c]{$\#d_{w2r}$} & \makecell[c]{$\#d_{w2w}$} & \makecell[c]{$\#d_{r2r}$} & \makecell[c]{$\#d_{r2w}$} & \makecell[c]{$\#d_{w}$} & \makecell[c]{$\#d_{w2r}$} & \makecell[c]{$\#d_{w2w}$} & \makecell[c]{$\#d_{r2r}$} & \makecell[c]{$\#d_{r2w}$}  \\ 
    \hline 
    BigDansing & 509 & 417 & 2 & 58 & 462 & 4,920 &  2,188  & 1,521 & 2 & 3,026  & 3,358  & 176 & 0 &  3 & 213 & 2,873 & 112   & 418  & 19 & 1,338 & 375 & 49 & 200 & 4 & 448    \\ \hline
    Baran & 509 & 221 & 48 & 0 & 3 & 4,920 &  2,773  & 214 & 0 & 230  & 3,358  & 2,378 & 837 & 0 & 1 & 2,873 & 2,011 & 108  & 0 & 3  & 375 & 27 & 173 & 0 & 2    \\ \hline
    HoloClean & 509 & 248 & 187 & 10,603 & 0 & 4,920 &  1,712  & 3,208 & 6,933 & 1  & 3,358  & 121 & 121 & 2,483 & 744 & 2,873 & 1,217   & 813  & 3,335 & 1,920 & 375 & 8 & 254 & 7,371 & 19    \\ \hline
  \end{tabular}
  }
  \vspace*{-0.1in}
\end{table*}

In this section, we evaluate 12 mainstream data repair algorithms to address the following key questions: 
\textbf{Q1:} To what degree can existing automatic data repair algorithms mitigate errors in real-world datasets?   
\textbf{Q2:} Can algorithms with suboptimal repair performance enhance the performance of downstream analysis models? 
\textbf{Q3:} How effective is the proposed general optimization strategy?
\textbf{Q4:} How effectively can these algorithms manage high error rates and different error types?

%\subsection{Experiment Settings}
% \noindent \textbf{Dataset.} We evaluate these 13 algorithms on 3 real datasets and 1 synthetic dataset that are described in {\color{blue} tablex}
%\noindent 
\underline{Datasets}.
We conduct repair experiments over five real-world datasets, as outlined in Table ~\ref{tab:Dataset}.  
\emph{Hospital} and \emph{Flights} \cite{Rekatsinas17holoclean} include a high degree of duplicate tuples and correlated columns.
% The data errors of \emph{Hospital} are scarce while the degree of trustworthy contextual information of \emph{Flights} is lower with a high error rate. 
\emph{Beers} is a real-world dataset sourced through web scraping and manually cleaned by the dataset owner~\cite{Mahdavi20baran}.
% It provides information on different beers, containing 2,410 tuples and 11 attributes. 
\emph{Rayyan} is another real-world dataset cleansed by its owners ~\cite{raha19mahdavi}. It contains all the error types, thus making it hard to repair.
\emph{Tax} is a large dataset describing tax payment records~\cite{Arocena15bart}. It contains various data error types with 200,000 tuples and 15 attributes. 
Rich with insights into social resource allocation and commercial values, these datasets are valuable for academic inquiry and \emph{data-driven} decision-making. 
Four kinds of error types, namely missing value (MV), typo (T), violated attribute dependency (VAD), and formatting issue (FI), exist in the datasets, and are primarily encountered in real-world scenarios~\cite{Mahdavi20baran,Arocena15bart,raha19mahdavi,Pham21spade}. 
They stem from either incorrect value allocation, termed semantic errors, or the presence of out-of-domain values, known as syntactic errors~\cite{Mahdavi20baran,Pham21spade}.
For each dataset, the corresponding clean version exists, applied to get evaluation metrics and further experiments. 
To discover denial constraints (DCs), we initially employ two widely-used DC discovery methods DCFinder~\cite{dcfinder19Pena} and Hydra~\cite{hydra17Bleifu}.
We then manually check all discovered rules, deciding whether to accept, modify, or deny each rule. 
The final applied rules can be viewed in the code repository. 
% {\color{red}The used rules are listed in the table.} 

%\noindent 
%\underline{Error Generation}.
%This data generation approach covers the comprehensive set of errors observed 
 
To comprehensively explore various error rates and error types scenarios \emph{not covered} by real-world datasets, we generate dirty data by adding erroneous values randomly into clean data. 
For semantic errors, we generate errors by randomly selecting alternatives from the domain.
Regarding syntactic errors, we introduce typos, both \emph{explicit} and \emph{implicit} missing values, and Gaussian noise, aligning with T, MV, and FI but with a broader range. 
These errors are generated using the publicly available code from BigDaMa~\cite{errorgurl}.

\vspace{0.04in}
\underline{Evaluation metrics}. 
We employ the proposed metric \emph{EDR}, and running time \emph{runtime} (in seconds), CPU 
(in percent), and memory (in Mb) usage to measure the effectiveness and efficiency of data repair algorithms. 
For comparison, we report 
\emph{F1 score} in real scenarios, considering its wide application in previous studies~\cite{Chu13holistic, Chu16datacleaning, Beskales13relative}.
We have also applied a \emph{hybrid distance metric}, denoted by $hybrid\_dis=w_1* MSE + w_2*Jac\_dis$, where $MSE$ represents mean squared error, and $Jac\_dis$ denotes Jaccard distance. 
The weights $w_1$ and $w_2$ are the proportion of values calculated using $MSE$ and $Jac\_dis$, respectively.
%\emph{Precision} is defined as the ratio of correctly fixed errors to all fixed data and \emph{recall} represents the ratio of correctly fixed data errors to all data errors. 
% {\color{blue}We also present a case study to assess the performance of candidate \emph{generation} and \emph{decision} during the repair process.} 
% In the \emph{generation} stage, we calculate \emph{acc@CG}, which represents the ratio of candidate data sets \emph{containing} clean data to all candidate data sets.
% In the \emph{decision} stage, we calculate \emph{acc@CD}, denoting the ratio of correct decisions to all decisions in the candidate data \emph{containing} clean data.
For downstream data analysis tasks, we employ prevail \emph{F1} score for classification, \emph{mean squared error} (MSE) for regression, \emph{silhouette score} for clustering, and \textit{accuracy} for k nearest neighbor query (kNN), where k=1 in the experiments. 
% {\color{blue}For evaluating error detection effectiveness, we rely on the widely recognized metrics of precision, recall, and F1 score.
% Due to space limitations, our presentation in the results focuses exclusively on f1 score for error detection (\emph{ED\_F1}).
% Considering that the error detection goals of these data repair algorithms may encompass rule violations or outliers, for which explicit detection results are not guaranteed, we maintain a consistent evaluation approach based on the disparities between repaired and original cells.}

%As demonstrated in the following experiments, \emph{EDR} can provide more straightforward insight into error changes and better reflect error reduction performance.
%Additionally, for evaluating efficiency, we include the running time (\emph{runtime}) in seconds.
% Since some repairs may target correct cells with identical values, which usually have no positive influence on the original data.

%\noindent 
%\underline{Hyper-parameter Settings}.

%\noindent 
\underline{Implementation details}.
%Our experiment evaluates 12 representative data repair algorithms in Section~\ref{sec:methods}.
% These algorithms are carefully chosen from both academic literature and practical projects.
We re-implement \emph{seven} pieces of codes for Holistic~\cite{Chu13holistic}, BigDansing~\cite{Khayyat15bigdansing}, Horizon~\cite{Rezig21horizon}, Daisy~\cite{Giannakopoulou20relaxation}, Scare \cite{Yakout13scared}, Unified~\cite{Chiang11unified}, and Relative~\cite{Beskales13relative}, due to the unavailable source codes.
We use publicly accessible codebases for other five methods.
For \textsf{MLNClean} and \textsf{Baran}, which require manual cleaning, a consistent minimum of 20 tuples is used.
To feed detection results into \textsf{Baran} and \textsf{Scare}, we use \textsf{Raha}~\cite{raha19mahdavi}.
For other algorithms, we adopt their initial detection methods as their repair processes are closely tied to them.
For \textsf{Holistic, BigDansing, Nadeef} and \textsf{Relative}, we employ the original cardinality functions used in the paper. 
The other hyper-parameter settings are adopted as specified in the source codes or papers. 
Each reported result represents the average of \emph{three} repeated experiments.  All experiments are conducted on a Linux server with an Intel (R) Xeon (R) Gold 6326 CPU @ 2.90GHz and 512GB RAM, running Ubuntu 20.04.6 LTS.
\begin{figure*}[htbp]
% \vspace*{-0.13in}
\centering
\includegraphics[width=0.9\linewidth]{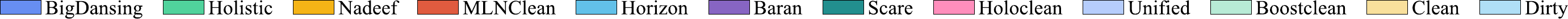}
% \vspace*{-0.05in}
\\
% \includegraphics[width=\linewidth]{bar-legend-12.pdf}\vspace*{-0.02in}
% \\
\hspace*{-0.04in}
\includegraphics[height=0.096\linewidth]{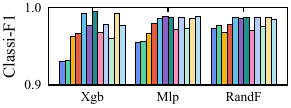}
\hspace*{-0.03in}
\includegraphics[height=0.096\linewidth]{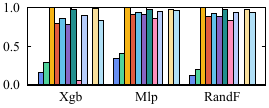}
\hspace*{-0.03in}
\includegraphics[height=0.096\linewidth]{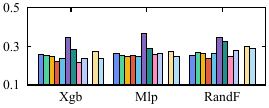}
\hspace*{-0.03in}
\includegraphics[height=0.096\linewidth]{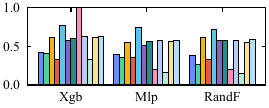}
\hspace*{-0.07in}
\vspace*{-0.03in}
\\
\hspace*{-0.04in}
\includegraphics[height=0.096\linewidth]{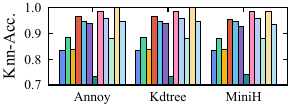}
\hspace*{-0.03in}
\includegraphics[height=0.096\linewidth]{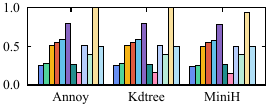}
\hspace*{-0.03in}
\includegraphics[height=0.096\linewidth]{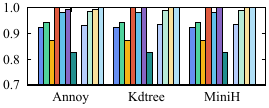}
\hspace*{-0.03in}
\includegraphics[height=0.096\linewidth]{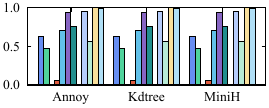}
\hspace*{-0.07in}
\vspace*{-0.1in}
\\
\hspace*{-0.07in}
\subfigure[\textit{Hospital}]{\raisebox{-0.2cm}{
\includegraphics[height=0.096\linewidth]{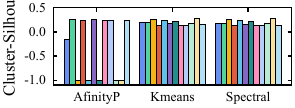}}
}\hspace*{-0.09in}
\subfigure[\textit{Flights}]{\raisebox{-0.2cm}{
\includegraphics[height=0.096\linewidth]{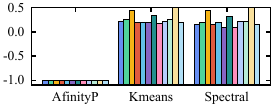}}
}\hspace*{-0.09in}
\subfigure[\textit{Beers}]{\raisebox{-0.2cm}{
\includegraphics[height=0.096\linewidth]{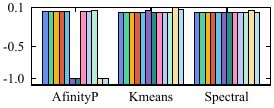}}
}\hspace*{-0.09in}
\subfigure[\textit{Rayyan}]{\raisebox{-0.2cm}{
\includegraphics[height=0.096\linewidth]{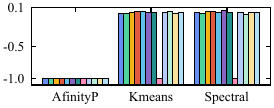}}
\hspace*{-0.07in}
}
% \hspace*{-0.05in}
\vspace*{-0.18in}
\caption{Performance of classification, kNN, and cluster on repaired data.}
\label{fig:three-tasks-error_rate}
\vspace*{-0.15in}
\end{figure*}

\vspace{-0.13in}
\subsection{Repair Performance on Real Scenarios}
The first set of experiments evaluates the performance of all data repair algorithms over five real-world datasets. 
Table \ref{tab:repair_real_data} lists the corresponding experimental results in terms of \emph{EDR}, \emph{F1} score, and \emph{Hybrid\_dis} (Hybrid). The best values for each metric within the same dataset are identified using \textbf{bold} formatting. 
We also label the rank in the circle of algorithms within each dataset. 
As none of the methods could finish the repair process in 24 hours (labeled with ``n/a'') except \textsf{MLNClean}, we partition \emph{Tax} into subsets, where the larger ones encompass the entirety of the smaller ones.
The smallest subset is used to evaluate the repair performance of algorithms, the others are for the repair cost study.
For further illustration, Table~\ref{tab:case_study} reports the 
specific values of $\#d_{w}$, $\#d_{w2r}$, $\#d_{r2r}$ and $\#d_{r2w}$ of each dataset over relatively good algorithm from each group. 
Based on the repaired data, we then conduct a set of experiments under \emph{four} data analysis tasks: classification and regression (with XGBoost~\cite{Chen16xgboost}, MLP~\cite{tang2015extreme}, and RandomForest models~\cite{liaw2002randomf}), as well as clustering (with Affinity Propagation~\cite{frey2007clusterap}, Spectral Clustering~\cite{ng2001spectral}, and K-means models~\cite{lloyd1982kmeans}) and kNN query (with Annoy~\cite{li2019annoy}, MiniHash~\cite{broder1997minihash}, and KD-tree models~\cite{bentley1975kdtree}).
For classification, kNN, and clustering, we select `HospitalName', `flight', `style', and `article\_title' for the datasets of Hospital, Flights, Beers, and Rayyan as the target column, respectively.
For regression, `Score' and `ibv' 
are chosen for the Hospital and Rayyan datasets, as the sole two featuring numerical attributes. 
These attributes exhibit a strong correlation with other values and are important features in the data.

\underline{Data repair performance.} 
As shown in Table~\ref{tab:repair_real_data}, most data repair methods exhibit \emph{negative} values for the EDR metric.
This phenomenon can be attributed to the fact that, there are more incorrect repairs on correct values compared to the correct repair of wrong ones, according to the definition of EDR in Eq.~\ref{eq:edr}. 
Across datasets, most algorithms exhibit better EDR value on the Hospital dataset. This can be attributed to its abundance of redundancies, substantial cosntraints, and simpler contexts. While other datasets prove challenging to repair due to either high error rate (on Flights), or fewer redundancies with more complex contexts (on Beers, and Rayyan).
As for Tax, the data repair algorithms tend to introduce a larger amount of errors, possibly due to its scare errors and large size. 
% It leads to the introduction of more errors,  due to the potentially insufficient correction ability.

Among these methods, \textit{Baran} shows superior \emph{EDR} performance on most datasets except \emph{Hospital}.
It can be attributed to its utilization of more information(like manually cleaned data and precise error detection results), and comprehensive candidate generation strategies, which increase the likelihood of including latent clean data in the candidate sets to a large extent.
For other \emph{data-driven} methods, \textsf{BoostClean} and \textsf{Scare} perform much worse due to the simple strategies for generating candidates, and the dependence on fully clean attributes, respectively.
Moreover, most of the \emph{cstr-driven} algorithms display inferior performance except \textsf{MLNClean}. 
This observation highlights that, leveraging models like the Markov network to learn instantiated constraints may be promising.
% while solely relying on the given rules to instantiate rules may not be robust enough.
Besides, \textsf{Daisy} does not repair any of the data because the union of all the conditional probabilities potentially assigns the highest probability to the initial dirty tuples, leaving them unchanged. 
For \emph{hybrid-driven} methods, \textsf{HoloClean} and \textsf{Unified} perform much better on \emph{Hospital} than other datasets, due to substantial redundancies.
% make them more effectively leverage data distribution information
\textsf{HoloClean} is more sensitive to data than \textsf{Unified}.
The reason is that,  \textsf{HoloClean} employs Naive Bayesian model to identify possible correct data, which assumes attribute independence, being challenging in practice.
% , especially when the attribute count is not large.
\textsf{Relative} can not finish the repair in all cases due to its exponential level of time complexity. 
% Thirdly, in the candidate decision phase, by leveraging the trained \emph{classifier}, it can more accurately identify the correct data in the candidate sets, especially on \emph{Rayyan} with complex contents like long texts and numbers with various formats.

Note that, in the rest of experiments,
we exclude \textsf{Daisy} and \textsf{Relative}, since they either fail to produce results within 24 hours or have no repair impact on the datasets. The \emph{Tax} dataset is omitted, since only \emph{MLNClean} can complete the repair process.

\underline{Downstream task performance.} 
Figure ~\ref{fig:three-tasks-error_rate} shows the performance of classification, kNN, and clustering tasks on repaired data.
While Table~\ref{tab:regress_real_data} illustrates the performance of regression.
Firstly, it can also be observed that, with proper selections of repair algorithms, regardless of the applied model, the downstream task performance can always be enhanced over the dirty data.
Besides, among four tasks, performance on regression is rarely influenced by the repair data, probably due to the few numerical values in the data.
Notably, we can see that applying clean data to train the model using clean data is not always the best for downstream tasks, as demonstrated by the performance of Xgboost models on Flights.
Also, as evidenced by the RandomForest model on Rayyan, downstream analysis models trained on purely clean data may perform worse than models trained using dirty data.

\begin{figure*}[htbp]
\vspace*{-0.03in}
\centering
\includegraphics[width=0.96\linewidth]{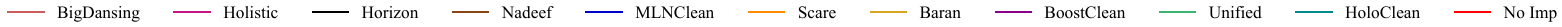}
\vspace*{-0.05in}
\\
% \hspace*{-0.08in}
\subfigure[\emph{Hospital}]{\raisebox{-0.1cm}{
\includegraphics[height=0.22\linewidth]{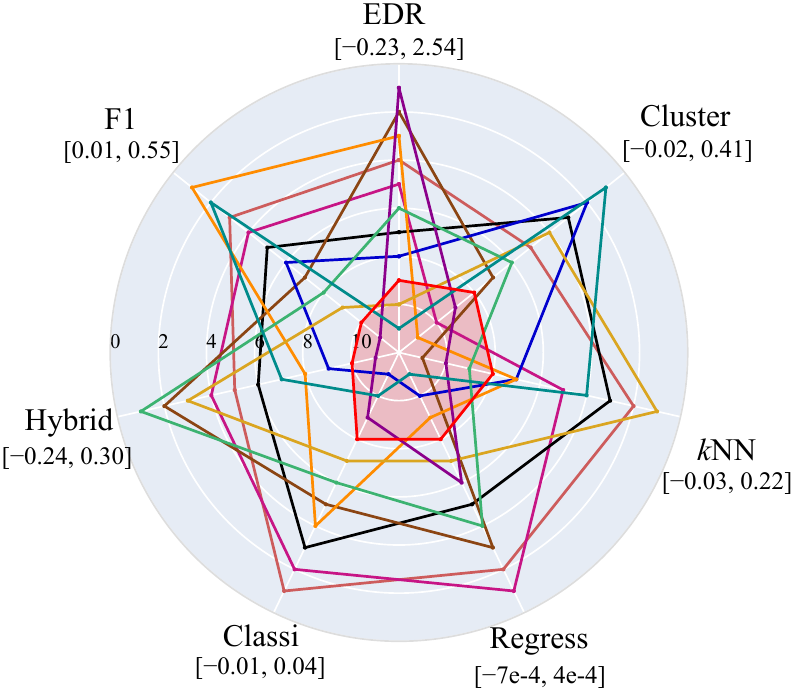}}
}
\subfigure[\emph{Beers}]{\raisebox{-0.1cm}{
\includegraphics[height=0.22\linewidth]{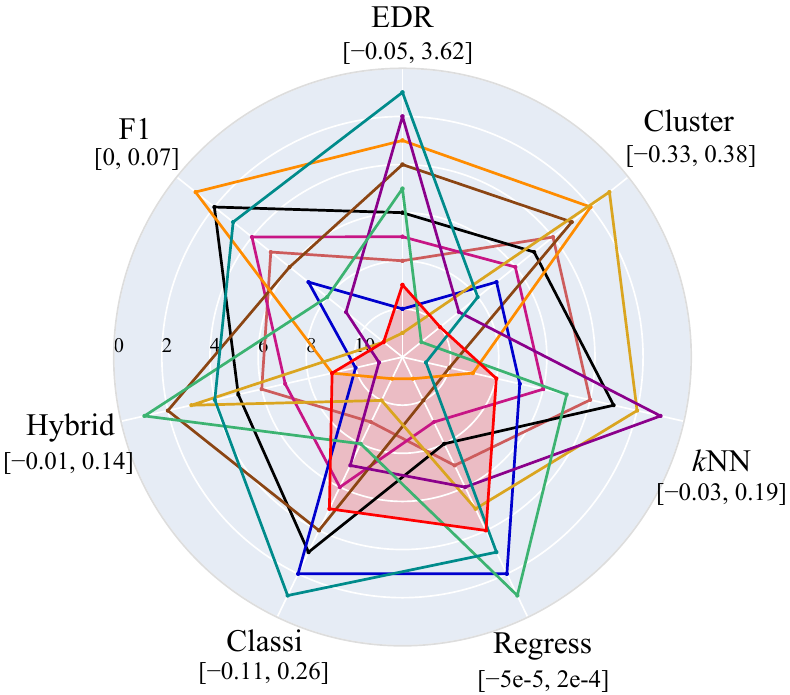}}
}
\subfigure[\emph{Flights}]{\raisebox{-0.1cm}{
\includegraphics[height=0.22\linewidth]{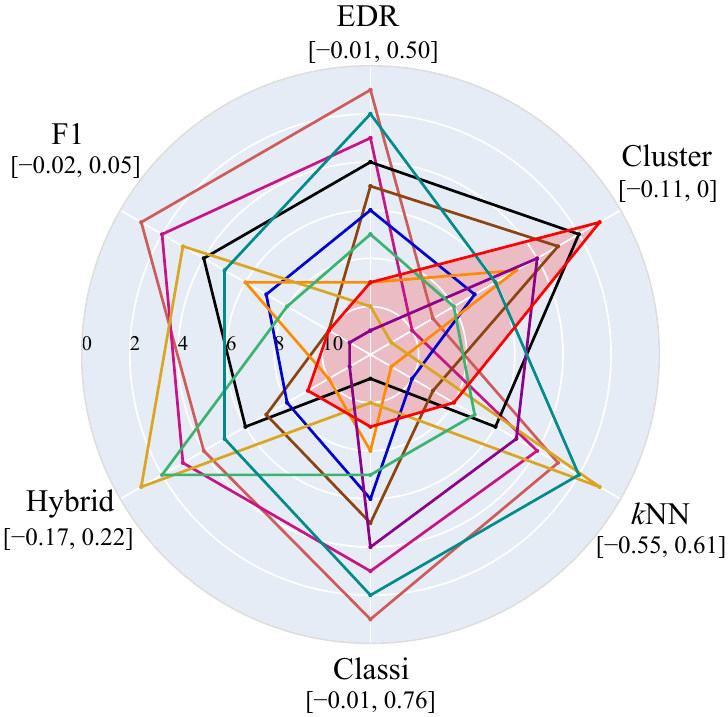}}
}
\subfigure[\emph{Rayyan}]{\raisebox{-0.1cm}{
\includegraphics[height=0.22\linewidth]{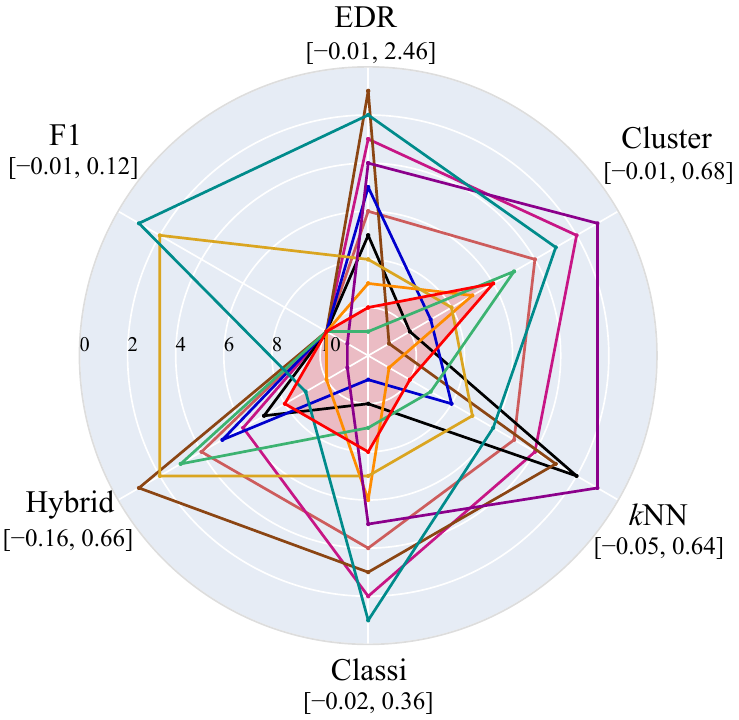}}
}
\vspace*{-0.18in}
\caption{Optimization gain ranking.}
\label{fig:opt_radar}
\vspace*{-0.12in}
\end{figure*}

\underline{Repair metric evaluation.}  
As shown in Table~\ref{tab:repair_real_data}, \emph{EDR} provides a simpler and more intuitive repair performance evaluation. 
With higher \emph{EDR} values denoting superior error reduction and negative (resp. positive) fluctuations indicating error increments (resp. decrements).
While it is hard to interpret the actual error reduction degree from F1 score and Hybrid\_dis.
Notably, EDR shares a 0.64 mean overlap rate with the F1 score top 5/11 methods, showing 0.8, 0.6, 0.8, 0.4, and 0.6 scores for Hospital, Flights, Beers, Rayyan, and Tax-10k. It also registers a 0.84 overlap rate with Hybrid\_dis, posting 1.0, 0.8, 0.8, 0.8, and 0.8 across the same datasets.
This phenomenon suggests that while there are some discrepancies in the rankings produced by these metrics, they are highly correlated in their evaluation of relative repair performance.

% We can observe from Table~\ref{tab:repair_real_data} that,
% the algorithms with good \emph{EDR} values cannot be well identified by the traditional metrics of \emph{precision},  \emph{recall}, and \emph{F1} score. 
%This simplicity makes it easier to conclude the error reduction degree, compared to interpreting from various \emph{precision}, \emph{recall}, and \emph{F1 scores}. 
%{\color{red} {\color{brown} Moreover,  
% In other words, \emph{precision}, \emph{recall} and  \emph{F1} scores are insufficient to comprehensively reflect the actual error reduction degree, compared to \emph{EDR}. 

In particular, when considering the F1 score for Flights, \textsf{HoloClean} achieves \emph{F1 score} as high as 0.47, placing it in the top 3 performance.
But the fact is that, it tends to introduce more errors than it eliminates.
This is because most of its correctly repaired cells are from the initial right data, as shown in Table~\ref{tab:case_study}.
It indicates that, the inclusion of $\#d_{r2r}$ may significantly improve F1 score, while it actually has no impact on the data error rate, thus making it accurate in evaluating the quality of repaired data.
Similar situations also occur on Beers and Rayyan when assessing \textsf{HoloClean}. 
Meanwhile, $\#d_{w2w}$ also influences the F1 score calculation.
\textsf{Baran} achieves a low F1 score on Tax-10k, due to incorrect repairs of 173 original wrong cells, constituting a large proportion of all repaired cells.
However, even with an F1 score as low as 0.0634, \textsf{Baran} can still decrease errors in the data.
These instances suggest that, the F1 score falls short in evaluating the actual error change condition.
With a high F1 score, there may also be an increase in the data error rate due to a large $\#d_{r2r}$.
A low F1 score may be attributed to large $\#d_{w2w}$, and thus the data errors can still be reduced.

\begin{table}[t]
  \fontsize{6.9pt}{0.8\baselineskip}\selectfont
  \centering
  \caption{Performance of regression on repaired data.}
  \vspace*{-0.15in}
  \label{tab:regress_real_data}
  \setlength{\tabcolsep}{0.3mm}
  % \renewcommand{\arraystretch}{1}
 % \resizebox{\columnwidth}{!}{
  \begin{tabular}{|c|c|c|c|c|c|c|}
  \hline
    \multirow{2}{*}{\textbf{Algos.}} 
    & \multicolumn{3}{c|}{\textbf{\makecell[c]{Hospital}}} & \multicolumn{3}{c|}{\textbf{\makecell[c]{Beers}}}    \\ \cline{2-7}
    & \makecell[c]{XGB($\times\ e^{-32}$)} & \makecell[c]{MLP} & \makecell[c]{RandF($\times\ e^{-32}$)} & \makecell[c]{XGB($\times\ e^{-32}$)} & \makecell[c]{MLP} & \makecell[c]{RandF($\times\ e^{-32}$)}   \\ 
    \hline 
    All clean & 2.9108 	 & 0.0019   & 1.3207 & 2.7342  & 0.0002 & 1.2741    \\ \cline{1-7}
    No repair & 2.9108 	 & 0.0026  & 1.3207 & 2.7342  & 0.0002 & 1.2741      \\
    \hline
    BigDansing & 2.9108   & 0.0031 & 1.3207 & 2.7342  & 0.0010  & 1.2741  \\ \cline{1-7}
    Holistic  & 2.9108   & 0.0028 & 1.3207 & 2.7342  & 0.0010 & 1.2741    \\ \cline{1-7}
    Horizon   & 2.9108  & 0.0027 & 1.3207 & 2.7342 & 0.0010  & 1.2741 \\  \cline{1-7}
    Nadeef    & 2.9108   & 0.0025 & 1.3207 & 2.7342  & 0.0002  & 1.2741   \\ \cline{1-7}
    MLNClean  & 2.9108   & 0.0022 & 1.3207 & 2.7342  & 0.0005  & 1.2741    \\ \cline{1-7}
    Unified  & 2.9108   & 0.0026  & 1.3207 & 2.7342  & 0.0004 & 1.2741\\ \cline{1-7}
    HoloClean  & 2.9108  & 0.0019 & 1.3207 & 2.7342  & 0.0009 & 1.2741   \\ \hline
    Scare  & 2.9108    & 0.0020   & 1.3207 & 2.7342  & 0.0008 & 1.2741    \\ \cline{1-7}
    Baran & 2.9108    & 0.0027 & 1.3207  & 2.7342  & 0.0010  & 1.2741 \\ \cline{1-7}
    BoostClean   & 2.9108  & 0.0024  &  1.3207 & 2.7342  & 0.0010 & 1.2741   \\ \hline
  \end{tabular}
  % }
  \vspace{-0.14in}
\end{table}

\begin{table}[t]
\centering
\caption{Anal. of repair metric and downstream performance.}
\fontsize{8.1pt}{0.9\baselineskip}\selectfont
\vspace*{-0.15in}
\setlength{\tabcolsep}{1.7mm}
% \resizebox{0.95\columnwidth}{!}{%
\begin{tabular}{|c|l|l|r|r|}
\cline{1-5}
\multicolumn{1}{|c}{\textbf{Task}}  &  \multicolumn{1}{|c|}{\textbf{Corr. indicator}}                                       & \multicolumn{1}{c|}{\textbf{EDR}} & \multicolumn{1}{c|}{\makecell[c]{\textbf{F1 score}}} & \multicolumn{1}{c|}{\makecell[c]{\textbf{Hybrid\_dis}}} \\ \cline{1-5}
\multicolumn{1}{|c|}{\multirow{2}{*}{Classi.}}  & Mean \emph{rank corr.}     & \textbf{0.5311}          & 0.2423                  & 0.0816                \\ \cline{2-5}
\multicolumn{1}{|c|}{}                          & Mean \emph{top-5 overlap}. & \textbf{0.5733}          & 0.5200                  & 0.4400                \\ \cline{1-5}
\multicolumn{1}{|c|}{\multirow{2}{*}{Regress.}} & Mean \emph{rank corr.}     & 0.1909                   & \textbf{0.3394}         & 0.0244                         \\ \cline{2-5}
\multicolumn{1}{|c|}{}                          & Mean \emph{top-5 overlap}. & \textbf{0.6000}          & \textbf{0.6000}                 & 0.5000               \\ \cline{1-5}
\multicolumn{1}{|c|}{\multirow{2}{*}{Cluster}}  & Mean \emph{rank corr.}     & \textbf{0.1158}          & 0.0974                  & $-$0.0893              \\ \cline{2-5}
\multicolumn{1}{|c|}{}                          & Mean \emph{top-5 overlap}. & \textbf{0.4133}          & \textbf{0.4133}                  & \textbf{0.4133}                \\ \cline{1-5}
\multicolumn{1}{|c|}{\multirow{2}{*}{kNN}}      & Mean \emph{rank corr.}     & \textbf{0.1920}          & 0.0278                  & 0.0456                \\ \cline{2-5}
\multicolumn{1}{|c|}{}                          & Mean \emph{top-5 overlap}. & \textbf{0.5200}          & 0.4000                  & 0.4400                \\ \cline{1-5}
\end{tabular}
% }
\label{tab:rep-down}
\vspace*{-0.17in}
\end{table}

Furthermore, according to the rank of the repair tools in Table~\ref{tab:repair_real_data}, and the performance rank of downstream tasks on different repaired data in Figure~\ref{fig:three-tasks-error_rate} and Table~\ref{tab:regress_real_data}, we evaluate the mean
Pearson correlation coefficient and the mean Top-5 methods’ overlap rate between repair and downstream performance (across all evaluated models on the repaired dataset, i.e., Hospital, Flights, Beers, and Rayyan).
As shown in Table~\ref{tab:rep-down}, 
EDR displays the strongest correlation with downstream performance, and the highest overlap rate among repair metrics in most cases.
This suggests that EDR is a more reliable indicator of data quality when performing data repairs aimed at enhancing downstream task performance.
Thus, it is advisable to accord greater importance to EDR when assessing repair tools in data analysis tasks.

\begin{table}[t]
\centering
% \vspace*{-0.08in}
\fontsize{8.4pt}{0.9\baselineskip}\selectfont
\setlength{\tabcolsep}{2mm}
\caption{Overhead of the optimization strategy.}
\vspace*{-0.1in}
% \resizebox{0.95\columnwidth}{!}{%
\begin{tabular}{|l|l|l|l|l|}
\hline
           \textbf{Overhead} & \textbf{Hospital} & \textbf{Flights} & \textbf{Beers} & \textbf{Rayyan} \\ \hline
Max CPU usage  & 4840\% & 5120\% & 2432\% & 5690\%  \\ \hline
Max Memory usage    & 331Mb & 307Mb & 387Mb & 252Mb \\ \hline
Runtime      & 294s & 88s & 208s & 107s \\ \hline
\end{tabular}%
% }
\label{tab:opt_overhead}
\vspace*{-0.15in}
\end{table}

\begin{table}[t]
\fontsize{8.1pt}{0.9\baselineskip}\selectfont
\centering
% \caption{Identified condition of the optimization strategy. Iden-W and Iden-R indicates the wrong and right identified cell amount compared with revised cell amount, respectively.}
\caption{Frequency of incorrect prevention v.s. effective opt.}
\vspace*{-0.15in}
\setlength{\tabcolsep}{0.5mm}
\begin{tabular}{|r|rr|rr|rr|rr|}
\hline
           \multirow{2}{*}{\textbf{Algos.}}  & \multicolumn{2}{c|}{\textbf{Hospital}} & \multicolumn{2}{c|}{\textbf{Flights}} & \multicolumn{2}{c|}{\textbf{Beers}} & \multicolumn{2}{c|}{\textbf{Rayyan}}                                                  \\ \cline{2-9}
            & Iden-W & Iden-R & Iden-W & Iden-R & Iden-W & Iden-R & Iden-W & Iden-R\\ \hline
Big.  & 0.29 & \textbf{0.52} & 0.04 & \textbf{0.40}  & 0 & \textbf{0.55} & 0 & \textbf{0.72}  \\ \hline
Holi.    & 0.31 & \textbf{0.51} & 0.05 & \textbf{0.40}  & 0 & \textbf{0.55} & 0 & \textbf{0.81}  \\ \hline
Nadeef      & 0.02 & \textbf{0.91} & 0 & \textbf{0.54}   & 0 & \textbf{0.81}  & 0 & \textbf{0.71}  \\ \hline
MLNC.    & 0.28 & \textbf{0.30} & 0.01 & \textbf{0.36}  & 0.01 & \textbf{0.20}  & 0 & \textbf{0.81}  \\ \hline
Hori.     & 0.34 & \textbf{0.47} & 0.06 & \textbf{0.19}  & 0 & \textbf{0.56}  & 0 & \textbf{0.98}  \\ \hline
Baran & 0.55 & 0.24 & 0.03 & 0.03 & 0.02 & \textbf{0.04}  & 0.02 & \textbf{0.36}  \\ \hline
Scared      & 0.07  & \textbf{0.79} & 0  & 0 & 0  & \textbf{0.92}  & 0 & \textbf{0.46}   \\ \hline
Holo.   & 0  & \textbf{0.44} & 0.01 & 0 & 0.01 & \textbf{0.57}  & 0 & \textbf{0.96}  \\ \hline
Uni.     & 0 & \textbf{0.19} & 0.08 & \textbf{0.14} & 0.01 & \textbf{0.86}  & 0 & 0  \\ \hline
Boost.  & 0.11 & \textbf{0.65} &  0 & \textbf{0.50} & 0 & \textbf{0.78}  & 0 & \textbf{0.12}  \\  \hline
\end{tabular}%
\label{tab:opt_iden}
\vspace*{-0.17in}
\end{table}

\begin{table*}[ht]
% \vspace*{-0.03in}
  \small
  \centering
  \caption{Maximum calculation resource usage of the algorithms across varying data sizes.}
  \vspace*{-0.15in}
  \label{tab:algos_overhead}
  \setlength{\tabcolsep}{0.68mm}
  % \renewcommand{\arraystretch}{1}
  % \resizebox{\linewidth}{!}{
  \begin{tabular}{|c|r|r|r|r|r|r|r|r|r|r|r|r|r|r|r|}
  \hline
    \multirow{2}{*}{\textbf{Algos.}} 
    & \multicolumn{3}{c|}{\textbf{\makecell[c]{Tax-10k}}} & \multicolumn{3}{c|}{\textbf{\makecell[c]{Tax-20k}}} & \multicolumn{3}{c|}{\textbf{\makecell[c]{Tax-30k}}} & \multicolumn{3}{c|}{\textbf{\makecell[c]{Tax-40k}}} & \multicolumn{3}{c|}{\textbf{\makecell[c]{Tax-50k}}}  \\ \cline{2-16}
    & Runtime & CPU & Memory
    & Runtime & CPU & Memory
    & Runtime & CPU & Memory
    & Runtime & CPU & Memory
    & Runtime & CPU & Memory \\ 
    \hline
    BigDansing & 2,691 & 109\% &6,557Mb & n/a & n/a & n/a & n/a & n/a & n/a & n/a & n/a & n/a & n/a & n/a & n/a    \\ \hline
    Holistic & 45,261 & 109\% &47,670Mb & n/a & n/a & n/a & n/a & n/a & n/a & n/a & n/a & n/a & n/a & n/a & n/a    \\ \hline
    Horizon & 1,675 & 110\% & 118Mb & 5,757 & 111\% & 162Mb & 11,378 & 109\% & 186Mb & 18,613 & 110\% &249Mb & 27,124 & 110\% & 288Mb   \\ \hline
    Nadeef & 821 & 110\% & 3,004Mb & 4,421 & 111\%& 6,281Mb & 7,497 & 112\%& 6,763Mb & 15,492 & 110\% & 8,744Mb & 19,571 & 110\% & 9,839Mb   \\ \hline
    MLNClean & 66 & 110\% & 4,796Mb &122 &110\% &5,250Mb &226 & 110\% & 5,735Mb &329 & 112\% & 7,776Mb &395 & 113\% &8,671Mb    \\ \hline
    Scare & 38,096 & 5760\% & 8,693Mb & 23,705 & n/a & n/a & n/a & n/a & n/a & n/a & n/a & n/a & n/a & n/a & n/a    \\ \hline
    Baran & 49,346 & 6400\% & 2,070Mb & n/a & n/a & n/a & n/a & n/a & n/a & n/a & n/a & n/a & n/a & n/a & n/a    \\ \hline
    BoostClean & 502 & 1180\% &6919Mb &1,233 & 1090\% &15,820Mb &1,729 & 1140\% & 25,379Mb &2,708 & 1020\% & 37,712Mb &3,939 & 1100\%& 51,345Mb   \\ \hline
    HoloClean & 935 & 4820\% & 191,019Mb & n/a* & n/a* & n/a* & n/a* & n/a* & n/a* & n/a* & n/a* & n/a* & n/a* & n/a* & n/a*    \\ \hline
  \end{tabular}
  % }
  \vspace*{-0.13in}
\end{table*}

\begin{figure*}[htbp]
\centering
\includegraphics[width=0.93\linewidth]{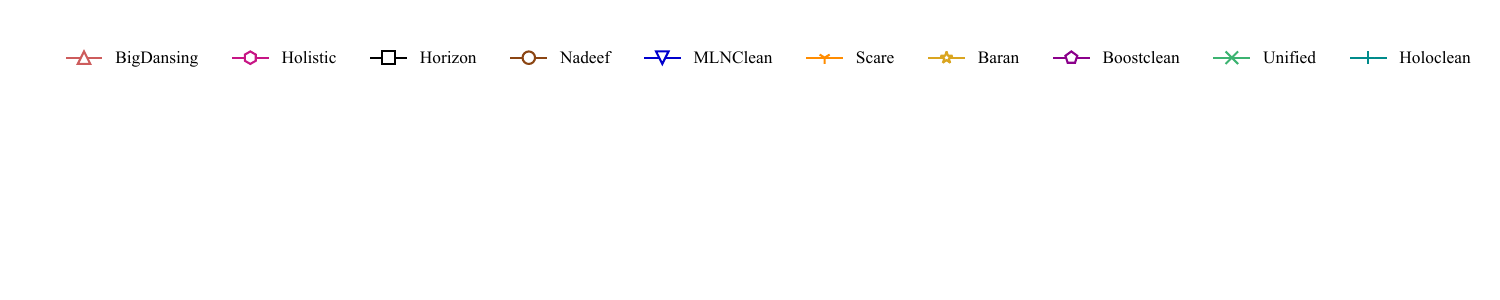}
\vspace*{-0.1in}
\\
\hspace*{-0.08in}
\subfigure[\emph{Hospital}, semantic error]{\raisebox{-0.2cm}{
\includegraphics[width=0.25\linewidth]{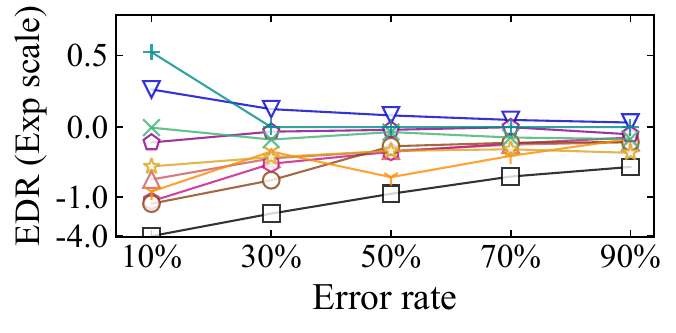}}
~\label{ETi-hospital}}\hspace*{-0.06in}
\subfigure[\emph{Flights}, semantic error]{\raisebox{-0.2cm}{
\includegraphics[width=0.25\linewidth]{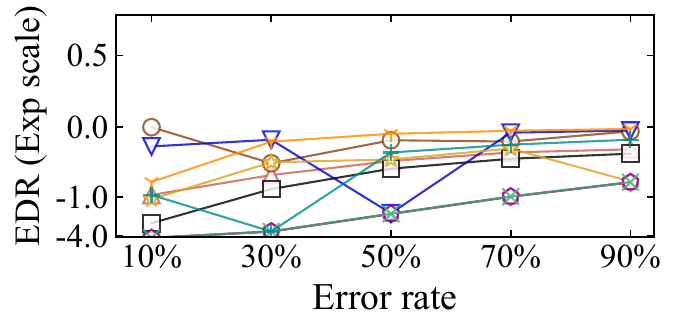}}
~\label{ETi-flights}}\hspace*{-0.06in}
\subfigure[\emph{Beers}, semantic error]{\raisebox{-0.2cm}{
\includegraphics[width=0.25\linewidth]{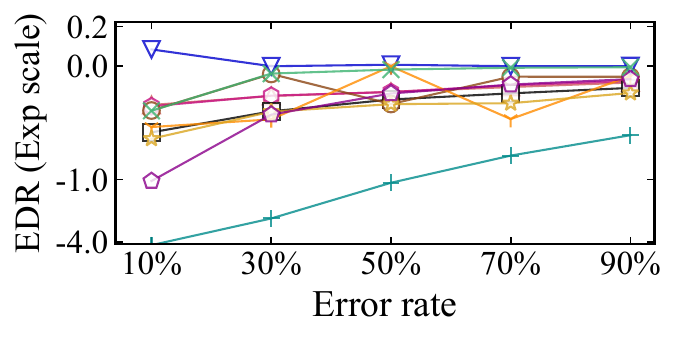}}
}\hspace*{-0.06in}
\subfigure[\emph{Rayyan}, semantic error]{\raisebox{-0.2cm}{
\includegraphics[width=0.25\linewidth]{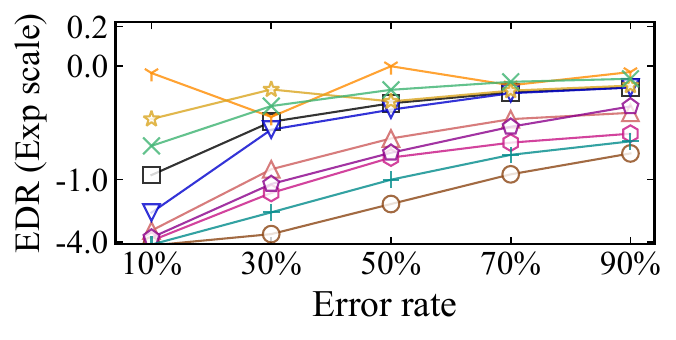}}
~\label{ETi-rayyan}}
\vspace*{-0.13in}
\\
\hspace*{-0.08in}
\subfigure[\emph{Hospital}, syntactic error]{\raisebox{-0.2cm}{
\includegraphics[width=0.25\linewidth]{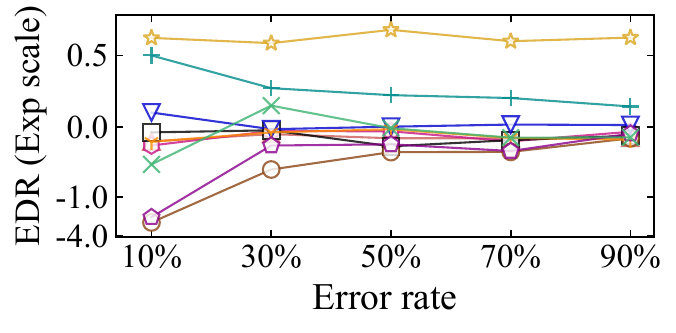}}
~\label{ETo-hospital}}\hspace*{-0.06in}
\subfigure[\emph{Flights}, syntactic error]{\raisebox{-0.2cm}{
\includegraphics[width=0.25\linewidth]{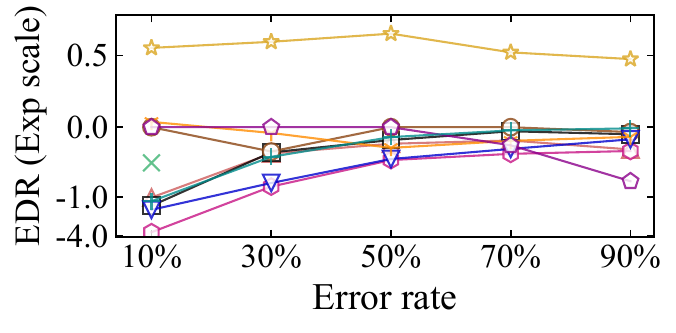}}
~\label{ETo-flights}}\hspace*{-0.06in}
\subfigure[\emph{Beers}, syntactic error]{\raisebox{-0.2cm}{
\includegraphics[width=0.25\linewidth]{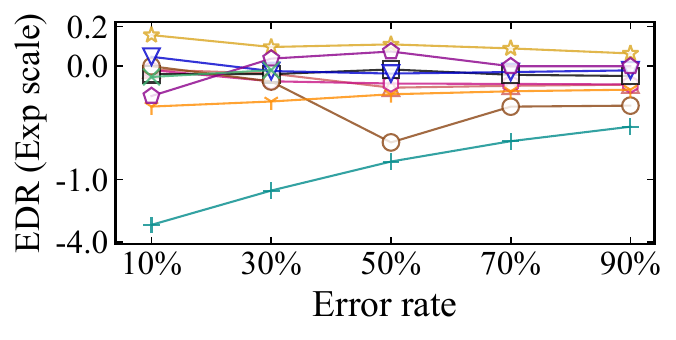}}
~\label{ETo-beers}}\hspace*{-0.06in}
\subfigure[\emph{Rayyan}, syntactic error]{\raisebox{-0.2cm}{
\includegraphics[width=0.25\linewidth]{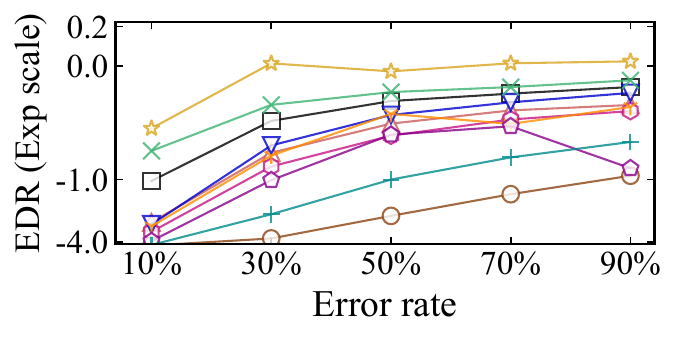}}
~\label{ETo-rayyan}}
\vspace*{-0.17in}
\caption{Data repair performance vs. different error rates.}
\label{fig:error_types}
\vspace*{-0.1in}
\end{figure*}

\begin{figure*}[htbp]
\centering
\includegraphics[width=0.93\linewidth]{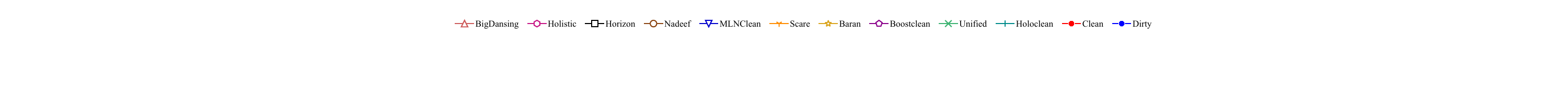}\vspace*{-0.05in}
\\
\hspace*{-0.05in}
\subfigure[\textit{Hospital}, semantic errors]{\raisebox{-0.2cm}{
\includegraphics[width=0.242\linewidth]{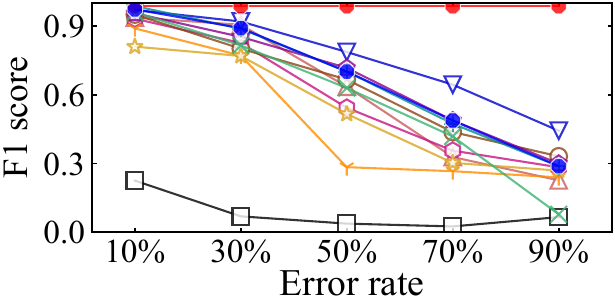}}
}\hspace*{-0.03in}
\subfigure[\textit{Hospital}, syntactic errors]{\raisebox{-0.2cm}{
\includegraphics[width=0.242\linewidth]{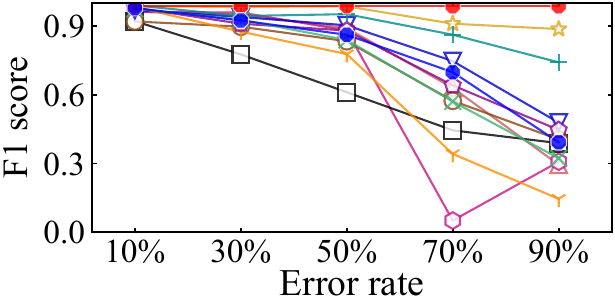}}
}\hspace*{-0.03in}
\subfigure[\textit{Rayyan}, semantic errors]{\raisebox{-0.2cm}{
\includegraphics[width=0.242\linewidth]{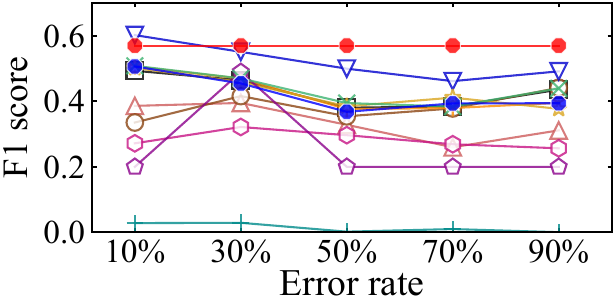}}
}\hspace*{-0.03in}
\subfigure[\textit{Rayyan}, syntactic errors]{\raisebox{-0.2cm}{
\includegraphics[width=0.242\linewidth]{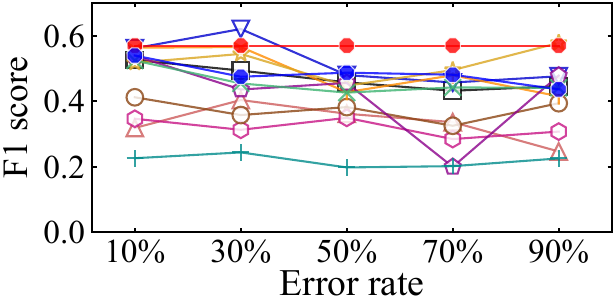}}
}
\vspace*{-0.18in}
\caption{Classification performance vs. different error rates.}
\label{fig:classi-error_rate}
\vspace*{-0.13in}
\end{figure*}

\underline{Effect of optimization strategy.}  Firstly, to fully represent the effect of the optimization strategy, we provide the ranks of advancement for both repair metrics and downstream performance.
Besides, we also show the range for the absolute values of improvement, all of which is demonstrated in Figure \ref{fig:opt_radar}.
The term 'No Imp' represents the rank where no enhancements are observed. In other words, points that are positioned outside the 'No Imp' range signal improvements.
Notably, in a majority of instances, the optimization strategy results in observable improvements.
Secondly, we report the frequency of instances where the optimization strategy blocks the correct alteration of initially incorrect values, and successfully prevents incorrect modifications of initially correct values.
We denote the former and latter frequency as Iden-W and Iden-R, respectively, as depicted in Table~\ref{tab:opt_iden}.
It is noticeable that, despite the presence of incorrect preventative actions, the instances of Iden-R exceed those of Iden-W in most scenarios. This outcome suggests that, on balance, the optimization strategy serves its intended purpose effectively.
Finally, to report the extra cost of the optimization strategy, we have reported its maximum CPU (in percent, e.g. 4840\% indicates the use of 48 CPU cores) and memory usage (in Mb), as well as the end-to-end runtime across four datasets, as outlined in Table ~\ref{tab:opt_overhead}.
Though the max CPU usage appears high, the memory usage and runtime are within reasonable limits, ensuring that the optimization strategy remains practical applications.

% experimental results of the \emph{EDR} performance after integrating the optimization strategy into data repair algorithms. 
% It also lists the \emph{EDR} gain, compared to the one without the optimization strategy. 
% We can observe that in most conditions,  with the optimization strategy, these methods can eliminate data errors more or less instead of increasing data errors.
% Specifically, on \emph{Flights}, all methods exhibit marginal improvements. This results from their  minimal incorrect repairs to correct data along with the substantial erroneous values inherent in \emph{Flights}.
% The \emph{rule-driven} methods benefit from the optimization strategy in almost all cases except on \emph{Flights}. It indicates the importance of combining current state-of-the-art error detection methods with \emph{rule-driven} methods to eliminate errors.
% Regarding \emph{data-driven} and \emph{hybrid-driven} algorithms, though the performance is enhanced in most cases, \textsf{Baran}, \textsf{HoloClean} and \textsf{Unified} suffer \emph{EDR} decreasing over \emph{Hospital}.
% It is attributed to identifying the wrong data with the right one, showing the limitations of the optimization strategy. 

\underline{Scalable performance.}  
Table~\ref{tab:algos_overhead} displays the execution time (in seconds), max CPU (in percent), and Memory usage (in Mb) of data repair algorithms on \emph{Tax}. 
It is observed that only half of the algorithms can finish the experiments, indicating the significant time cost challenges.
Among them, \textsf{Horizon} consumes the least resources, showing almost a linear increase in resource usage as the data size increases.
This verifies the effectiveness of its value-based modeling approaches. 
\textsf{BoostClean} also scales but has a notably higher CPU and Memory usage.
\textsf{MLNClean} appears to be the most efficient in terms of runtime for smaller datasets, although the memory usage is still significant.
\textsf{HoloClean} and Holistic have extremely high memory usage even for the smallest data size.
It is important to note that, \textsf{BigDansing} consumes much less resources than \textsf{Holistic}, validating its scaling strategies.
\textsf{Unified} cannot complete tasks, largely due to the extensive time consumption in assessing both data and constraint cost. 
For \textsf{Baran}, the challenge in its scalability arises from the substantial cost of generating diverse candidates.
Its high CPU usage also indicates the validity of the parallel strategy.

\begin{figure*}[htbp]
\vspace*{-0.22in}
\centering
\includegraphics[width=0.93\linewidth]{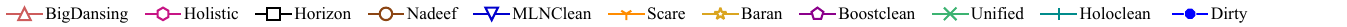}\vspace*{-0.05in}
\\
\hspace*{-0.08in}
\subfigure[\textit{Classification}, Titanic]{\raisebox{-0.2cm}{
\includegraphics[width=0.24\linewidth]{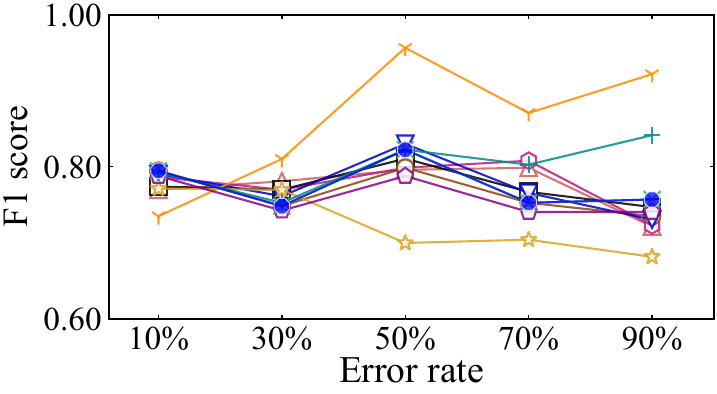}}
}%\hspace*{0.01in}
\subfigure[\textit{Classification}, Restaurants]{\raisebox{-0.2cm}{
\includegraphics[width=0.24\linewidth]{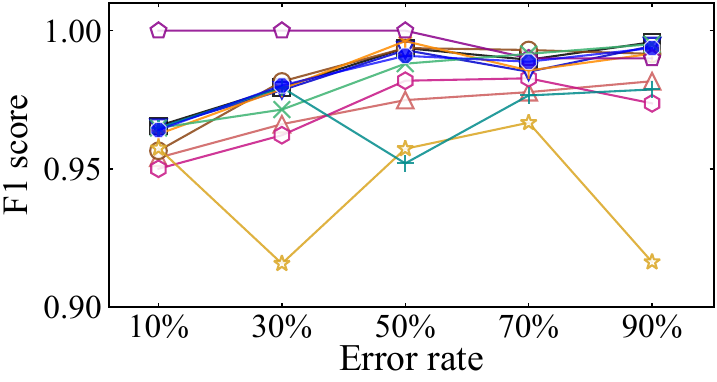}}
}
\subfigure[\textit{Regression}, Airfoil]{\raisebox{-0.2cm}{
\includegraphics[width=0.24\linewidth]{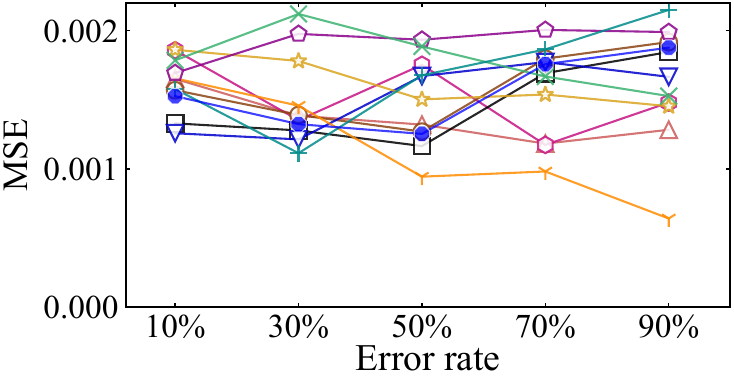}}
}
\subfigure[\textit{Regression}, Printer]{\raisebox{-0.2cm}{
\includegraphics[width=0.24\linewidth]{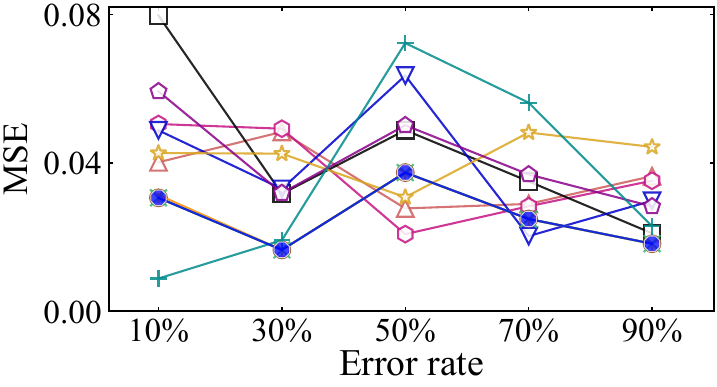}}
}
\vspace*{-0.12in}
\\
\hspace*{-0.08in}
\subfigure[\textit{kNN}, Bank]{\raisebox{-0.2cm}{
\includegraphics[width=0.24\linewidth]{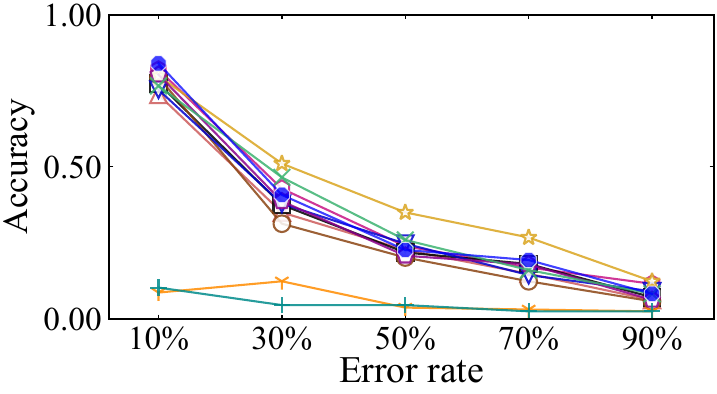}}
}\hspace*{-0.02in}
\subfigure[\textit{kNN}, Adult]{\raisebox{-0.2cm}{
\includegraphics[width=0.238\linewidth]{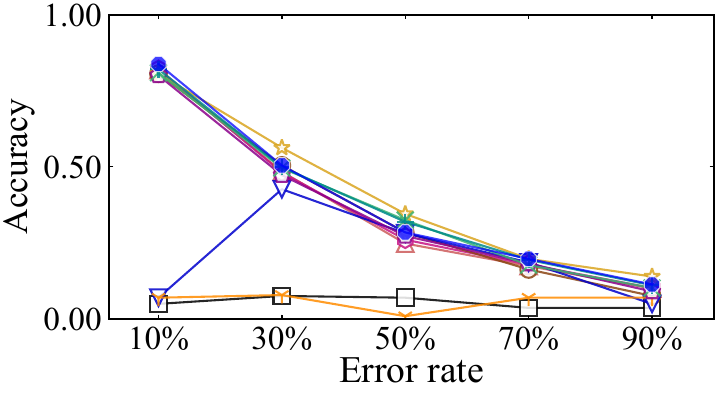}}
} %\hspace*{-0.02in}
\subfigure[\textit{Clustering}, Dress]{\raisebox{-0.2cm}{
\includegraphics[width=0.24\linewidth]{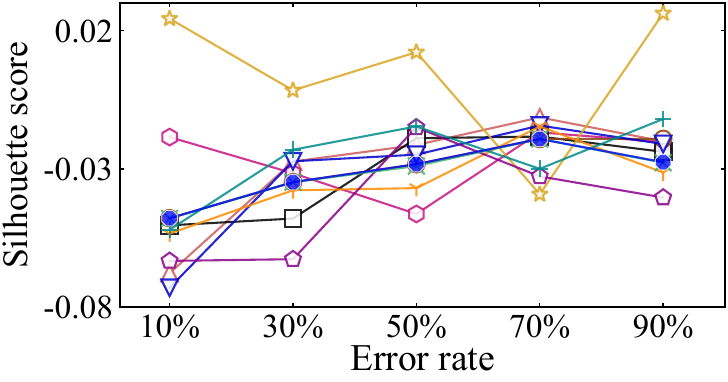}}
} \hspace*{-0.05in}
\subfigure[\textit{Clustering}, Adult]{\raisebox{-0.2cm}{
\includegraphics[width=0.24\linewidth]{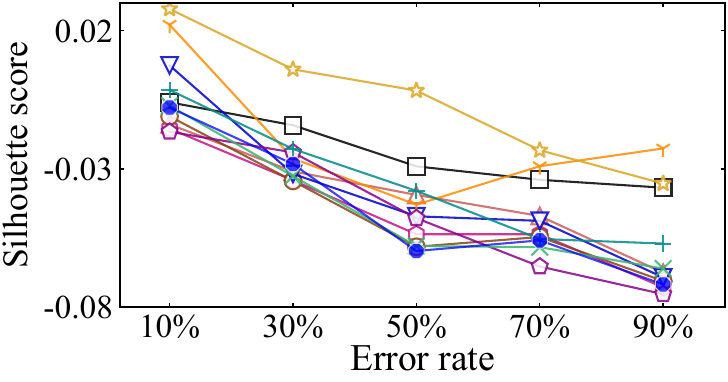}}
}
\vspace*{-0.18in}
\caption{Real workloads v.s. error rates.}
\label{fig:workloads-error_rate}
\vspace*{-0.18in}
\end{figure*}

\vspace{-0.1in}
\subsection{Effects of Varying Errors }

The second set of experiments is to fully verify the effect of data error \emph{both} on the repair performance \emph{and} downstream effects of data repair algorithms, considering the \emph{semantic} and \emph{syntactic} errors across various error rates. We only report \emph{EDR} metric for its superiority. 
Regarding downstream tasks, we focus solely on the classification tasks, given their extensive applications.
%Additionally, we validate the efficiency of the optimization strategy in these complex error scenarios.

% We introduce  \emph{inner} and \emph{outer} errors into the datasets at varying levels of error rates respectively.
% For the experiment concerning error rate, we introduce hybrid errors, including both \emph{inner} and \emph{outer} errors.
% Assuming that each cause of error occurs independently of others, the proportion of \emph{inner} errors to \emph{outer} errors is 1:4.
% For the experiment concerning error types, we evaluate the impact of inner and outer errors on the data repair algorithms separately.
% , which may not be included in the current real-world datasets.
% We assume that each cause of error occurs randomly and independently of others.

\vspace*{0.02in}
%\underline{Error repair performance.} 
% Overall performance
For repair performance, Figure~\ref{fig:error_types} illustrates the experimental results when the error rate changes from 10\% to 90\%. 
The vertical axis is in an exponential scale. 
Remarkably, an increase in the error rate usually corresponds to the reduced correction effectiveness of these algorithms.  
Existing data repair algorithms struggle to eliminate \emph{semantic} errors, while \emph{syntactic} errors can be reduced more or less. 
It is because that, semantic errors involve similar value patterns and inter-value relations as the clean data, which may make it hard to detect and decide the correct candidate.
% Except \textsf{Baran}, minimal error reduction is observed across almost all methods on \emph{Flights} with syntactic errors, likely due to a large number of distinct values. %is further escalated by \emph{outer} errors.
% The discrepancies across other datasets are similar as the former experimental results.
Concerning syntactic errors, \textsf{Baran} excels across all cases. Even with 90\% error rates in Hospital, it can rectify over half errors, demonstrating its strength in handling syntactic errors. However, it underperforms in addressing semantic errors, possibly due to its feature extraction strategies' imperfections and reliance on minimal manually cleaned data.
\emph{Cstr-driven} methods appear to encounter challenges in significantly reducing errors across varying error rates.
For \emph{hybrid-driven} methods, consistent with previous results, \textsf{HoloClean} performs much better on \emph{Hospital} than other datasets, while \textsf{Unified} demonstrates greater robustness.
Regarding semantic errors, \emph{cstr-driven} methods exhibit greater robustness.
In particular, \textsf{MLNClean} demonstrates superior performance compared to syntactic errors, especially on Hospital and Beers, showing the effectiveness of its strategy to learn the trustworthy degree of instantiated constraints.

For the classification task, Figure~\ref{fig:classi-error_rate} illustrates the corresponding experimental results. 
We employ MLP on Hospital and Rayyan, as MLP generally performs better than XGBoost.
% , and data repair algorithms tend to achieve their best and worst performance on Hospital and \emph{Rayyan} respectively.
We can observe that, the classification performance decreases almost linearly with an increasing error rate in most cases, without the sharp drop expected.
Semantic errors generally have larger negative impacts on MLP than syntactic ones.
In terms of semantic errors, except for data repaired by \textsf{MLNClean}, the repaired dataset scarcely helps improve the classification performance of the trained model beyond that achieved with the original dirty dataset.
It is attributed to the further disruption of the data distribution by the repair process, making it harder for the model to learn feature-label relations.

% \begin{figure*}[tbp]
% \centering
% \includegraphics[width=0.85\linewidth]{curve-legend-12.pdf}\vspace*{-0.08in}
% \\
% \hspace*{-0.08in}
% \subfigure[\textit{Hospital}, inner error]{\raisebox{-0.2cm}{
% \includegraphics[height=0.1298\linewidth]{regress-inner-mlpr-hospital.pdf}}
% }\hspace*{-0.07in}
% \subfigure[\textit{Hospital}, outer error]{\raisebox{-0.2cm}{
% \includegraphics[height=0.1298\linewidth]{regress-outer-mlpr-hospital.pdf}}
% }\hspace*{-0.05in}
% \subfigure[\textit{Beers}, inner error]{\raisebox{-0.2cm}{
% \includegraphics[height=0.1298\linewidth]{regress-inner-mlpr-beers.pdf}}
% }\hspace*{-0.07in} 
% \subfigure[\textit{Beers}, outer error]{\raisebox{-0.2cm}{
% \includegraphics[height=0.1298\linewidth]{regress-outer-mlpr-beers.pdf}}
% }\hspace*{-0.05in} 
% \vspace*{-0.18in}
% \caption{Regression performance v.s. different error rates.}
% \label{fig:regress-error_types}
% \vspace*{-0.04in}
% \end{figure*}

\vspace{-0.12in}
\subsection{Repair Effect on Downstream Tasks}
To further answer the critical question of how data repair algorithms impact the downstream task, we conduct a set of experiments on datasets with real workloads.

Following previous related works~\cite{Li21cleanml,Czarnowski12cluster, li2014query, Abdelaal23rein}, for classification, we employ \emph{Restaurants} and \emph{Titanic} datasets; for regression, \emph{Airfoil} and \emph{Printer} datasets are used; clustering analysis is conducted using \emph{Adult} and \emph{Dress} datasets; and k-nearest neighbor (kNN) queries perform on \emph{Adult} and \emph{Bank} datasets.
We employ Xgboost, Spectral clustering, and KD-tree models to conduct this experiment, selected based on their median performance in previous experiments, thereby proving generality.
Due to the lack of clean data, we treat the initial data as correct, and introduce dirty data by adding previously mentioned \emph{semantic} and \emph{syntactic} errors.
Since errors can be introduced by independent factors in reality, thus the proportion of each specific error (i.e., implicit missing, explicit missing, typos, gaussian noise, and misallocated values) is equal.
These errors are both injected into features and labels.
With a 1:4 ratio of semantic to syntactic errors, we illustrate the performance of \emph{four} workloads on \emph{both} the repaired data \emph{and} the initial dirty data, given the absence of corresponding clean data.
Besides, for cstr-driven and some hybrid-driven methods, we adopt the aforementioned DC discovery tools, i.e., DCFinder~\cite{dcfinder19Pena} and
Hydra~\cite{hydra17Bleifu} to discover constraints, which we then manually check.

As shown in Figure~\ref{fig:workloads-error_rate}, aligning with earlier results, data repair enhances performance in \emph{almost every} condition across all 4 tasks and 7 datasets, with few negligible deviations at a 10\% error rate.
With proper algorithm selection, data repair can enhance downstream model performance compared to the original dirty data.
Typically, the performance of models with repaired data fluctuates between that with original dirty data.
Among them, \textsf{Baran} can consistently augment downstream task performance across nearly all datasets, with the exception of Restaurants. Coupled with \textsf{MLNClean}, it boosts model performance except in the case of the Adult dataset with the kNN task. 
While \textsf{Nadeef} provides less substantial improvements, contributing only to the performance enhancement on Restaurants and Adult within the clustering task. 
Other methods also enhance model performance on more than two datasets, confirming the need for data repair algorithms in practical scenarios. Interestingly, on the Restaurant and Dress datasets, as the error rate increases, performance improves.
This counter intuitive phenomenon may occur due to various factors, such as the model possibly leveraging noise to avoid overfitting, thereby gaining a more generalized performance on varied datasets.

\vspace{-0.12in}
\subsection{Discussion} %\label{sec:discussion}
It is observed that existing automatic data repair algorithms can effectively reduce syntactic errors given adequate redundancies but struggle to eliminate semantic errors. 
However, with proper methodology, data repair can \textit{benefit} downstream tasks in the majority of situations. 
We summarize some takeaways for data repair algorithms regarding the scenarios based on the experiments.
%Subsequently, we delve into a discussion on practical guidelines and existing challenges in this area. Furthermore, we underscore potential avenues for additional exploration within this domain.

\underline{Recommendations.}
%Based on the experimental results above, 
\textsf{Baran}, 
\textsf{MLNClean},  \textsf{Horizon}, and \textsf{BigDansing} are better alternatives in real-life scenarios. 

\begin{itemize}[leftmargin=*,itemsep=1pt,topsep=1pt]
\setlength{\itemsep}{0pt}
\setlength{\parsep}{0pt}
\setlength{\parskip}{0pt}

\item Dominant syntactic errors: \textsf{Baran} emerges as the foremost recommendation, demonstrating its unique efficacy in dealing with syntactic errors. Previous experiments have consistently illustrated its superior performance over other algorithms.

% Both empirical error reduction and downstream task performance 
% Baran is particularly effective in handling syntactic errors.
% \vspace{-0.08in}
\item Dominant semantic errors: 
To solve semantic errors, the use of \textsf{MLNClean} is recommended for its beneficial application of Markov Logic Networks in learning the credibility of constraints, showing significant performance in managing semantic errors.
% \textsf{MLNClean} is suggested. Its utilization of Markov logic networks to learn the trustworthiness of constraints, has proved to be beneficial in handling semantic errors.}
% In situations with strict time constraints that prohibit the adoption of \textsf{Baran}, \textsf{MLNClean} presents a viable alternative capable of achieving adequate performance across error reduction and enhancing downstream data analysis results.
\item Large-scale Datasets: \textsf{Horizon} deserves consideration.
In situations characterized by large volumes of data, the value-based approach of \textsf{Horizon} can handle large datasets without excessive computation resource consumption.

\item User and Domain Expert Involvement:
Considering that \textsf{HoloClean} demands considerable human input in the form of configurations, it can be beneficial to incorporate this approach as a way to efficiently leverage user and expert insights.
% If strict adherence to rules is required, \textsf{MLNClean} and \textsf{BigDansing} stand as suitable options. 
% However, it is acknowledged that their error reduction performance might not consistently meet expectations. 
% To mitigate the risk of correct values being changed into incorrect ones, combining additional error detection tools with these algorithms is promising.

\end{itemize}

\underline{Guidelines.}
%Furthermore, based on the experiment results, 
Based on the experiments, we provide four guidelines regarding errors, optimization strategies, and data analysis models. 

i) \emph{The optimization strategy is suggested.}
The key cause of subpar performance in current data repair algorithms is incorrect data repair. Integrating the sota error detection methods can reduce the risk of incorrectly repairing accurate data. 
% This strategy focuses primarily on ensuring a baseline level of performance. 
% In cases where original repair performance is already satisfactory, this strategy may have unintended negative effects.

ii) \emph{EDR should be paid more attention.}
EDR provides an intuitive and quantifiable assessment of data quality enhancements.
By focusing on it, a more profound understanding of variations in data quality can be achieved. 
Furthermore, it exhibits the \emph{strongest} correlation with the performance of downstream tasks, augmenting forecasting performances with a higher degree of trustworthiness.

iii) \emph{It is always worthy of repairing data for downstream tasks.} 
Data repair remains a critical operation regardless of the error rate. 
Via carefully selecting appropriate data repair algorithms, it is possible to eliminate up to half of the errors even when the error rate is as high as 90\%. 
It can significantly benefit the downstream tasks.

iv) \emph{Semantic errors should be paid more attention.} \emph{Semantic} errors may detrimentally affect downstream models more than \emph{syntactic} ones. 
The repair algorithms even exacerbate this disruption.
Hence, resolving \emph{semantic} errors should be prioritized.

v) \emph{Choice of data analysis model also matters.}
Different models make different assumptions on data distribution. If improving the data quality is challenging, taking the time to systematically evaluate different models may lead to better performance, like the  F1 score discrepancies between \textit{XGBoost} and \textit{MLP} in the experiments. 
% However, in the practical scenarios, we suggest that data repair algorithms should be applied with error detection methods together, especially when pursuing the error reduction performance using \emph{rule-driven} algorithms.
% In this way, a large number of cells with right values could avoid being changed into wrong ones, this could significantly improve the \emph{error reduction performance}.

% \emph{For downstream tasks.} 
% First of all, data is always worthy of conducting repair.
% Previous experiments underscore that irrespective of the error rate within the data, data repair could consistently lead to significant performance enhancements.
% Besides, mitigating inner errors should be given special attention.
% The distortion introduced into the inner data distribution magnifies negative impacts, surpassing those incurred by outer errors.
% Then, model selection is equally important as the data quality, different data models the feature distribution to the label from different perspectives, thus leading to huge differences.
% Finally, if the data contains more category text data than the regression data in the regression task, model selection for data repair might be redundant. 
% Given the high influence of dirty data, common repair methods, such as utilizing mode or mean values, may prove effective.

% \noindent \underline{\textbf{Tendencies.}}
% We summarize current tendencies as follows:
\vspace{-0.1in}
\section{Future Directions} \label{sec:discussion}

In this section, we underscore research challenges and potential avenues for further exploration within the data repair domain.

\textbf{Research  challenges.}
Within the development of various downstream models, especially the arising of large language model (LLM)~\cite{huang2023reasoning}, more complex data scenarios and problems are appearing.
We summarize current challenges as follows.

\emph{Effective and efficient data repair.}
Existing repair tools struggle with complex datasets due to the infinite error domain versus a single correct value. Efficiency issues persist as well, with solutions like \emph{Baran} unable to process 20K records within a day, demonstrating the pressing demand for improved data repair algorithms.

\emph{Proper metrics evaluating data quality for models.}
Our experiments show no necessarily negative correlation between the downstream task performance and data error rate. With the rise of LLMs, considering data quality issues like hallucination~\cite{manakul2023selfcheckgpt,wei2022chain} gains significance. 
This necessitates the exploration of new metrics to thoroughly assess data quality \emph{w.r.t.} applied models.

\textbf{Potential directions.}
% At present, most of the data repair methods lie in error rate reduction within datasets. With the development of various downstream models, especially the arising of LLM, more complex data scenarios and problems are appearing. 
% We summarize the future directions as follows.
current data repair methods mainly focus on reducing errors, but the advent of LLMs and subsequent complex data scenarios point towards a shift in future directions:

\emph{Combining rule discovery and data information for data repair.}
Existing sota algorithms like \textsf{Baran}~\cite{Mahdavi20baran,Koh17influencefunction, miao21influence} leverage data information. However, their susceptibility to unrelated values highlights the importance of combining rule discovery and data information, especially given the high human costs in the big data era~\cite{Mei23discoveredit}.
% It is noteworthy that, the existing state-of-the-art algorithms like \textsf{Baran}~\cite{Mahdavi20baran}, usually employ data information in the repair process.
% These methods harness the strong learning abilities of machine learning models, enabling them to assess the reliability of potential candidates with rich features, which is beyond the scope of traditional methods.
% However, these algorithms are easily influenced by unrelated values, since values in the data are usually determined by few cells, which is the strength of rules. 
% While in the big data era with high human cost, efficient rule discovery is also an important task~\cite{Mei23discoveredit}.
% Thus, combining rule discovery and data information for data repair is a future trend for data repair.

\emph{LLM for data repair.}
Current automated data repair algorithms struggle to generate candidate solutions beyond the available data~\cite{Mahdavi20baran, Chu15Katara, Fan12master}. Embracing LLMs, with their reasoning ability and extensive knowledge, can overcome this by generating semantically significant and grammatically valid repair candidates~\cite{wang2022diffusiondb, chen2023seed, tang2023verifai, Guha23dcfairness}, making it a fascinating research direction.

\vspace{-0.08in}
\section{Related Work} \label{sec:relatedwork}

%\noindent {\textbf{Data Cleaning.}} 

% The data cleaning process primarily comprises two steps: error detection and repair~\cite{Chu16datacleaning}.
% The error detection involves two work lines, i.e., \emph{cstr-driven} algorithms~\cite{ Abiteboul1995Foundations, Beskales13relative, Chu13holistic, Rekatsinas17holoclean, Ge22mlnclean, Giannakopoulou20relaxation, Rezig21horizon, Fan08Cfds, Ebaid13nadeef, Benjelloun09Swoosh, Boukerche20outlier, Equille11DiscoveryGP}, and \emph{data-driven} ones~\cite{Heidari19holodetect, Neutatz19ed2, raha19mahdavi, Pham21spade, Wang19unidetect, Huang18autodetect, Krishnan16activeclean, miao2021generative, miao23imputationsurvey}.
% And error 
Data cleaning mainly comprises two steps: error detection and repair~\cite{Chu16datacleaning}.
The error detection involves two work lines, i.e., \emph{cstr-driven} algorithms~\cite{ Abiteboul1995Foundations, Beskales13relative, Chu13holistic, Rekatsinas17holoclean, Fan08Cfds, Ebaid13nadeef, Benjelloun09Swoosh, Boukerche20outlier, Equille11DiscoveryGP}, and \emph{data-driven} ones~\cite{Heidari19holodetect, Neutatz19ed2, raha19mahdavi, Pham21spade, Wang19unidetect, Huang18autodetect, Krishnan16activeclean, miao2021generative, miao23imputationsurvey}.
The error repair process typically employs intrinsic data properties like integrity constraints~\cite{Chu16datacleaning,Chu13holistic,Hao17novelcost, Geerts13Llunatic, Rezig21horizon, Fan08Cfds} and the data distribution~\cite{Yakout13scared, Mahdavi20baran, Song16cvtolerant}, or external information like master data~\cite{Fan12master}, knowledge base~\cite{Chu15Katara}, and downstream data analysis models~\cite{Krishnan17boostclean, Hara19sgdcleaning, miao23imputationsurvey} to correct the values.
Deduplication is also an essential operation in data preparation~\cite{Dong18DIandML},
leveraging meta-information~\cite{Ivan1969TheoryForRL, Galhardas01DeclarativeDC, Hassanzadeh09EvaluatingDD}, entity matching rules~\cite{Singh2017EMRule, Fan09RMRules}, ML models~\cite{Mudgal18deepem, Ebraheem18Representationser, Li20ditto, Paganelli22bertER}, and clustering algorithms~\cite{Wu20ZeroER} to deduplicate the original data.  
This survey primarily concentrates on evaluating final data repair results.

\vspace{-0.08in}
\section{Conclusions} \label{sec:conclusions}
In this paper, we comprehensively examine 12 mainstream data repair algorithms through a new taxonomy in a unified framework. 
We propose an effective optimization strategy to improve all of these algorithms. 
We also introduce a novel metric for fair data repair evaluation, highlighting the limitations of existing metrics in evaluating error reduction. 
Through thorough testing on five real-world datasets and analysis across four common downstream tasks encompassing 7 practical workloads, we assess the algorithms' efficacy in various complex scenarios and downstream tasks. We conclude a series of key insights, observations, and future research directions in the field of data repair.
Moving forward, we intend to further enhance the overall data repair performance in practice.

% In this paper, we comprehensively evaluate 12 representative data repair algorithms based on a novel taxonomy within a unified framework. 
% We propose an effective optimization strategy to improve all of these algorithms. We also present a novel metric to fairly evaluate the data repair performance, and reveal the deficiencies of traditional metrics in error reduction evaluation. 
% Extensive experiments over five real-world datasets and four common downstream tasks (with 7 real workloads) reflect the performance of 12 data repair algorithms under different complex scenarios and downstream tasks. 
% We conclude a series of interesting findings and observations, as well as promising research directions and principles in data repair domain. 
%Importantly, our study focuses exclusively on core algorithms based on main memory. 
% Moving forward, we intend to further enhance the overall data repair performance in practice.

\balance
\vspace{-0.08in}
\begin{acks}
 This work is supported by Leading Goose R\&D Program of Zhejiang (No.2024C01109), the NSFC (No.62372404), Beijing Life Science Academy (No.2024900CB0080, No.2023000CB0020), Henan Institute of Chinese Engineering Development Strategies (No.2023HENZDB01), and the Fundamental Research Funds for the Central Universities (No.226-2024-00030).  Xiangyu Zhao is partially supported by Research Impact Fund (No.R1015-23), APRC-CityU New Research Initiatives (No.9610565), CityU-HKIDS Early Career Research Grant (No.9360163), Huawei (Huawei Innovation Research Program), Tencent (CCF-Tencent Open Fund, Tencent Rhino-Bird Focused Research Program), Ant Group (CCF-Ant Research Fund, Ant Group Research Fund), CCF-BaiChuan-Ebtech Foundation Model Fund, and Kuaishou. 
 Xiaoye Miao is the corresponding author.
 % of the work.
  % Huawei (Huawei Innovation Research Program), Tencent (CCF-Tencent Open Fund, Tencent Rhino-Bird Focused Research Program), Ant Group (CCF-Ant Research Fund, Ant Group Research Fund), CCF-BaiChuan-Ebtech Foundation Model Fund, and Kuaishou
\end{acks}

%\clearpage
\bibliographystyle{ACM-Reference-Format}
\bibliography{7.References}


\begin{thebibliography}{87}

%%% the Web address.

\ifx \showCODEN    \undefined \def \showCODEN     #1{\unskip}     \fi
\ifx \showDOI      \undefined \def \showDOI       #1{#1}\fi
\ifx \showISBNx    \undefined \def \showISBNx     #1{\unskip}     \fi
\ifx \showISBNxiii \undefined \def \showISBNxiii  #1{\unskip}     \fi
\ifx \showISSN     \undefined \def \showISSN      #1{\unskip}     \fi
\ifx \showLCCN     \undefined \def \showLCCN      #1{\unskip}     \fi
\ifx \shownote     \undefined \def \shownote      #1{#1}          \fi
\ifx \showarticletitle \undefined \def \showarticletitle #1{#1}   \fi
\ifx \showURL      \undefined \def \showURL       {\relax}        \fi
% invisible to TeX
\providecommand\bibfield[2]{#2}
\providecommand\bibinfo[2]{#2}
\providecommand\natexlab[1]{#1}
\providecommand\showeprint[2][]{arXiv:#2}

\bibitem[\protect\citeauthoryear{Abdelaal, Hammacher, and
  Sch{\"{o}}ning}{Abdelaal et~al\mbox{.}}{2023}]%
        {Abdelaal23rein}
\bibfield{author}{\bibinfo{person}{Mohamed Abdelaal},
  \bibinfo{person}{Christian Hammacher}, {and} \bibinfo{person}{Harald
  Sch{\"{o}}ning}.} \bibinfo{year}{2023}\natexlab{}.
\newblock \showarticletitle{{REIN:} {A} comprehensive benchmark framework for
  data cleaning methods in {ML} pipelines}. In
  \bibinfo{booktitle}{\emph{EDBT}}. \bibinfo{pages}{499--511}.
\newblock


\bibitem[\protect\citeauthoryear{Abedjan, Chu, Deng, Fernandez, Ilyas, Ouzzani,
  Papotti, Stonebraker, and Tang}{Abedjan et~al\mbox{.}}{2016}]%
        {Ziawasch16detecting}
\bibfield{author}{\bibinfo{person}{Ziawasch Abedjan}, \bibinfo{person}{Xu Chu},
  \bibinfo{person}{Dong Deng}, \bibinfo{person}{Raul~Castro Fernandez},
  \bibinfo{person}{Ihab~F. Ilyas}, \bibinfo{person}{Mourad Ouzzani},
  \bibinfo{person}{Paolo Papotti}, \bibinfo{person}{Michael Stonebraker}, {and}
  \bibinfo{person}{Nan Tang}.} \bibinfo{year}{2016}\natexlab{}.
\newblock \showarticletitle{Detecting data errors: Where are we and what needs
  to be done?}
\newblock \bibinfo{journal}{\emph{Proceedings of the VLDB Endowment}}
  \bibinfo{volume}{9}, \bibinfo{number}{12} (\bibinfo{year}{2016}),
  \bibinfo{pages}{993–1004}.
\newblock


\bibitem[\protect\citeauthoryear{Abiteboul, Hull, and Vianu}{Abiteboul
  et~al\mbox{.}}{1995}]%
        {Abiteboul1995Foundations}
\bibfield{author}{\bibinfo{person}{Serge Abiteboul}, \bibinfo{person}{Richard
  Hull}, {and} \bibinfo{person}{Victor Vianu}.}
  \bibinfo{year}{1995}\natexlab{}.
\newblock \bibinfo{booktitle}{\emph{Foundations of Databases: The Logical
  Level}}.
\newblock \bibinfo{publisher}{Addison-Wesley Longman Publishing Cooperation,
  Inc.}
\newblock


\bibitem[\protect\citeauthoryear{Arocena, Glavic, Mecca, Miller, Papotti, and
  Santoro}{Arocena et~al\mbox{.}}{2015}]%
        {Arocena15bart}
\bibfield{author}{\bibinfo{person}{Patricia~C. Arocena}, \bibinfo{person}{Boris
  Glavic}, \bibinfo{person}{Giansalvatore Mecca}, \bibinfo{person}{Ren\'{e}e~J.
  Miller}, \bibinfo{person}{Paolo Papotti}, {and} \bibinfo{person}{Donatello
  Santoro}.} \bibinfo{year}{2015}\natexlab{}.
\newblock \showarticletitle{Messing up with BART: Error generation for
  evaluating data-cleaning algorithms}.
\newblock \bibinfo{journal}{\emph{Proceedings of the VLDB Endowment}}
  \bibinfo{volume}{9}, \bibinfo{number}{2} (\bibinfo{year}{2015}),
  \bibinfo{pages}{36–47}.
\newblock


\bibitem[\protect\citeauthoryear{Benjelloun, Garcia-Molina, Menestrina, Su,
  Whang, and Widom}{Benjelloun et~al\mbox{.}}{2009}]%
        {Benjelloun09Swoosh}
\bibfield{author}{\bibinfo{person}{Omar Benjelloun}, \bibinfo{person}{Hector
  Garcia-Molina}, \bibinfo{person}{David Menestrina}, \bibinfo{person}{Qi Su},
  \bibinfo{person}{Steven~Euijong Whang}, {and} \bibinfo{person}{Jennifer
  Widom}.} \bibinfo{year}{2009}\natexlab{}.
\newblock \showarticletitle{Swoosh: A generic approach to entity resolution}.
\newblock \bibinfo{journal}{\emph{The VLDB Journal}} \bibinfo{volume}{18},
  \bibinfo{number}{1} (\bibinfo{year}{2009}), \bibinfo{pages}{255–276}.
\newblock


\bibitem[\protect\citeauthoryear{Bentley}{Bentley}{1975}]%
        {bentley1975kdtree}
\bibfield{author}{\bibinfo{person}{Jon~Louis Bentley}.}
  \bibinfo{year}{1975}\natexlab{}.
\newblock \showarticletitle{Multidimensional binary search trees used for
  associative searching}.
\newblock \bibinfo{journal}{\emph{Commun. ACM}} \bibinfo{volume}{18},
  \bibinfo{number}{9} (\bibinfo{year}{1975}), \bibinfo{pages}{509--517}.
\newblock


\bibitem[\protect\citeauthoryear{Berti-Equille, Dasu, and
  Srivastava}{Berti-Equille et~al\mbox{.}}{2011}]%
        {Equille11DiscoveryGP}
\bibfield{author}{\bibinfo{person}{Laure Berti-Equille},
  \bibinfo{person}{Tamraparni Dasu}, {and} \bibinfo{person}{Divesh
  Srivastava}.} \bibinfo{year}{2011}\natexlab{}.
\newblock \showarticletitle{Discovery of complex glitch patterns: A novel
  approach to quantitative data cleaning}. In \bibinfo{booktitle}{\emph{ICDE}}.
  \bibinfo{pages}{733–744}.
\newblock


\bibitem[\protect\citeauthoryear{Beskales, Ilyas, Golab, and
  Galiullin}{Beskales et~al\mbox{.}}{2013}]%
        {Beskales13relative}
\bibfield{author}{\bibinfo{person}{George Beskales}, \bibinfo{person}{Ihab~F.
  Ilyas}, \bibinfo{person}{Lukasz Golab}, {and} \bibinfo{person}{Artur
  Galiullin}.} \bibinfo{year}{2013}\natexlab{}.
\newblock \showarticletitle{On the relative trust between inconsistent data and
  inaccurate constraints}. In \bibinfo{booktitle}{\emph{ICDE}}.
  \bibinfo{pages}{541--552}.
\newblock


\bibitem[\protect\citeauthoryear{BigDaMa}{BigDaMa}{2023}]%
        {errorgurl}
\bibfield{author}{\bibinfo{person}{BigDaMa}.} \bibinfo{year}{2023}\natexlab{}.
\newblock \bibinfo{title}{Error generator}.
\newblock
  \bibinfo{howpublished}{\url{https://github.com/BigDaMa/error-generator}}.
\newblock
\newblock
\shownote{Accessed: 2024-06-14.}


\bibitem[\protect\citeauthoryear{Bleifu\ss{}, Kruse, and Naumann}{Bleifu\ss{}
  et~al\mbox{.}}{2017}]%
        {hydra17Bleifu}
\bibfield{author}{\bibinfo{person}{Tobias Bleifu\ss{}},
  \bibinfo{person}{Sebastian Kruse}, {and} \bibinfo{person}{Felix Naumann}.}
  \bibinfo{year}{2017}\natexlab{}.
\newblock \showarticletitle{Efficient denial constraint discovery with hydra}.
\newblock \bibinfo{journal}{\emph{Proceedings of the VLDB Endowment}}
  \bibinfo{volume}{11}, \bibinfo{number}{3} (\bibinfo{year}{2017}),
  \bibinfo{pages}{311–323}.
\newblock


\bibitem[\protect\citeauthoryear{Borji}{Borji}{2023}]%
        {borji2023midjourney}
\bibfield{author}{\bibinfo{person}{Ali Borji}.}
  \bibinfo{year}{2023}\natexlab{}.
\newblock \showarticletitle{Generated faces in the wild: Quantitative
  comparison of stable diffusion, midjourney and DALL-E 2}.
\newblock \bibinfo{journal}{\emph{ArXiv Preprint ArXiv:2210.00586}}
  (\bibinfo{year}{2023}).
\newblock


\bibitem[\protect\citeauthoryear{Boukerche, Zheng, and Alfandi}{Boukerche
  et~al\mbox{.}}{2020}]%
        {Boukerche20outlier}
\bibfield{author}{\bibinfo{person}{Azzedine Boukerche}, \bibinfo{person}{Lining
  Zheng}, {and} \bibinfo{person}{Omar Alfandi}.}
  \bibinfo{year}{2020}\natexlab{}.
\newblock \showarticletitle{Outlier detection: Methods, models, and
  classification}.
\newblock \bibinfo{journal}{\emph{Comput. Surveys}} \bibinfo{volume}{53},
  \bibinfo{number}{3}, Article \bibinfo{articleno}{55} (\bibinfo{year}{2020}),
  \bibinfo{numpages}{37}~pages.
\newblock


\bibitem[\protect\citeauthoryear{Broder}{Broder}{1997}]%
        {broder1997minihash}
\bibfield{author}{\bibinfo{person}{Andrei~Z Broder}.}
  \bibinfo{year}{1997}\natexlab{}.
\newblock \showarticletitle{On the resemblance and containment of documents}.
  In \bibinfo{booktitle}{\emph{SEQUENCES}}. \bibinfo{pages}{21--29}.
\newblock


\bibitem[\protect\citeauthoryear{Chen and Guestrin}{Chen and Guestrin}{2016}]%
        {Chen16xgboost}
\bibfield{author}{\bibinfo{person}{Tianqi Chen} {and} \bibinfo{person}{Carlos
  Guestrin}.} \bibinfo{year}{2016}\natexlab{}.
\newblock \showarticletitle{XGBoost: A scalable tree boosting system}. In
  \bibinfo{booktitle}{\emph{KDD}}. \bibinfo{pages}{785–794}.
\newblock


\bibitem[\protect\citeauthoryear{CHen, Cao, Madden, Fan, Tang, Gu, Shang, Liu,
  Cafarella, and Kraska}{CHen et~al\mbox{.}}{2023}]%
        {chen2023seed}
\bibfield{author}{\bibinfo{person}{Zui CHen}, \bibinfo{person}{Lei Cao},
  \bibinfo{person}{Sam Madden}, \bibinfo{person}{Ju Fan}, \bibinfo{person}{Nan
  Tang}, \bibinfo{person}{Zihui Gu}, \bibinfo{person}{Zeyuan Shang},
  \bibinfo{person}{Chunwei Liu}, \bibinfo{person}{Michael Cafarella}, {and}
  \bibinfo{person}{Tim Kraska}.} \bibinfo{year}{2023}\natexlab{}.
\newblock \showarticletitle{SEED: Simple, efficient, and effective data
  management via large language models}.
\newblock \bibinfo{journal}{\emph{ArXiv Preprint ArXiv:2310.00749}}
  (\bibinfo{year}{2023}).
\newblock


\bibitem[\protect\citeauthoryear{Chiang and Miller}{Chiang and Miller}{2011}]%
        {Chiang11unified}
\bibfield{author}{\bibinfo{person}{Fei Chiang} {and}
  \bibinfo{person}{Ren{\'{e}}e~J. Miller}.} \bibinfo{year}{2011}\natexlab{}.
\newblock \showarticletitle{A unified model for data and constraint repair}. In
  \bibinfo{booktitle}{\emph{ICDE}}. \bibinfo{pages}{446--457}.
\newblock


\bibitem[\protect\citeauthoryear{Chu, Ilyas, Krishnan, and Wang}{Chu
  et~al\mbox{.}}{2016}]%
        {Chu16datacleaning}
\bibfield{author}{\bibinfo{person}{Xu Chu}, \bibinfo{person}{Ihab~F. Ilyas},
  \bibinfo{person}{Sanjay Krishnan}, {and} \bibinfo{person}{Jiannan Wang}.}
  \bibinfo{year}{2016}\natexlab{}.
\newblock \showarticletitle{Data cleaning: Overview and emerging challenges}.
  In \bibinfo{booktitle}{\emph{SIGMOD}}. \bibinfo{pages}{2201–2206}.
\newblock


\bibitem[\protect\citeauthoryear{Chu, Ilyas, and Papotti}{Chu
  et~al\mbox{.}}{2013}]%
        {Chu13holistic}
\bibfield{author}{\bibinfo{person}{Xu Chu}, \bibinfo{person}{I.~F. Ilyas},
  {and} \bibinfo{person}{P. Papotti}.} \bibinfo{year}{2013}\natexlab{}.
\newblock \showarticletitle{Holistic data cleaning: Putting violations into
  context}. In \bibinfo{booktitle}{\emph{ICDE}}. \bibinfo{pages}{458--469}.
\newblock


\bibitem[\protect\citeauthoryear{Chu, Morcos, Ilyas, Ouzzani, Papotti, Tang,
  and Ye}{Chu et~al\mbox{.}}{2015}]%
        {Chu15Katara}
\bibfield{author}{\bibinfo{person}{Xu Chu}, \bibinfo{person}{John Morcos},
  \bibinfo{person}{Ihab~F. Ilyas}, \bibinfo{person}{Mourad Ouzzani},
  \bibinfo{person}{Paolo Papotti}, \bibinfo{person}{Nan Tang}, {and}
  \bibinfo{person}{Yin Ye}.} \bibinfo{year}{2015}\natexlab{}.
\newblock \showarticletitle{KATARA: A data cleaning system powered by knowledge
  bases and crowdsourcing}. In \bibinfo{booktitle}{\emph{SIGMOD}}.
  \bibinfo{pages}{1247–1261}.
\newblock


\bibitem[\protect\citeauthoryear{Czarnowski}{Czarnowski}{2012}]%
        {Czarnowski12cluster}
\bibfield{author}{\bibinfo{person}{Ireneusz Czarnowski}.}
  \bibinfo{year}{2012}\natexlab{}.
\newblock \showarticletitle{Cluster-based instance selection for machine
  classification}.
\newblock \bibinfo{journal}{\emph{Knowledge and Information Systems}}
  \bibinfo{volume}{30}, \bibinfo{number}{5} (\bibinfo{year}{2012}),
  \bibinfo{pages}{113--133}.
\newblock


\bibitem[\protect\citeauthoryear{Dong and Rekatsinas}{Dong and
  Rekatsinas}{2018}]%
        {Dong18DIandML}
\bibfield{author}{\bibinfo{person}{Xin~Luna Dong} {and}
  \bibinfo{person}{Theodoros Rekatsinas}.} \bibinfo{year}{2018}\natexlab{}.
\newblock \showarticletitle{Data integration and machine learning: A natural
  synergy}. In \bibinfo{booktitle}{\emph{SIGMOD}}.
  \bibinfo{pages}{1645–1650}.
\newblock


\bibitem[\protect\citeauthoryear{Ebaid, Elmagarmid, Ilyas, Ouzzani,
  Quiane-Ruiz, Tang, and Yin}{Ebaid et~al\mbox{.}}{2013}]%
        {Ebaid13nadeef}
\bibfield{author}{\bibinfo{person}{Amr Ebaid}, \bibinfo{person}{Ahmed
  Elmagarmid}, \bibinfo{person}{Ihab~F. Ilyas}, \bibinfo{person}{Mourad
  Ouzzani}, \bibinfo{person}{Jorge-Arnulfo Quiane-Ruiz}, \bibinfo{person}{Nan
  Tang}, {and} \bibinfo{person}{Si Yin}.} \bibinfo{year}{2013}\natexlab{}.
\newblock \showarticletitle{NADEEF: A generalized data cleaning system}.
\newblock \bibinfo{journal}{\emph{Proceedings of the VLDB Endowment}}
  \bibinfo{volume}{6}, \bibinfo{number}{12} (\bibinfo{year}{2013}),
  \bibinfo{pages}{1218–1221}.
\newblock


\bibitem[\protect\citeauthoryear{Ebraheem, Thirumuruganathan, Joty, Ouzzani,
  and Tang}{Ebraheem et~al\mbox{.}}{2018}]%
        {Ebraheem18Representationser}
\bibfield{author}{\bibinfo{person}{Muhammad Ebraheem},
  \bibinfo{person}{Saravanan Thirumuruganathan}, \bibinfo{person}{Shafiq Joty},
  \bibinfo{person}{Mourad Ouzzani}, {and} \bibinfo{person}{Nan Tang}.}
  \bibinfo{year}{2018}\natexlab{}.
\newblock \showarticletitle{Distributed representations of tuples for entity
  resolution}.
\newblock \bibinfo{journal}{\emph{Proceedings of the VLDB Endowment}}
  \bibinfo{volume}{11}, \bibinfo{number}{11} (\bibinfo{year}{2018}),
  \bibinfo{pages}{1454–1467}.
\newblock


\bibitem[\protect\citeauthoryear{Fan, Geerts, Jia, and Kementsietsidis}{Fan
  et~al\mbox{.}}{2008}]%
        {Fan08Cfds}
\bibfield{author}{\bibinfo{person}{Wenfei Fan}, \bibinfo{person}{Floris
  Geerts}, \bibinfo{person}{Xibei Jia}, {and} \bibinfo{person}{Anastasios
  Kementsietsidis}.} \bibinfo{year}{2008}\natexlab{}.
\newblock \showarticletitle{Conditional functional dependencies for capturing
  data inconsistencies}.
\newblock \bibinfo{journal}{\emph{ACM Transactions on Database Systems}}
  \bibinfo{volume}{33}, \bibinfo{number}{2}, Article \bibinfo{articleno}{6}
  (\bibinfo{year}{2008}), \bibinfo{numpages}{48}~pages.
\newblock


\bibitem[\protect\citeauthoryear{Fan, Jia, Li, and Ma}{Fan
  et~al\mbox{.}}{2009}]%
        {Fan09RMRules}
\bibfield{author}{\bibinfo{person}{Wenfei Fan}, \bibinfo{person}{Xibei Jia},
  \bibinfo{person}{Jianzhong Li}, {and} \bibinfo{person}{Shuai Ma}.}
  \bibinfo{year}{2009}\natexlab{}.
\newblock \showarticletitle{Reasoning about record matching rules}.
\newblock \bibinfo{journal}{\emph{Proceedings of the VLDB Endowment}}
  \bibinfo{volume}{2}, \bibinfo{number}{1} (\bibinfo{year}{2009}),
  \bibinfo{pages}{407–418}.
\newblock


\bibitem[\protect\citeauthoryear{Fan, Li, Ma, Tang, and Yu}{Fan
  et~al\mbox{.}}{2012}]%
        {Fan12master}
\bibfield{author}{\bibinfo{person}{Wenfei Fan}, \bibinfo{person}{Jianzhong Li},
  \bibinfo{person}{Shuai Ma}, \bibinfo{person}{Nan Tang}, {and}
  \bibinfo{person}{Wenyuan Yu}.} \bibinfo{year}{2012}\natexlab{}.
\newblock \showarticletitle{Towards certain fixes with editing rules and master
  data}.
\newblock \bibinfo{journal}{\emph{The VLDB Journal}} \bibinfo{volume}{21},
  \bibinfo{number}{2} (\bibinfo{year}{2012}), \bibinfo{pages}{213–238}.
\newblock


\bibitem[\protect\citeauthoryear{Fellegi and Sunter}{Fellegi and
  Sunter}{1969}]%
        {Ivan1969TheoryForRL}
\bibfield{author}{\bibinfo{person}{Ivan~P. Fellegi} {and}
  \bibinfo{person}{Alan~B. Sunter}.} \bibinfo{year}{1969}\natexlab{}.
\newblock \showarticletitle{A theory for record linkage}.
\newblock \bibinfo{journal}{\emph{J. Amer. Statist. Assoc.}}
  \bibinfo{volume}{64}, \bibinfo{number}{328} (\bibinfo{year}{1969}),
  \bibinfo{pages}{1183--1210}.
\newblock


\bibitem[\protect\citeauthoryear{Frey and Dueck}{Frey and Dueck}{2007}]%
        {frey2007clusterap}
\bibfield{author}{\bibinfo{person}{Brendan~J Frey} {and}
  \bibinfo{person}{Delbert Dueck}.} \bibinfo{year}{2007}\natexlab{}.
\newblock \showarticletitle{Clustering by passing messages between data
  points}.
\newblock \bibinfo{journal}{\emph{Science}} \bibinfo{volume}{315},
  \bibinfo{number}{5814} (\bibinfo{year}{2007}), \bibinfo{pages}{972--976}.
\newblock


\bibitem[\protect\citeauthoryear{Galhardas, Florescu, Shasha, Simon, and
  Saita}{Galhardas et~al\mbox{.}}{2001}]%
        {Galhardas01DeclarativeDC}
\bibfield{author}{\bibinfo{person}{Helena Galhardas}, \bibinfo{person}{Daniela
  Florescu}, \bibinfo{person}{Dennis Shasha}, \bibinfo{person}{Eric Simon},
  {and} \bibinfo{person}{Cristian-Augustin Saita}.}
  \bibinfo{year}{2001}\natexlab{}.
\newblock \showarticletitle{Declarative data cleaning: Language, model, and
  algorithms}.
\newblock \bibinfo{journal}{\emph{Proceedings of the VLDB Endowment}},
  \bibinfo{pages}{371–380}.
\newblock


\bibitem[\protect\citeauthoryear{Ge, Gao, Miao, Yao, and Wang}{Ge
  et~al\mbox{.}}{2022}]%
        {Ge22mlnclean}
\bibfield{author}{\bibinfo{person}{Congcong Ge}, \bibinfo{person}{Yunjun Gao},
  \bibinfo{person}{Xiaoye Miao}, \bibinfo{person}{Bin Yao}, {and}
  \bibinfo{person}{Haobo Wang}.} \bibinfo{year}{2022}\natexlab{}.
\newblock \showarticletitle{A hybrid data cleaning framework using markov logic
  networks}.
\newblock \bibinfo{journal}{\emph{IEEE Transactions on Knowledge and Data
  Engineering}} \bibinfo{volume}{34}, \bibinfo{number}{5}
  (\bibinfo{year}{2022}), \bibinfo{pages}{2048--2062}.
\newblock


\bibitem[\protect\citeauthoryear{Geerts, Mecca, Papotti, and Santoro}{Geerts
  et~al\mbox{.}}{2013}]%
        {Geerts13Llunatic}
\bibfield{author}{\bibinfo{person}{Floris Geerts},
  \bibinfo{person}{Giansalvatore Mecca}, \bibinfo{person}{Paolo Papotti}, {and}
  \bibinfo{person}{Donatello Santoro}.} \bibinfo{year}{2013}\natexlab{}.
\newblock \showarticletitle{The LLUNATIC data-cleaning framework}.
\newblock \bibinfo{journal}{\emph{Proceedings of the VLDB Endowment}}
  \bibinfo{volume}{6}, \bibinfo{number}{9} (\bibinfo{year}{2013}),
  \bibinfo{pages}{625–636}.
\newblock


\bibitem[\protect\citeauthoryear{Giannakopoulou, Karpathiotakis, and
  Ailamaki}{Giannakopoulou et~al\mbox{.}}{2020}]%
        {Giannakopoulou20relaxation}
\bibfield{author}{\bibinfo{person}{Stella Giannakopoulou},
  \bibinfo{person}{Manos Karpathiotakis}, {and} \bibinfo{person}{Anastasia
  Ailamaki}.} \bibinfo{year}{2020}\natexlab{}.
\newblock \showarticletitle{Cleaning denial constraint violations through
  relaxation}. In \bibinfo{booktitle}{\emph{SIGMOD}}.
  \bibinfo{pages}{805–815}.
\newblock


\bibitem[\protect\citeauthoryear{Guha, Khan, Stoyanovich, and Schelter}{Guha
  et~al\mbox{.}}{2023}]%
        {Guha23dcfairness}
\bibfield{author}{\bibinfo{person}{Shubha Guha}, \bibinfo{person}{Falaah~Arif
  Khan}, \bibinfo{person}{Julia Stoyanovich}, {and} \bibinfo{person}{Sebastian
  Schelter}.} \bibinfo{year}{2023}\natexlab{}.
\newblock \showarticletitle{Automated data cleaning can hurt fairness in
  machine learning-based decision making}. In \bibinfo{booktitle}{\emph{ICDE}}.
  \bibinfo{pages}{3747--3754}.
\newblock


\bibitem[\protect\citeauthoryear{Hao, Tang, Li, He, Ta, and Feng}{Hao
  et~al\mbox{.}}{2017}]%
        {Hao17novelcost}
\bibfield{author}{\bibinfo{person}{Shuang Hao}, \bibinfo{person}{Nan Tang},
  \bibinfo{person}{Guoliang Li}, \bibinfo{person}{Jian He}, \bibinfo{person}{Na
  Ta}, {and} \bibinfo{person}{Jianhua Feng}.} \bibinfo{year}{2017}\natexlab{}.
\newblock \showarticletitle{A novel cost-based model for data repairing}.
\newblock \bibinfo{journal}{\emph{IEEE Transactions on Knowledge and Data
  Engineering}} \bibinfo{volume}{29}, \bibinfo{number}{4}
  (\bibinfo{year}{2017}), \bibinfo{pages}{727–742}.
\newblock


\bibitem[\protect\citeauthoryear{Hara, Nitanda, and Maehara}{Hara
  et~al\mbox{.}}{2019}]%
        {Hara19sgdcleaning}
\bibfield{author}{\bibinfo{person}{Satoshi Hara}, \bibinfo{person}{Atsushi
  Nitanda}, {and} \bibinfo{person}{Takanori Maehara}.}
  \bibinfo{year}{2019}\natexlab{}.
\newblock \showarticletitle{Data cleansing for models trained with SGD}. In
  \bibinfo{booktitle}{\emph{NeurIPS}}. \bibinfo{pages}{4213–4222}.
\newblock


\bibitem[\protect\citeauthoryear{Hasan and Mahdavi}{Hasan and Mahdavi}{2021}]%
        {Hasan21wikierrorcorrection}
\bibfield{author}{\bibinfo{person}{Md~Kamrul Hasan} {and}
  \bibinfo{person}{Mohammad Mahdavi}.} \bibinfo{year}{2021}\natexlab{}.
\newblock \showarticletitle{Automatic Error Correction Using the Wikipedia Page
  Revision History}. In \bibinfo{booktitle}{\emph{CIKM}}.
  \bibinfo{pages}{3073–3077}.
\newblock


\bibitem[\protect\citeauthoryear{Hassanzadeh, Chiang, Lee, and
  Miller}{Hassanzadeh et~al\mbox{.}}{2009}]%
        {Hassanzadeh09EvaluatingDD}
\bibfield{author}{\bibinfo{person}{Oktie Hassanzadeh}, \bibinfo{person}{Fei
  Chiang}, \bibinfo{person}{Hyun~Chul Lee}, {and} \bibinfo{person}{Ren\'{e}e~J.
  Miller}.} \bibinfo{year}{2009}\natexlab{}.
\newblock \showarticletitle{Framework for evaluating clustering algorithms in
  duplicate detection}.
\newblock \bibinfo{journal}{\emph{Proceedings of the VLDB Endowment}}
  \bibinfo{volume}{2}, \bibinfo{number}{1} (\bibinfo{year}{2009}),
  \bibinfo{pages}{1282–1293}.
\newblock


\bibitem[\protect\citeauthoryear{Heidari, McGrath, Ilyas, and
  Rekatsinas}{Heidari et~al\mbox{.}}{2019}]%
        {Heidari19holodetect}
\bibfield{author}{\bibinfo{person}{Alireza Heidari}, \bibinfo{person}{Joshua
  McGrath}, \bibinfo{person}{Ihab~F. Ilyas}, {and} \bibinfo{person}{Theodoros
  Rekatsinas}.} \bibinfo{year}{2019}\natexlab{}.
\newblock \showarticletitle{HoloDetect: Few-shot learning for error detection}.
  In \bibinfo{booktitle}{\emph{SIGMOD}}. \bibinfo{pages}{829–846}.
\newblock


\bibitem[\protect\citeauthoryear{Hu, Wang, Jiao, Sankaran, Catlett, and
  Work}{Hu et~al\mbox{.}}{2019}]%
        {hu2019automatic}
\bibfield{author}{\bibinfo{person}{Yue Hu}, \bibinfo{person}{Yanbing Wang},
  \bibinfo{person}{Canwen Jiao}, \bibinfo{person}{Rajesh Sankaran},
  \bibinfo{person}{Charles~E Catlett}, {and} \bibinfo{person}{Daniel~B Work}.}
  \bibinfo{year}{2019}\natexlab{}.
\newblock \showarticletitle{Automatic data cleaning via tensor factorization
  for large urban environmental sensor networks}. In
  \bibinfo{booktitle}{\emph{Proceedings of the NeurIPS Workshop on Tackling
  Climate Change with Machine Learning}}.
\newblock


\bibitem[\protect\citeauthoryear{Huang and Chang}{Huang and Chang}{2023}]%
        {huang2023reasoning}
\bibfield{author}{\bibinfo{person}{Jie Huang} {and} \bibinfo{person}{Kevin
  Chen-Chuan Chang}.} \bibinfo{year}{2023}\natexlab{}.
\newblock \showarticletitle{Towards reasoning in large language models: A
  survey}.
\newblock \bibinfo{journal}{\emph{ArXiv Preprint ArXiv:2212.10403}}
  (\bibinfo{year}{2023}).
\newblock


\bibitem[\protect\citeauthoryear{Huang and He}{Huang and He}{2018}]%
        {Huang18autodetect}
\bibfield{author}{\bibinfo{person}{Zhipeng Huang} {and} \bibinfo{person}{Yeye
  He}.} \bibinfo{year}{2018}\natexlab{}.
\newblock \showarticletitle{Auto-detect: Data-driven error detection in
  tables}. In \bibinfo{booktitle}{\emph{SIGMOD}}. \bibinfo{pages}{1377–1392}.
\newblock


\bibitem[\protect\citeauthoryear{Ider and Schmietendorf}{Ider and
  Schmietendorf}{2018}]%
        {Kadir20dataprivacy}
\bibfield{author}{\bibinfo{person}{Kadir Ider} {and} \bibinfo{person}{Andreas
  Schmietendorf}.} \bibinfo{year}{2018}\natexlab{}.
\newblock \showarticletitle{Data privacy for AI fraud detection models}. In
  \bibinfo{booktitle}{\emph{ICDS}}. \bibinfo{pages}{102–107}.
\newblock


\bibitem[\protect\citeauthoryear{Ilyas and Chu}{Ilyas and Chu}{2015}]%
        {Ilyas15trendsdc}
\bibfield{author}{\bibinfo{person}{Ihab~F. Ilyas} {and} \bibinfo{person}{Xu
  Chu}.} \bibinfo{year}{2015}\natexlab{}.
\newblock \showarticletitle{Trends in cleaning relational data: Consistency and
  deduplication}.
\newblock \bibinfo{journal}{\emph{Foundations and Trends in Databases}}
  \bibinfo{volume}{5}, \bibinfo{number}{4} (\bibinfo{year}{2015}),
  \bibinfo{pages}{281–393}.
\newblock


\bibitem[\protect\citeauthoryear{Ilyas and Chu}{Ilyas and Chu}{2019}]%
        {Ilyas19datacleaning}
\bibfield{author}{\bibinfo{person}{Ihab~F. Ilyas} {and} \bibinfo{person}{Xu
  Chu}.} \bibinfo{year}{2019}\natexlab{}.
\newblock \bibinfo{booktitle}{\emph{Data cleaning}}.
\newblock \bibinfo{publisher}{Association for Computing Machinery}.
\newblock


\bibitem[\protect\citeauthoryear{Khayyat, Ilyas, Jindal, Madden, Ouzzani,
  Papotti, Quian\'{e}-Ruiz, Tang, and Yin}{Khayyat et~al\mbox{.}}{2015}]%
        {Khayyat15bigdansing}
\bibfield{author}{\bibinfo{person}{Zuhair Khayyat}, \bibinfo{person}{Ihab~F.
  Ilyas}, \bibinfo{person}{Alekh Jindal}, \bibinfo{person}{Samuel Madden},
  \bibinfo{person}{Mourad Ouzzani}, \bibinfo{person}{Paolo Papotti},
  \bibinfo{person}{Jorge-Arnulfo Quian\'{e}-Ruiz}, \bibinfo{person}{Nan Tang},
  {and} \bibinfo{person}{Si Yin}.} \bibinfo{year}{2015}\natexlab{}.
\newblock \showarticletitle{BigDansing: A system for big data cleansing}. In
  \bibinfo{booktitle}{\emph{SIGMOD}}. \bibinfo{pages}{1215–1230}.
\newblock


\bibitem[\protect\citeauthoryear{Koh and Liang}{Koh and Liang}{2017}]%
        {Koh17influencefunction}
\bibfield{author}{\bibinfo{person}{Pang~Wei Koh} {and} \bibinfo{person}{Percy
  Liang}.} \bibinfo{year}{2017}\natexlab{}.
\newblock \showarticletitle{Understanding black-box predictions via influence
  functions}. In \bibinfo{booktitle}{\emph{ICML}}.
  \bibinfo{pages}{1885–1894}.
\newblock


\bibitem[\protect\citeauthoryear{Krishnan, Franklin, Goldberg, and Wu}{Krishnan
  et~al\mbox{.}}{2017}]%
        {Krishnan17boostclean}
\bibfield{author}{\bibinfo{person}{Sanjay Krishnan},
  \bibinfo{person}{Michael~J. Franklin}, \bibinfo{person}{Ken Goldberg}, {and}
  \bibinfo{person}{Eugene Wu}.} \bibinfo{year}{2017}\natexlab{}.
\newblock \showarticletitle{BoostClean: Automated error detection and repair
  for machine learning}.
\newblock \bibinfo{journal}{\emph{ArXiv Preprint ArXiv:1711.01299}}
  (\bibinfo{year}{2017}).
\newblock


\bibitem[\protect\citeauthoryear{Krishnan, Haas, Franklin, and Wu}{Krishnan
  et~al\mbox{.}}{2016a}]%
        {Krishnan16survey}
\bibfield{author}{\bibinfo{person}{Sanjay Krishnan}, \bibinfo{person}{Daniel
  Haas}, \bibinfo{person}{Michael~J. Franklin}, {and} \bibinfo{person}{Eugene
  Wu}.} \bibinfo{year}{2016}\natexlab{a}.
\newblock \showarticletitle{Towards reliable interactive data cleaning: A user
  survey and recommendations}. In \bibinfo{booktitle}{\emph{HILDA}}. Article
  \bibinfo{articleno}{9}, \bibinfo{numpages}{5}~pages.
\newblock


\bibitem[\protect\citeauthoryear{Krishnan, Wang, Wu, Franklin, and
  Goldberg}{Krishnan et~al\mbox{.}}{2016b}]%
        {Krishnan16activeclean}
\bibfield{author}{\bibinfo{person}{Sanjay Krishnan}, \bibinfo{person}{Jiannan
  Wang}, \bibinfo{person}{Eugene Wu}, \bibinfo{person}{Michael~J. Franklin},
  {and} \bibinfo{person}{Ken Goldberg}.} \bibinfo{year}{2016}\natexlab{b}.
\newblock \showarticletitle{ActiveClean: Interactive data cleaning for
  statistical modeling}.
\newblock \bibinfo{journal}{\emph{Proceedings of the VLDB Endowment}}
  \bibinfo{volume}{9}, \bibinfo{number}{12} (\bibinfo{year}{2016}),
  \bibinfo{pages}{948–959}.
\newblock


\bibitem[\protect\citeauthoryear{Li, Hay, Miklau, and Wang}{Li
  et~al\mbox{.}}{2014}]%
        {li2014query}
\bibfield{author}{\bibinfo{person}{Chao Li}, \bibinfo{person}{Michael Hay},
  \bibinfo{person}{Gerome Miklau}, {and} \bibinfo{person}{Yue Wang}.}
  \bibinfo{year}{2014}\natexlab{}.
\newblock \showarticletitle{A data- and workload-aware algorithm for range
  queries under differential privacy}.
\newblock \bibinfo{journal}{\emph{Proceedings of the VLDB Endowment}}
  \bibinfo{volume}{7}, \bibinfo{number}{5} (\bibinfo{year}{2014}),
  \bibinfo{pages}{341–352}.
\newblock


\bibitem[\protect\citeauthoryear{Li, Rao, Blase, Zhang, Chu, and Zhang}{Li
  et~al\mbox{.}}{2021}]%
        {Li21cleanml}
\bibfield{author}{\bibinfo{person}{Peng Li}, \bibinfo{person}{Xi Rao},
  \bibinfo{person}{Jennifer Blase}, \bibinfo{person}{Yue Zhang},
  \bibinfo{person}{Xu Chu}, {and} \bibinfo{person}{Ce Zhang}.}
  \bibinfo{year}{2021}\natexlab{}.
\newblock \showarticletitle{CleanML: A study for evaluating the impact of data
  cleaning on {ML} classification tasks}. In \bibinfo{booktitle}{\emph{ICDE}}.
  \bibinfo{pages}{13--24}.
\newblock


\bibitem[\protect\citeauthoryear{Li, Zhang, Sun, Wang, Li, Zhang, and Lin}{Li
  et~al\mbox{.}}{2019}]%
        {li2019annoy}
\bibfield{author}{\bibinfo{person}{Wen Li}, \bibinfo{person}{Ying Zhang},
  \bibinfo{person}{Yifang Sun}, \bibinfo{person}{Wei Wang},
  \bibinfo{person}{Mingjie Li}, \bibinfo{person}{Wenjie Zhang}, {and}
  \bibinfo{person}{Xuemin Lin}.} \bibinfo{year}{2019}\natexlab{}.
\newblock \showarticletitle{Approximate nearest neighbor search on high
  dimensional data—experiments, analyses, and improvement}.
\newblock \bibinfo{journal}{\emph{IEEE Transactions on Knowledge and Data
  Engineering}} \bibinfo{volume}{32}, \bibinfo{number}{8}
  (\bibinfo{year}{2019}), \bibinfo{pages}{1475--1488}.
\newblock


\bibitem[\protect\citeauthoryear{Li, Li, Suhara, Doan, and Tan}{Li
  et~al\mbox{.}}{2020}]%
        {Li20ditto}
\bibfield{author}{\bibinfo{person}{Yuliang Li}, \bibinfo{person}{Jinfeng Li},
  \bibinfo{person}{Yoshihiko Suhara}, \bibinfo{person}{AnHai Doan}, {and}
  \bibinfo{person}{Wang-Chiew Tan}.} \bibinfo{year}{2020}\natexlab{}.
\newblock \showarticletitle{Deep entity matching with pre-trained language
  models}.
\newblock \bibinfo{journal}{\emph{Proceedings of the VLDB Endowment}}
  \bibinfo{volume}{14}, \bibinfo{number}{1} (\bibinfo{year}{2020}),
  \bibinfo{pages}{50–60}.
\newblock


\bibitem[\protect\citeauthoryear{Liaw, Wiener, et~al\mbox{.}}{Liaw
  et~al\mbox{.}}{2002}]%
        {liaw2002randomf}
\bibfield{author}{\bibinfo{person}{Andy Liaw}, \bibinfo{person}{Matthew
  Wiener}, {et~al\mbox{.}}} \bibinfo{year}{2002}\natexlab{}.
\newblock \showarticletitle{Classification and regression by randomForest}.
\newblock \bibinfo{journal}{\emph{R news}} \bibinfo{volume}{2},
  \bibinfo{number}{3} (\bibinfo{year}{2002}), \bibinfo{pages}{18--22}.
\newblock


\bibitem[\protect\citeauthoryear{Lloyd}{Lloyd}{1982}]%
        {lloyd1982kmeans}
\bibfield{author}{\bibinfo{person}{Stuart Lloyd}.}
  \bibinfo{year}{1982}\natexlab{}.
\newblock \showarticletitle{Least squares quantization in PCM}.
\newblock \bibinfo{journal}{\emph{IEEE Transactions on Information Theory}}
  \bibinfo{volume}{28}, \bibinfo{number}{2} (\bibinfo{year}{1982}),
  \bibinfo{pages}{129--137}.
\newblock


\bibitem[\protect\citeauthoryear{Mahdavi and Abedjan}{Mahdavi and
  Abedjan}{2020}]%
        {Mahdavi20baran}
\bibfield{author}{\bibinfo{person}{Mohammad Mahdavi} {and}
  \bibinfo{person}{Ziawasch Abedjan}.} \bibinfo{year}{2020}\natexlab{}.
\newblock \showarticletitle{Baran: Effective error correction via a unified
  context representation and transfer learning}.
\newblock \bibinfo{journal}{\emph{Proceedings of the VLDB Endowment}}
  \bibinfo{volume}{13}, \bibinfo{number}{12} (\bibinfo{year}{2020}),
  \bibinfo{pages}{1948–1961}.
\newblock


\bibitem[\protect\citeauthoryear{Mahdavi, Abedjan, Castro~Fernandez, Madden,
  Ouzzani, Stonebraker, and Tang}{Mahdavi et~al\mbox{.}}{2019}]%
        {raha19mahdavi}
\bibfield{author}{\bibinfo{person}{Mohammad Mahdavi}, \bibinfo{person}{Ziawasch
  Abedjan}, \bibinfo{person}{Raul Castro~Fernandez}, \bibinfo{person}{Samuel
  Madden}, \bibinfo{person}{Mourad Ouzzani}, \bibinfo{person}{Michael
  Stonebraker}, {and} \bibinfo{person}{Nan Tang}.}
  \bibinfo{year}{2019}\natexlab{}.
\newblock \showarticletitle{Raha: A configuration-free error detection system}.
  In \bibinfo{booktitle}{\emph{SIGMOD}}. \bibinfo{pages}{865–882}.
\newblock


\bibitem[\protect\citeauthoryear{Manakul, Liusie, and Gales}{Manakul
  et~al\mbox{.}}{2023}]%
        {manakul2023selfcheckgpt}
\bibfield{author}{\bibinfo{person}{Potsawee Manakul}, \bibinfo{person}{Adian
  Liusie}, {and} \bibinfo{person}{Mark~JF Gales}.}
  \bibinfo{year}{2023}\natexlab{}.
\newblock \showarticletitle{Selfcheckgpt: Zero-resource black-box hallucination
  detection for generative large language models}.
\newblock \bibinfo{journal}{\emph{ArXiv Preprint ArXiv:2303.08896}}
  (\bibinfo{year}{2023}).
\newblock


\bibitem[\protect\citeauthoryear{Mei, Song, Fang, Wei, Fang, and Long}{Mei
  et~al\mbox{.}}{2023}]%
        {Mei23discoveredit}
\bibfield{author}{\bibinfo{person}{Yinan Mei}, \bibinfo{person}{Shaoxu Song},
  \bibinfo{person}{Chenguang Fang}, \bibinfo{person}{Ziheng Wei},
  \bibinfo{person}{Jingyun Fang}, {and} \bibinfo{person}{Jiang Long}.}
  \bibinfo{year}{2023}\natexlab{}.
\newblock \showarticletitle{Discovering editing rules by deep reinforcement
  learning}. In \bibinfo{booktitle}{\emph{ICDE}}. \bibinfo{pages}{355--367}.
\newblock


\bibitem[\protect\citeauthoryear{Miao, Gao, Guo, and Liu}{Miao
  et~al\mbox{.}}{2018}]%
        {miao18incompletesurvey}
\bibfield{author}{\bibinfo{person}{Xiaoye Miao}, \bibinfo{person}{Yunjun Gao},
  \bibinfo{person}{Su Guo}, {and} \bibinfo{person}{Wanqi Liu}.}
  \bibinfo{year}{2018}\natexlab{}.
\newblock \showarticletitle{Incomplete data management: A survey}.
\newblock \bibinfo{journal}{\emph{Frontiers of Computer Science}}
  \bibinfo{volume}{12}, \bibinfo{number}{1} (\bibinfo{year}{2018}),
  \bibinfo{pages}{4–25}.
\newblock


\bibitem[\protect\citeauthoryear{Miao, Wu, Chen, Gao, Wang, and Yin}{Miao
  et~al\mbox{.}}{2021a}]%
        {miao21influence}
\bibfield{author}{\bibinfo{person}{Xiaoye Miao}, \bibinfo{person}{Yangyang Wu},
  \bibinfo{person}{Lu Chen}, \bibinfo{person}{Yunjun Gao}, \bibinfo{person}{Jun
  Wang}, {and} \bibinfo{person}{Jianwei Yin}.}
  \bibinfo{year}{2021}\natexlab{a}.
\newblock \showarticletitle{Efficient and effective data imputation with
  influence functions}.
\newblock \bibinfo{journal}{\emph{Proceedings of the VLDB Endowment}}
  \bibinfo{volume}{15}, \bibinfo{number}{3} (\bibinfo{year}{2021}),
  \bibinfo{pages}{624–632}.
\newblock


\bibitem[\protect\citeauthoryear{Miao, Wu, Chen, Gao, and Yin}{Miao
  et~al\mbox{.}}{2023}]%
        {miao23imputationsurvey}
\bibfield{author}{\bibinfo{person}{Xiaoye Miao}, \bibinfo{person}{Yangyang Wu},
  \bibinfo{person}{Lu Chen}, \bibinfo{person}{Yunjun Gao}, {and}
  \bibinfo{person}{Jianwei Yin}.} \bibinfo{year}{2023}\natexlab{}.
\newblock \showarticletitle{An experimental survey of missing data imputation
  algorithms}.
\newblock \bibinfo{journal}{\emph{IEEE Transactions on Knowledge and Data
  Engineering}} \bibinfo{volume}{35}, \bibinfo{number}{7}
  (\bibinfo{year}{2023}), \bibinfo{pages}{6630--6650}.
\newblock


\bibitem[\protect\citeauthoryear{Miao, Wu, Wang, Gao, Mao, and Yin}{Miao
  et~al\mbox{.}}{2021b}]%
        {miao2021generative}
\bibfield{author}{\bibinfo{person}{Xiaoye Miao}, \bibinfo{person}{Yangyang Wu},
  \bibinfo{person}{Jun Wang}, \bibinfo{person}{Yunjun Gao},
  \bibinfo{person}{Xudong Mao}, {and} \bibinfo{person}{Jianwei Yin}.}
  \bibinfo{year}{2021}\natexlab{b}.
\newblock \showarticletitle{Generative semi-supervised learning for
  multivariate time series imputation}. In \bibinfo{booktitle}{\emph{AAAI}}.
  \bibinfo{pages}{8983--8991}.
\newblock


\bibitem[\protect\citeauthoryear{Mudgal, Li, Rekatsinas, Doan, Park, Krishnan,
  Deep, Arcaute, and Raghavendra}{Mudgal et~al\mbox{.}}{2018}]%
        {Mudgal18deepem}
\bibfield{author}{\bibinfo{person}{Sidharth Mudgal}, \bibinfo{person}{Han Li},
  \bibinfo{person}{Theodoros Rekatsinas}, \bibinfo{person}{AnHai Doan},
  \bibinfo{person}{Youngchoon Park}, \bibinfo{person}{Ganesh Krishnan},
  \bibinfo{person}{Rohit Deep}, \bibinfo{person}{Esteban Arcaute}, {and}
  \bibinfo{person}{Vijay Raghavendra}.} \bibinfo{year}{2018}\natexlab{}.
\newblock \showarticletitle{Deep learning for entity matching: A design space
  exploration}. In \bibinfo{booktitle}{\emph{SIGMOD}}.
  \bibinfo{pages}{19–34}.
\newblock


\bibitem[\protect\citeauthoryear{Neutatz, Mahdavi, and Abedjan}{Neutatz
  et~al\mbox{.}}{2019}]%
        {Neutatz19ed2}
\bibfield{author}{\bibinfo{person}{Felix Neutatz}, \bibinfo{person}{Mohammad
  Mahdavi}, {and} \bibinfo{person}{Ziawasch Abedjan}.}
  \bibinfo{year}{2019}\natexlab{}.
\newblock \showarticletitle{ED2: A case for active learning in error
  detection}. In \bibinfo{booktitle}{\emph{CIKM}}.
  \bibinfo{pages}{2249–2252}.
\newblock


\bibitem[\protect\citeauthoryear{Ng, Jordan, and Weiss}{Ng
  et~al\mbox{.}}{2001}]%
        {ng2001spectral}
\bibfield{author}{\bibinfo{person}{Andrew Ng}, \bibinfo{person}{Michael
  Jordan}, {and} \bibinfo{person}{Yair Weiss}.}
  \bibinfo{year}{2001}\natexlab{}.
\newblock \showarticletitle{On spectral clustering: Analysis and an algorithm}.
  In \bibinfo{booktitle}{\emph{NeurIPS}}. \bibinfo{pages}{849--856}.
\newblock


\bibitem[\protect\citeauthoryear{Ouyang, Wu, Jiang, Almeida, Wainwright,
  Mishkin, Zhang, Agarwal, Slama, Ray, Schulman, Hilton, Kelton, Miller,
  Simens, Askell, Welinder, Christiano, Leike, and Lowe}{Ouyang
  et~al\mbox{.}}{2022}]%
        {ouyang2022trainingllm}
\bibfield{author}{\bibinfo{person}{Long Ouyang}, \bibinfo{person}{Jeff Wu},
  \bibinfo{person}{Xu Jiang}, \bibinfo{person}{Diogo Almeida},
  \bibinfo{person}{Carroll~L. Wainwright}, \bibinfo{person}{Pamela Mishkin},
  \bibinfo{person}{Chong Zhang}, \bibinfo{person}{Sandhini Agarwal},
  \bibinfo{person}{Katarina Slama}, \bibinfo{person}{Alex Ray},
  \bibinfo{person}{John Schulman}, \bibinfo{person}{Jacob Hilton},
  \bibinfo{person}{Fraser Kelton}, \bibinfo{person}{Luke Miller},
  \bibinfo{person}{Maddie Simens}, \bibinfo{person}{Amanda Askell},
  \bibinfo{person}{Peter Welinder}, \bibinfo{person}{Paul Christiano},
  \bibinfo{person}{Jan Leike}, {and} \bibinfo{person}{Ryan Lowe}.}
  \bibinfo{year}{2022}\natexlab{}.
\newblock \showarticletitle{Training language models to follow instructions
  with human feedback}.
\newblock \bibinfo{journal}{\emph{ArXiv Preprint ArXiv:2203.02155}}
  (\bibinfo{year}{2022}).
\newblock


\bibitem[\protect\citeauthoryear{Paganelli, Buono, Baraldi, and
  Guerra}{Paganelli et~al\mbox{.}}{2022}]%
        {Paganelli22bertER}
\bibfield{author}{\bibinfo{person}{Matteo Paganelli},
  \bibinfo{person}{Francesco~Del Buono}, \bibinfo{person}{Andrea Baraldi},
  {and} \bibinfo{person}{Francesco Guerra}.} \bibinfo{year}{2022}\natexlab{}.
\newblock \showarticletitle{Analyzing how BERT performs entity matching}.
\newblock \bibinfo{journal}{\emph{Proceedings of the VLDB Endowment}}
  \bibinfo{volume}{15}, \bibinfo{number}{8} (\bibinfo{year}{2022}),
  \bibinfo{pages}{1726–1738}.
\newblock


\bibitem[\protect\citeauthoryear{Papenbrock, Ehrlich, Marten, Neubert, Rudolph,
  Sch\"{o}nberg, Zwiener, and Naumann}{Papenbrock et~al\mbox{.}}{2015}]%
        {Papenbrock15FdEvaluation}
\bibfield{author}{\bibinfo{person}{Thorsten Papenbrock}, \bibinfo{person}{Jens
  Ehrlich}, \bibinfo{person}{Jannik Marten}, \bibinfo{person}{Tommy Neubert},
  \bibinfo{person}{Jan-Peer Rudolph}, \bibinfo{person}{Martin Sch\"{o}nberg},
  \bibinfo{person}{Jakob Zwiener}, {and} \bibinfo{person}{Felix Naumann}.}
  \bibinfo{year}{2015}\natexlab{}.
\newblock \showarticletitle{Functional dependency discovery: An experimental
  evaluation of seven algorithms}.
\newblock \bibinfo{journal}{\emph{Proceedings of the VLDB Endowment}}
  \bibinfo{volume}{8}, \bibinfo{number}{10} (\bibinfo{year}{2015}),
  \bibinfo{pages}{1082–1093}.
\newblock


\bibitem[\protect\citeauthoryear{Pena, de~Almeida, and Naumann}{Pena
  et~al\mbox{.}}{2019}]%
        {dcfinder19Pena}
\bibfield{author}{\bibinfo{person}{Eduardo H.~M. Pena},
  \bibinfo{person}{Eduardo~C. de Almeida}, {and} \bibinfo{person}{Felix
  Naumann}.} \bibinfo{year}{2019}\natexlab{}.
\newblock \showarticletitle{Discovery of approximate (and exact) denial
  constraints}.
\newblock \bibinfo{journal}{\emph{Proceedings of the VLDB Endowment}}
  \bibinfo{volume}{13}, \bibinfo{number}{3} (\bibinfo{year}{2019}),
  \bibinfo{pages}{266–278}.
\newblock


\bibitem[\protect\citeauthoryear{Pham, Knoblock, Chen, Vu, and Pujara}{Pham
  et~al\mbox{.}}{2021}]%
        {Pham21spade}
\bibfield{author}{\bibinfo{person}{Minh Pham}, \bibinfo{person}{Craig~A.
  Knoblock}, \bibinfo{person}{Muhao Chen}, \bibinfo{person}{Binh Vu}, {and}
  \bibinfo{person}{Jay Pujara}.} \bibinfo{year}{2021}\natexlab{}.
\newblock \showarticletitle{SPADE: A semi-supervised probabilistic approach for
  detecting errors in tables}. In \bibinfo{booktitle}{\emph{IJCAI}}.
  \bibinfo{pages}{3543–3551}.
\newblock


\bibitem[\protect\citeauthoryear{Rekatsinas, Chu, Ilyas, and R\'{e}}{Rekatsinas
  et~al\mbox{.}}{2017}]%
        {Rekatsinas17holoclean}
\bibfield{author}{\bibinfo{person}{Theodoros Rekatsinas}, \bibinfo{person}{Xu
  Chu}, \bibinfo{person}{Ihab~F. Ilyas}, {and} \bibinfo{person}{Christopher
  R\'{e}}.} \bibinfo{year}{2017}\natexlab{}.
\newblock \showarticletitle{HoloClean: Holistic data repairs with probabilistic
  inference}.
\newblock \bibinfo{journal}{\emph{Proceedings of the VLDB Endowment}}
  \bibinfo{volume}{10}, \bibinfo{number}{11} (\bibinfo{year}{2017}),
  \bibinfo{pages}{1190–1201}.
\newblock


\bibitem[\protect\citeauthoryear{Rezig, Ouzzani, Aref, Elmagarmid, Mahmood, and
  Stonebraker}{Rezig et~al\mbox{.}}{2021}]%
        {Rezig21horizon}
\bibfield{author}{\bibinfo{person}{El~Kindi Rezig}, \bibinfo{person}{Mourad
  Ouzzani}, \bibinfo{person}{Walid~G. Aref}, \bibinfo{person}{Ahmed~K.
  Elmagarmid}, \bibinfo{person}{Ahmed~R. Mahmood}, {and}
  \bibinfo{person}{Michael Stonebraker}.} \bibinfo{year}{2021}\natexlab{}.
\newblock \showarticletitle{Horizon: Scalable dependency-driven data cleaning}.
\newblock \bibinfo{journal}{\emph{Proceedings of the VLDB Endowment}}
  \bibinfo{volume}{14}, \bibinfo{number}{11} (\bibinfo{year}{2021}),
  \bibinfo{pages}{2546–2554}.
\newblock


\bibitem[\protect\citeauthoryear{Rissanen}{Rissanen}{1978}]%
        {rissanen1978modeling}
\bibfield{author}{\bibinfo{person}{Jorma Rissanen}.}
  \bibinfo{year}{1978}\natexlab{}.
\newblock \showarticletitle{Modeling by shortest data description}.
\newblock \bibinfo{journal}{\emph{Automatica}} \bibinfo{volume}{14},
  \bibinfo{number}{5} (\bibinfo{year}{1978}), \bibinfo{pages}{465--471}.
\newblock


\bibitem[\protect\citeauthoryear{Shin, Wu, Wang, De~Sa, Zhang, and R\'{e}}{Shin
  et~al\mbox{.}}{2015}]%
        {Shin15deepdive}
\bibfield{author}{\bibinfo{person}{Jaeho Shin}, \bibinfo{person}{Sen Wu},
  \bibinfo{person}{Feiran Wang}, \bibinfo{person}{Christopher De~Sa},
  \bibinfo{person}{Ce Zhang}, {and} \bibinfo{person}{Christopher R\'{e}}.}
  \bibinfo{year}{2015}\natexlab{}.
\newblock \showarticletitle{Incremental knowledge base construction using
  deepdive}.
\newblock \bibinfo{journal}{\emph{Proceedings of the VLDB Endowment}}
  \bibinfo{volume}{8}, \bibinfo{number}{11} (\bibinfo{year}{2015}),
  \bibinfo{pages}{1310–1321}.
\newblock


\bibitem[\protect\citeauthoryear{Singh, Meduri, Elmagarmid, Madden, Papotti,
  Quian{\'{e}}{-}Ruiz, Solar{-}Lezama, and Tang}{Singh et~al\mbox{.}}{2017}]%
        {Singh2017EMRule}
\bibfield{author}{\bibinfo{person}{Rohit Singh},
  \bibinfo{person}{Venkata~Vamsikrishna Meduri}, \bibinfo{person}{Ahmed~K.
  Elmagarmid}, \bibinfo{person}{Samuel Madden}, \bibinfo{person}{Paolo
  Papotti}, \bibinfo{person}{Jorge{-}Arnulfo Quian{\'{e}}{-}Ruiz},
  \bibinfo{person}{Armando Solar{-}Lezama}, {and} \bibinfo{person}{Nan Tang}.}
  \bibinfo{year}{2017}\natexlab{}.
\newblock \showarticletitle{Generating concise entity matching rules}. In
  \bibinfo{booktitle}{\emph{SIGMOD}}. \bibinfo{pages}{1635--1638}.
\newblock


\bibitem[\protect\citeauthoryear{Song, Zhu, and Wang}{Song
  et~al\mbox{.}}{2016}]%
        {Song16cvtolerant}
\bibfield{author}{\bibinfo{person}{Shaoxu Song}, \bibinfo{person}{Han Zhu},
  {and} \bibinfo{person}{Jianmin Wang}.} \bibinfo{year}{2016}\natexlab{}.
\newblock \showarticletitle{Constraint-variance tolerant data repairing}. In
  \bibinfo{booktitle}{\emph{SIGMOD}}. \bibinfo{pages}{877–892}.
\newblock


\bibitem[\protect\citeauthoryear{Tang, Deng, and Huang}{Tang
  et~al\mbox{.}}{2015}]%
        {tang2015extreme}
\bibfield{author}{\bibinfo{person}{Jiexiong Tang}, \bibinfo{person}{Chenwei
  Deng}, {and} \bibinfo{person}{Guang-Bin Huang}.}
  \bibinfo{year}{2015}\natexlab{}.
\newblock \showarticletitle{Extreme learning machine for multilayer
  perceptron}.
\newblock \bibinfo{journal}{\emph{IEEE Transactions on Neural Networks and
  Learning Systems}} \bibinfo{volume}{27}, \bibinfo{number}{4}
  (\bibinfo{year}{2015}), \bibinfo{pages}{809--821}.
\newblock


\bibitem[\protect\citeauthoryear{Tang, Yang, Fan, Cao, Luo, and Halevy}{Tang
  et~al\mbox{.}}{2023}]%
        {tang2023verifai}
\bibfield{author}{\bibinfo{person}{Nan Tang}, \bibinfo{person}{Chenyu Yang},
  \bibinfo{person}{Ju Fan}, \bibinfo{person}{Lei Cao}, \bibinfo{person}{Yuyu
  Luo}, {and} \bibinfo{person}{Alon Halevy}.} \bibinfo{year}{2023}\natexlab{}.
\newblock \showarticletitle{VerifAI: verified generative AI}.
\newblock \bibinfo{journal}{\emph{ArXiv Preprint ArXiv:2307.02796}}
  (\bibinfo{year}{2023}).
\newblock


\bibitem[\protect\citeauthoryear{Wang and He}{Wang and He}{2019}]%
        {Wang19unidetect}
\bibfield{author}{\bibinfo{person}{Pei Wang} {and} \bibinfo{person}{Yeye He}.}
  \bibinfo{year}{2019}\natexlab{}.
\newblock \showarticletitle{Uni-detect: A unified approach to automated error
  detection in tables}. In \bibinfo{booktitle}{\emph{SIGMOD}}.
  \bibinfo{pages}{811–828}.
\newblock


\bibitem[\protect\citeauthoryear{Wang, Montoya, Munechika, Yang, Hoover, and
  Chau}{Wang et~al\mbox{.}}{2022}]%
        {wang2022diffusiondb}
\bibfield{author}{\bibinfo{person}{Zijie~J Wang}, \bibinfo{person}{Evan
  Montoya}, \bibinfo{person}{David Munechika}, \bibinfo{person}{Haoyang Yang},
  \bibinfo{person}{Benjamin Hoover}, {and} \bibinfo{person}{Duen~Horng Chau}.}
  \bibinfo{year}{2022}\natexlab{}.
\newblock \showarticletitle{Diffusiondb: A large-scale prompt gallery dataset
  for text-to-image generative models}.
\newblock \bibinfo{journal}{\emph{ArXiv Preprint arXiv:2210.14896}}
  (\bibinfo{year}{2022}).
\newblock


\bibitem[\protect\citeauthoryear{Wei, Wang, Schuurmans, Bosma, Ichter, Xia,
  Chi, Le, and Zhou}{Wei et~al\mbox{.}}{2022}]%
        {wei2022chain}
\bibfield{author}{\bibinfo{person}{Jason Wei}, \bibinfo{person}{Xuezhi Wang},
  \bibinfo{person}{Dale Schuurmans}, \bibinfo{person}{Maarten Bosma},
  \bibinfo{person}{Brian Ichter}, \bibinfo{person}{Fei Xia},
  \bibinfo{person}{Ed Chi}, \bibinfo{person}{Quoc Le}, {and}
  \bibinfo{person}{Denny Zhou}.} \bibinfo{year}{2022}\natexlab{}.
\newblock \showarticletitle{Chain-of-thought prompting elicits reasoning in
  large language models}.
\newblock \bibinfo{journal}{\emph{NeurIPS}} (\bibinfo{year}{2022}),
  \bibinfo{pages}{24824--24837}.
\newblock


\bibitem[\protect\citeauthoryear{Wu, Chaba, Sawlani, Chu, and
  Thirumuruganathan}{Wu et~al\mbox{.}}{2020}]%
        {Wu20ZeroER}
\bibfield{author}{\bibinfo{person}{Renzhi Wu}, \bibinfo{person}{Sanya Chaba},
  \bibinfo{person}{Saurabh Sawlani}, \bibinfo{person}{Xu Chu}, {and}
  \bibinfo{person}{Saravanan Thirumuruganathan}.}
  \bibinfo{year}{2020}\natexlab{}.
\newblock \showarticletitle{ZeroER: Entity resolution using zero labeled
  examples}. In \bibinfo{booktitle}{\emph{SIGMOD}}.
  \bibinfo{pages}{1149–1164}.
\newblock


\bibitem[\protect\citeauthoryear{Wu, Miao, Li, He, Yuan, and Yin}{Wu
  et~al\mbox{.}}{2023}]%
        {wu23imputationtoolbox}
\bibfield{author}{\bibinfo{person}{Yangyang Wu}, \bibinfo{person}{Xiaoye Miao},
  \bibinfo{person}{Zilinghan Li}, \bibinfo{person}{Shilan He},
  \bibinfo{person}{Xinkai Yuan}, {and} \bibinfo{person}{Jianwei Yin}.}
  \bibinfo{year}{2023}\natexlab{}.
\newblock \showarticletitle{An efficient generative data imputation toolbox
  with adversarial learning}. In \bibinfo{booktitle}{\emph{ICDE}}.
  \bibinfo{pages}{3651--3654}.
\newblock


\bibitem[\protect\citeauthoryear{Yakout, Berti-\'{E}quille, and
  Elmagarmid}{Yakout et~al\mbox{.}}{2013}]%
        {Yakout13scared}
\bibfield{author}{\bibinfo{person}{Mohamed Yakout}, \bibinfo{person}{Laure
  Berti-\'{E}quille}, {and} \bibinfo{person}{Ahmed~K. Elmagarmid}.}
  \bibinfo{year}{2013}\natexlab{}.
\newblock \showarticletitle{Don't be scared: Use scalable automatic repairing
  with maximal likelihood and bounded changes}. In
  \bibinfo{booktitle}{\emph{SIGMOD}}. \bibinfo{pages}{553–564}.
\newblock


\bibitem[\protect\citeauthoryear{Zhang, Chai, Doan, Koutris, and Arcaute}{Zhang
  et~al\mbox{.}}{2020}]%
        {Zhang2020ManuallyDE}
\bibfield{author}{\bibinfo{person}{Haojun Zhang}, \bibinfo{person}{Chengliang
  Chai}, \bibinfo{person}{AnHai Doan}, \bibinfo{person}{Paris Koutris}, {and}
  \bibinfo{person}{Esteban Arcaute}.} \bibinfo{year}{2020}\natexlab{}.
\newblock \showarticletitle{Manually detecting errors for data cleaning using
  adaptive crowdsourcing strategies}. In \bibinfo{booktitle}{\emph{EDBT}}.
  \bibinfo{pages}{311--322}.
\newblock


\bibitem[\protect\citeauthoryear{Zhao, Zhou, Li, Tang, Wang, Hou, Min, Zhang,
  Zhang, Dong, et~al\mbox{.}}{Zhao et~al\mbox{.}}{2023}]%
        {zhao2023survey}
\bibfield{author}{\bibinfo{person}{Wayne~Xin Zhao}, \bibinfo{person}{Kun Zhou},
  \bibinfo{person}{Junyi Li}, \bibinfo{person}{Tianyi Tang},
  \bibinfo{person}{Xiaolei Wang}, \bibinfo{person}{Yupeng Hou},
  \bibinfo{person}{Yingqian Min}, \bibinfo{person}{Beichen Zhang},
  \bibinfo{person}{Junjie Zhang}, \bibinfo{person}{Zican Dong},
  {et~al\mbox{.}}} \bibinfo{year}{2023}\natexlab{}.
\newblock \showarticletitle{A survey of large language models}.
\newblock \bibinfo{journal}{\emph{ArXiv Preprint ArXiv:2303.18223}}
  (\bibinfo{year}{2023}).
\newblock


\end{thebibliography}

\end{document}